\documentclass{aa}

\usepackage{graphicx}
\usepackage{siunitx}
\usepackage{textcomp}
\usepackage{txfonts}
\usepackage[colorlinks=true,linkcolor=blue,citecolor=blue,urlcolor=blue]{hyperref}
\usepackage{tikz}
\usepackage{float}
\usepackage{import}

   \def\HeI{\ion{He}{i}\,$\lambda$\,}
 \def\HeII{\ion{He}{ii}\,$\lambda$\,}  
  \def\CIII{\ion{C}{iii}\,$\lambda$\,} \def\CIV{\ion{C}{iv}\,$\lambda$\,} \def\CIVd{\ion{C}{iv}\,$\lambda\lambda$\,}
    
 \def\NIII{\ion{N}{iii}\,$\lambda$\,}
  \def\NIIId{\ion{N}{iii}\,$\lambda\lambda$\,}
  \def\NIV{\ion{N}{iv}\,$\lambda$\,}
  \def\NIVd{\ion{N}{iv}\,$\lambda\lambda$\,}
    \def\NIVt{\ion{N}{iv}\,$\lambda\lambda\lambda$\,}
     \def\NVd{\ion{N}{v}\,$\lambda\lambda$\,}
   
    \def\OIVt{\ion{O}{iv}\,$\lambda\lambda\lambda$\,}
  \def\OVId{\ion{O}{vi}\,$\lambda\lambda$\,}

  \def\SiIVd{\ion{Si}{iv}\,$\lambda\lambda$\,}

\newcommand{\msun}{\ifmmode M_{\odot} \else M$_{\odot}$\fi}
\newcommand{\rsun}{\ifmmode R_{\odot} \else R$_{\odot}$\fi}
\newcommand{\lsun}{\ifmmode L_{\odot} \else L$_{\odot}$\fi}
\newcommand{\zsun}{\ifmmode Z_{\odot} \else $Z_{\odot}$\fi}
\newcommand{\xsun}{\ifmmode X_{\odot} \else $X_{\odot}$\fi}
\newcommand{\msunpyr}{\ifmmode{\,M_{\odot}\,\mbox{yr}^{-1}} \else{ M$_{\odot}$/yr}\fi}
\newcommand{\velo}{\ifmmode\varv\else$\varv$\fi}
\newcommand{\vinf}{\ifmmode\velo_\infty\else$\velo_\infty$\fi}
\newcommand{\rgal}{\ifmmode \,R_{\mathrm{gal}} \else R$_{\mathrm{gal}}$\fi}

\begin{document}

 \title{A low-metallicity massive contact binary undergoing slow Case A mass transfer: A detailed spectroscopic and orbital analysis of SSN~7 in NGC~346 in the SMC \thanks{Based on observations with the NASA/ESA {\em Hubble Space Telescope}, which is operated by the Association of Universities for Research in Astronomy, Inc., under NASA contract NAS 5-2655. Also based on observations collected at the European Organisation for Astronomical Research in the Southern Hemisphere.}}

\author{M. J. Rickard\inst{1,2}
\and
D. Pauli\inst{1}}

\institute{Institut f\"{u}r Physik und Astronomie, Universit\"{a}t Potsdam, Karl-Liebknecht-Str. 24/25, D-14476 Potsdam, Germany\\
 \email{matthew.rickard.18@ucl.ac.uk}
 \and
Department of Physics and Astronomy, University College London, Gower Street, London WC1E 6BT, UK}

\date{Received 1 February 2023 / Accepted 26 March 2023}

\abstract
{Most massive stars are believed to be born in close binary systems where they can exchange mass, which impacts the evolution of both binary components. Their evolution is of great interest in the search for the progenitors of gravitational waves. However, there are unknowns in the physics of mass transfer as observational examples are rare, especially at low metallicity. Nearby low-metallicity environments are particularly interesting hunting grounds for interacting systems as they act as the closest proxy for the early universe where we can resolve individual stars.}
{Using multi-epoch spectroscopic data, we complete a consistent spectral and orbital analysis of the early-type massive binary SSN~7 hosting a ON3\,If$^\ast$+O5.5\,V((f)) star. Using these detailed results, we  constrain an evolutionary scenario that can help us to understand binary evolution in low metallicity.}
{We were able to derive reliable radial velocities of the two components from the multi-epoch data, which were used to constrain the orbital parameters. The spectroscopic data covers the UV, optical, and near-IR, allowing a consistent analysis with the stellar atmosphere code, PoWR. Given the stellar and orbital parameters, we interpreted the results using binary evolutionary models.}
{The two stars in the system have comparable luminosities of ${\log (L_1/\lsun) = 5.75}$ and ${\log (L_2/\lsun) = 5.78}$ for the primary and secondary, respectively, but have different temperatures (${T_1=\SI{43.6}{kK}}$ and ${T_2=\SI{38.7}{kK}}$). The primary ($32\msun$) is less massive than the secondary ($55\,\msun$), suggesting mass exchange. The mass estimates are confirmed by the orbital analysis. The revisited orbital period is $\SI{3}{d}$. Our evolutionary models also predict mass exchange. Currently, the system is a contact binary undergoing a slow Case A phase, making it the most massive Algol-like system yet discovered.} 
% conclusions heading (optional), leave it empty if necessary-surpressed by including blank {} 
{ Following the initial mass function, massive stars are rare, and to find them in an Algol-like configuration is even more unlikely. To date, no comparable system to SSN~7 has been found, making it a unique object to study the efficiency of mass transfer in massive star binaries. This example increases our understanding of massive star binary evolution and the formation of gravitational wave progenitors.}

%%% WORDS 299 - NO MORE WORDS LEFT !!!! %%%%%

 \titlerunning{SSN 7 Paper}
\maketitle

\section{Introduction}

\begin{figure*}
        \center
        \begin{tikzpicture}
        \node[anchor=south west,inner sep=0] at (0,0) {\includegraphics[angle=0, origin=c, trim={0.5cm 0.02cm 0.02cm 1.3cm},clip, width = 0.49\hsize]{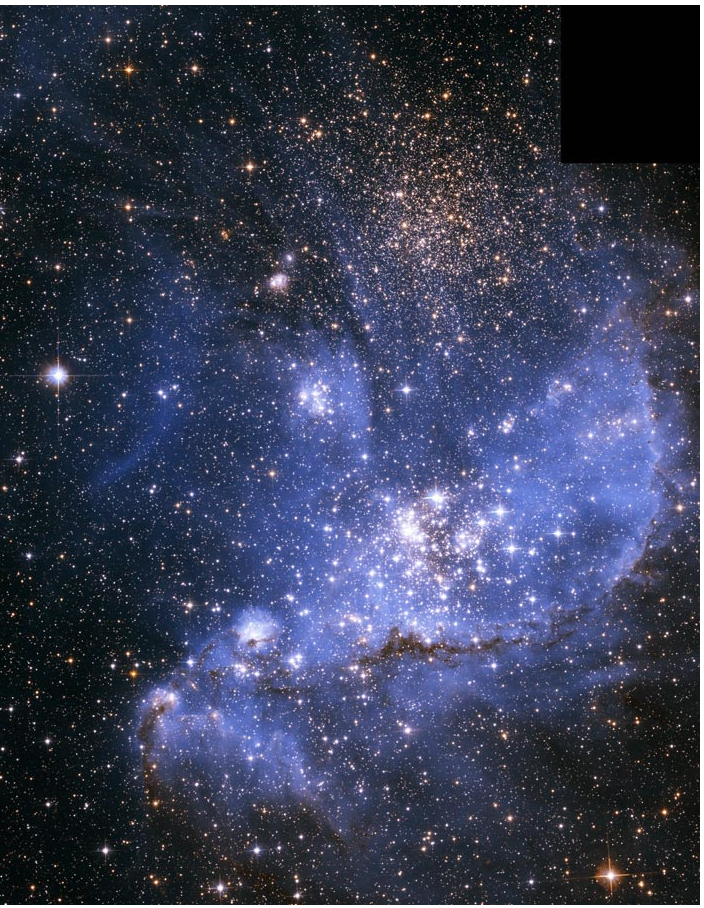}};
        \draw[green,line width = 0.5mm] (6.07,5.41) rectangle (6.32,5.66);
        \draw[red,line width = 0.5mm] (5.17,5.615) rectangle (5.42,5.865);
        % Draw direction on sky
        \draw[white,line width=0.35mm, -latex] (8.5,0.5) -- (8.5, 1.25);
        \draw[white] (8.5, 1.4) node {\tiny N};
        \draw[white,line width=0.35mm, -latex] (8.5,0.5) -- (7.75, 0.5);
        \draw[white] (7.6, 0.5) node {\tiny E};
        \end{tikzpicture}
        \begin{tikzpicture}
        \node[anchor=south west,inner sep=0] at (0,0) {\includegraphics[angle=0, origin=c, trim={0.5cm 0.02cm 0.02cm 1.3cm},clip, width = 0.49\hsize]{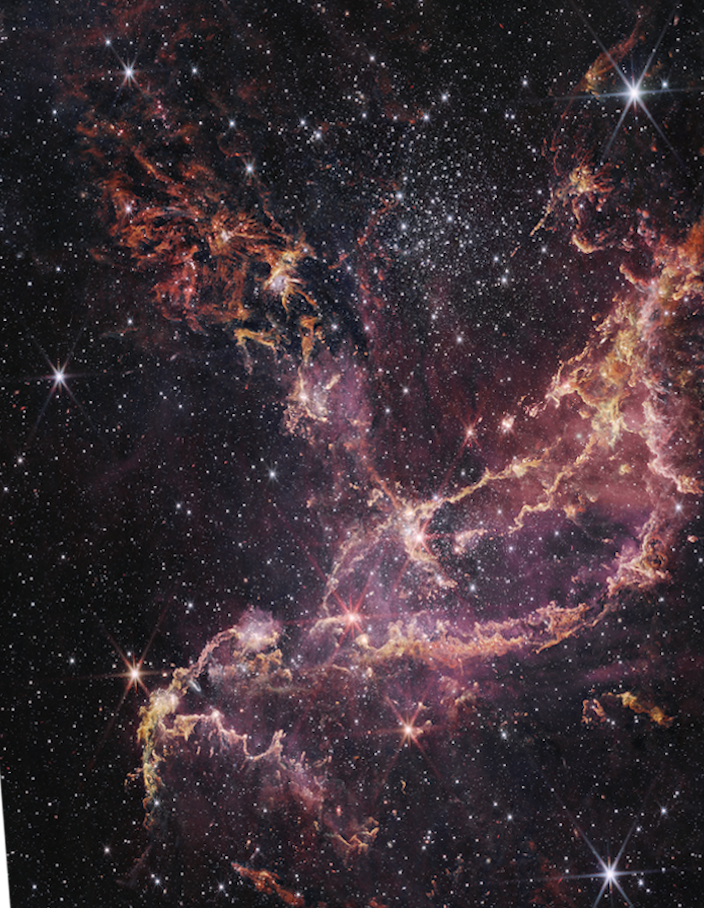}};
        \draw[green,line width = 0.5mm] (6.07,5.41) rectangle (6.32,5.66);
        \draw[red,line width = 0.5mm] (5.17,5.615) rectangle (5.42,5.865);
        % Draw direction on sky
        \draw[white,line width=0.35mm, -latex] (8.5,0.5) -- (8.5, 1.25);
        \draw[white] (8.5, 1.4) node {\tiny N};
        \draw[white,line width=0.35mm, -latex] (8.5,0.5) -- (7.75, 0.5);
        \draw[white] (7.6, 0.5) node {\tiny E};
        \end{tikzpicture}
        \caption{Optical (left) and infrared (right) image of NGC346 in the SMC. Each image shows an area on the sky of approximately 
        $3.5'\times 3.5'$ ($\approx 62$\,pc$\,\times$ 62\,pc at 
        d=61\,kpc). The positions of the main ionising sources of NGC~346 (see Sect.~\ref{sec:ionisation}), namely SSN\,7 (ON3\,If$^\ast$+O5.5\,V((f))) and SSN\,9 (O2\,III(f*)), are highlighted by red and green squares, respectively. The false-colour optical image was taken with the HST
        ACS. Credit: A.\,Nota (ESA/STScI). The false-colour infrared image was taken with the James Webb Space Telescopes (JWST) Near Infrared Camera (NIRCam). Credits: NASA, ESA, CSA, O. Jones (UK ATC), G. De Marchi (ESTEC), and M. Meixner (USRA). Image processing: A. Pagan (STScI), N. Habel (USRA), L. Lenkic (USRA), and L. Chu (NASA/Ames). }
        \label{fig:NGC_346_Obs}
    \end{figure*}
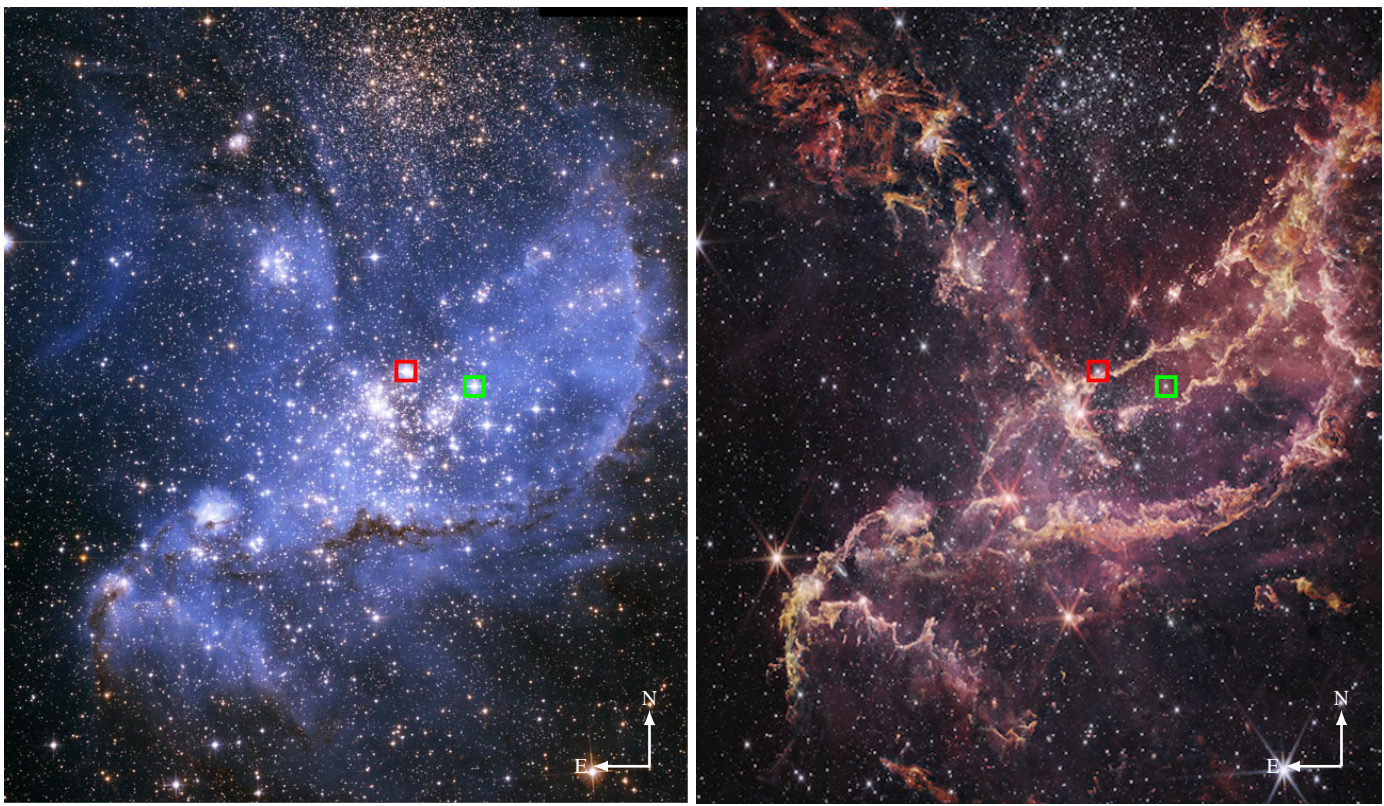

    Massive stars ($\mathrm{M} \gtrsim 8 \mathrm{M}_\odot$) have a pivotal role in the evolution of Galaxies. They influence their environments throughout their entire evolution via multiple feedback mechanisms. Core hydrogen-burning massive stars are hot, leading to ionisation feedback \citep{1999RvMP...71..173H, 2002ApJ...566..302M} and forming spectacular \ion{H}{II} regions. Their high ultraviolet (UV) luminosity drives powerful stellar winds, depositing mechanical feedback and heavy elements into the interstellar medium \citep{2013MNRAS.431.1337R}. When a massive star evolves, it expands and becomes a red supergiant (RSG) and cools down. These stars still have strong winds, powered by radiation pressure on dust grains. Most massive stars are believed to end their lives with a supernova explosion that rapidly deposits substantial amounts of energy in the surrounding medium \citep{2013MNRAS.431.1337R}. During these events, neutron stars \citep[NSs, ][]{2018RPPh...81e6902B, 2018EPJP..133..445V} or black holes \citep[BHs, ][]{2011ApJ...742...84O, 2021Sci...371.1046M} are formed. All these feedback mechanisms make massive stars, which are the drivers of the chemical evolution of the interstellar medium \citep{1957RvMP...29..547B, 2010ApJ...710.1557P, 2011PrPNP..66..346T, 2017Natur.551...80K, 2019PrPNP.107..109K} and the arbitrators of stellar formation;  massive stars can initiate or abort star formation \citep{2005ApJ...626..864M}.

    Massive stars  preferentially form in multiple systems \citep{2012Sci...337..444S, 2013A&A...550A.107S, 2014ApJS..215...15S, 2017ApJS..230...15M}. The evolution of stars in binary systems differs from single star evolution as the two components can interact and exchange mass, for instance  when one star in the binary (i.e. the more massive one) expands and fills its Roche lobe \citep{1994A&A...290..119P, 1998A&ARv...9...63V, 2001ApJ...552..664N,  2014ARA&A..52..487S}. This mass can then, depending on other factors such as rotation, accrete onto the companion star. Due to the fate of massive stars as either NSs or BHs, close massive star binaries are the progenitors of gravitational wave observations \citep{2016ApJ...832L..21A, 2019PhRvD.100l2002A}. However, the final mass of a star dictates its fate \citep{2003ApJ...591..288H}, and in a binary system the final mass depends on the previous evolution and interaction \citep{2020A&A...634A..79S, 2021ApJ...922..177M, 2022A&A...667A..58P}.

    The duration of mass transfer strongly depends on the evolutionary stage of the star \citep{1967AcA....17..355P, 1980A&A....88..230V, 1999A&A...350..148W}. If the expanding star is still core H-burning, it   first drops out of nuclear equilibrium, making the first phase of mass transfer rapid as it takes place on the thermal timescale (fast Case A). As soon as the star regains nuclear equilibrium, it   continues to expand on the nuclear timescale, overflowing its Roche lobe and transferring mass to the companion star (slow Case A). If the mass donor during the slow Case A mass transfer is less massive than the mass gainer, the systems are called Algol-like system \citep[named for $\beta\,\mathrm{Per}$, ][]{1971ARA&A...9..183P, 1989SSRv...50....1B, 2022AJ....163..203L}. Alternatively, stars that have finished core H-burning expand on the  thermal timescale, filling their Roche lobe quickly, and lose a large fraction of their envelope (Case B). As most of the mass transfer events are on short timescales, it is practically impossible to catch a system during the fast Case A or Case B mass transfer phase. As slow Case A mass transfer occurs on the nuclear timescale, it is possible to find observational counterparts.

    Metallicity ($Z$) has many effects on stellar evolution. For instance, stars are more compact and hotter, and line-driven winds get weaker for lower metallicity stars \citep{2007A&A...465.1003M}. Low-$Z$ environments are of particular interest when looking to understand the origin of gravitational waves as the observed mergers take place in high-redshift galaxies. However, it is only possible to study low-Z stars in nearby irregular dwarf galaxies, such as the Small Magellanic Cloud (SMC, ${Z_\mathrm{SMC}=1/7\,\zsun}$) as these stars can be resolved. To date, there are few detailed studies on massive stars in pre- and post-interaction binaries in the SMC  \citep{2007A&A...467.1181D, 2022A&A...659A...9P}.

    In this paper we  remedy this by studying in detail one of the most massive binaries in the low-$Z$ SMC galaxy, namely SSN~7 (\citealt{2007AJ....133...44S}, also known as MPG~435; \citealt{1989AJ.....98.1305M}, or NMC~26; \citealt{1986PASP...98.1133N}). Our target is located within the core of the young cluster NGC~346 \citep{2019AA...626A..50D}, which contains a large population of O stars \citep{1986PASP...98.1133N, 1989AJ.....98.1305M, 2022RICKARDa}. The cluster is nearby \citep[$d=61$\,kpc,][]{2005MNRAS.357..304H} and has been studied several times with different instruments. Previous studies have categorised SSN~7 as a spectroscopic binary with two moving components \citep[SB2][]{2019AA...626A..50D}, but it has never been analysed consistently with a stellar atmosphere code.

    The archival data, including spectra that cover the full range of the electromagnetic spectrum as well as photometric observations, are detailed in Section~\ref{sec:obs}. In Section~\ref{sec:orb_analysis} we explain  how the spectroscopic data is used to obtain radial velocity (RV) shifts, used to conduct an orbital analysis. Furthermore, this data is used in Section~\ref{sec:spec_analysis} to make a consistent spectroscopic analysis, yielding stellar and wind parameters. The empirically derived stellar and orbital parameters are put into an evolutionary context in Section~\ref{sec:stellar_evo_mod} and its implications are discussed in Section~\ref{sec:discussion}.

    \begin{table*}
        \centering
        \small
        \caption{List of spectroscopic observations of NGC~346 SSN~7.}
                \begin{tabular}{ccccccccc}\hline \hline \rule{0cm}{2.2ex}%
                        ID & Instrument & Wavelength &  Res. power & Exp. time  &  MJD$^{(a)}$ & \multicolumn{1}{c}{Program ID} & \multicolumn{1}{c}{PI} \\
                        & & [\AA] & [$\lambda \ / \ \delta \lambda$] &[s] & [d]  & &\\
                        \hline \rule{0cm}{2.4ex}%
                 1  & FUSE            & 920-1180   & $\sim20,000$ & 4464 & 52146.44 & P203           & K. Sembach      \\
2  & FUSE            & 920-1180   & $\sim20\,000$ & 6386 & 51828.57 & P117           & J. B. Hutchings \\
3  & HST FOS G130H   & 1140-1605  & $\sim1150$   & 2803 & 48948.09 & GO4110         & F.-P. Kudritzki \\
4  & HST FOS G130H   & 1140-1605  & $\sim1150$   & 2803 & 49287.81 & GO4110         & F.-P. Kudritzki \\
5  & HST STIS G140L  & 1160-1705  & $\sim1000$   & 225  & 58482.89 & GO15112        & L. Oskinova     \\
6  & HST FOS G190H   & 1560-2340  & $\sim1150$ & 1215 & 48948.15 & GO4110         & F.-P. Kudritzki \\
7  & HST FOS G190H   & 1560-2340  & $\sim1150$ & 1215 & 49287.85 & GO4110         & F.-P. Kudritzki \\
8  & HST STIS G230LB & 1640-3100  & $\sim700$    & 41   & 59363.76 & GO16230        & D. L. Massa     \\
9  & X-Shooter       & 2989-5560  & 6655         & 440  & 59216.11 & 1106.D-0775(A) & J. S. Vink      \\
10 & HST STIS G430L  & 3000-5550  & $\sim500$    & 118  & 59363.76 & GO16230        & D. L. Massa     \\
11 & UVES            & 3281-4562  & 40\,970        & 180  & 52249.01 & 68.D-0328(A)   & D. Minniti      \\
12 & GIRAFFE         & 3960-4570  & 6300         & 1725 & 53277.28 & 074.D-0011(A)  & C. J. Evans     \\
13 & GIRAFFE         & 3960-4570  & 6300         & 1725 & 53277.26 & 074.D-0011(A)  & C. J. Evans     \\
14 & GIRAFFE         & 3960-4570  & 6300         & 22   & 53277.26 & 074.D-0011(A)  & C. J. Evans     \\
15 & GIRAFFE         & 3960-4570  & 6300         & 50   & 53277.26 & 074.D-0011(A)  & C. J. Evans     \\
16 & GIRAFFE         & 3960-4570  & 6300         & 1725 & 53276.24 & 074.D-0011(A)  & C. J. Evans     \\
17 & GIRAFFE         & 4450-5080  & 7500         & 1725 & 53292.06 & 074.D-0011(A)  & C. J. Evans     \\
18 & GIRAFFE         & 4450-5080  & 7500         & 1725 & 53280.2  & 074.D-0011(A)  & C. J. Evans     \\
19 & UVES            & 4584-6687  & 42\,310        & 180  & 52249.01 & 68.D-0328(A)   & D. Minniti      \\
20 & MUSE            & 4620-9265  & 2989         & 315  & 57611.33 & 098.D-0211(A)  & W.-R. Hamann    \\
21 & MUSE            & 4620-9265  & 2989         & 315  & 57613.25 & 098.D-0211(A)  & W.-R. Hamann    \\
22 & MUSE            & 4620-9265  & 2989         & 315  & 57613.29 & 098.D-0211(A)  & W.-R. Hamann    \\
23 & MUSE            & 4620-9265  & 2989         & 315  & 57617.23 & 098.D-0211(A)  & W.-R. Hamann    \\
24 & MUSE            & 4620-9265  & 2989         & 315  & 57618.1  & 098.D-0211(A)  & W.-R. Hamann    \\
25 & MUSE            & 4620-9265  & 2989         & 315  & 57620.15 & 098.D-0211(A)  & W.-R. Hamann    \\
26 & MUSE            & 4620-9265  & 2989         & 315  & 57620.19 & 098.D-0211(A)  & W.-R. Hamann    \\
27 & MUSE            & 4620-9265  & 2989         & 315  & 57620.23 & 098.D-0211(A)  & W.-R. Hamann    \\
28 & MUSE            & 4620-9265  & 2989         & 315  & 57621.21 & 098.D-0211(A)  & W.-R. Hamann    \\
29 & MUSE            & 4620-9265  & 2989         & 315  & 57621.25 & 098.D-0211(A)  & W.-R. Hamann    \\
30 & MUSE            & 4620-9265  & 2989         & 315  & 57622.15 & 098.D-0211(A)  & W.-R. Hamann    \\
31 & X-Shooter       & 5337-10200 & 11\,333        & 510  & 59216.11 & 1106.D-0775(A) & J. S. Vink      \\
32 & X-Shooter       & 9940-24780 & 8031         & 300  & 59216.11 & 1106.D-0775(A) & J. S. Vink      \\
33 & X-Shooter       & 9940-24790 & 8031         & 300  & 59216.11 & 1106.D-0775(A) & J. S. Vink     \\
                        \hline
                \end{tabular}
                \label{tab:spectra}
            \rule{0cm}{2.8ex}%
            \tablefoot{
                \ignorespaces 
                $^{(a)}$ Mid-exposure time in $\mathrm{HJD}-2400000.5$. 
                For brevity, the spectra referred to in this paper use the ID numbers listed here. }
    \end{table*}

\section{Observations}
\label{sec:obs}
    
    NGC~346 SSN~7 has been observed numerous times across almost the entirety of the electromagnetic spectrum, ranging from the far-ultraviolet (FUV) to the infrared (IR). In this work we complement UV observations from the Hubble Space Telescopes (HST) UV Space Telescopes Imaging Spectrograph (STIS) and multi-epoch datacube observations obtained by the Multi Unit Spectroscopic Explorer (MUSE), which is mounted on the Very Large Telescope (VLT), with archival spectra from X-Shooter, GIRAFFE, the UV-visual Echelle Spectrograph (UVES), and far-ultraviolet (FUV) spectra from the Far Ultraviolet Spectroscopic Explorer (FUSE). A summary of all spectra employed in this work is given in Table~\ref{tab:spectra}, where  each spectrum is assigned  an ID number. 
    
    The UV (including the FUV) spectra are rectified by division through the synthetic model continuum, while the optical and near-IR spectra are rectified by hand. The photometric data used for the fitting of the spectral energy distribution (SED) is given in Table~\ref{tab:photometry}.
        
    \begin{table}
        \caption{Photometry of SSN~7, sorted by central wavelength.}\label{tab:photometry}
        \centering
            \small
        \begin{tabular}{ccc}
            \hline \hline \rule{0cm}{2.2ex}
            Filter & Magnitude & Reference\\
             \hline \rule{0cm}{2.4ex}
            \rule{0cm}{2.4ex}HST F255W & 11.98 & \citet{2022RICKARDa}\\
            %\rule{0cm}{2.2ex}HST F814W & 12.74 & \citet{2007AJ....133...44S}\\
            \rule{0cm}{2.2ex}B & 12.56 & \citet{2016MNRAS.463.4210N}\\
            \rule{0cm}{2.2ex}$\mathrm{G}_\mathrm{BP}$ & 12.39 & \citet{2022arXiv220800211G} \\
            \rule{0cm}{2.2ex}V & 12.56 & \citet{2016MNRAS.463.4210N}\\
            \rule{0cm}{2.2ex}HST F555W & 12.61 & \citet{2007AJ....133...44S}\\
            \rule{0cm}{2.2ex}$\mathrm{G}_\mathrm{RP}$ & 12.69 & \citet{2022arXiv220800211G} \vspace{0.3ex}\\
            \rule{0cm}{2.2ex}J & 12.96 & \citet{2010AJ....140..416B} \\
            \rule{0cm}{2.2ex}H & 13.00 & \citet{2010AJ....140..416B} \\
            \rule{0cm}{2.2ex}K & 13.07 & \citet{2010AJ....140..416B} \\
            \hline
        \end{tabular}
    \end{table}

    The core of NGC~346, including our target and other bright O stars (namely MPG~396 (O7V), MPG~417 (O7.5V), MPG~470 (O8.5III), and MPC~476 (O6V)), was observed with the \textit{Chandra} X-ray telescope \citep{cxo2000}. The core region has a combined X-ray flux, already corrected for reddening, of ${F_\mathrm{X,\,obs}=\SI{1.5e34}{erg\,s^{-1}\,cm^{-2}}}$ in the $\SIrange{0.3}{10.0}{keV}$ \citep{2002ApJ...580..225N}. We  use this as an upper constraint on the X-ray flux that we can include in our atmospheric models. However, we are cautious in our use of this value, as close O-type binaries in the Milky Way suggest that the majority of the X-rays might originate from colliding wind zones of close binaries \citep{rau1:16}.

\section{Orbital analysis}
\label{sec:orb_analysis}

    NGC~346 SSN~7 is a known SB2 binary;  both components show contributions in the \ion{He}{I} and \ion{He}{II} lines. The ephemeris of this system was   obtained by \citet{2004NewAR..48..727N} who obtained RVs by fitting helium lines, yielding an orbital period of $\SI{24.2}{d}$ and an eccentricity of $0.42$. The high eccentricity is surprising for an O-star system with an orbital period below $\SI{50}{d}$, as theory predicts that the orbit in such systems is quickly circularised. Unfortunately, we do not have access to their data. Still, we have 17 optical spectra, of which 11 (the MUSE spectra) cover half of the reported period, allowing us to make further restrictions to the estimate of the orbital period and the eccentricity, and to constrain a mass ratio.
    
    \begin{figure}[tb]
    \footnotesize
        \centering
        \includegraphics[width=0.97\hsize] {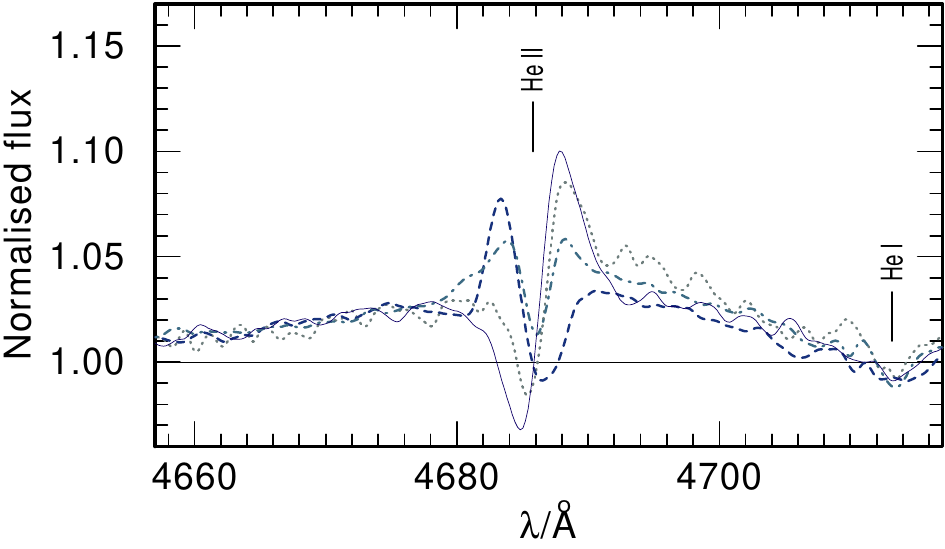}
        \caption{Multi-epoch spectral observations of SSN~7. The X-Shooter spectrum (ID 09) is shown as a solid line, the GIRAFFE spectrum (ID 17) as a dashed line, the GIRAFFE spectrum (ID 18) as a dot-dashed line, and the UVES spectrum (ID 19) as a dotted line. The shifting emission and absorption features are associated with the primary and secondary, respectively.}
        \label{fig:4686_RV_shift_obs}
    \end{figure}

\subsection{Line associations to individual binary components} % Matthew
\label{sec:analysis_components}
    
    To prepare for the extraction of the RVs of each component, we identified which lines originate from which star.
    The most prominent line, which is detected in almost all of the used optical spectra of SSN~7, is the \HeII{4686} line. In Fig.~\ref{fig:4686_RV_shift_obs} we illustrate how the morphology of the line changes within the individual observations. For clarity, only a handful of observations are shown. The emission feature of \HeII{4686} line shifts opposite to the absorption feature across multiple epochs. All lines that shift with the emission feature of the \HeII{4686} line are associated with the spectroscopic primary, and all lines that shift with the \HeII{4686} absorption line are associated with the spectroscopic secondary. 
    
    A full table of the considered lines is presented in Table~\ref{table:lines_considered}, including   which line can be assigned to the primary, secondary, or to both. Spectral lines that only show a contribution of either the primary or the secondary are scarce.

\subsection{Radial velocities}
\label{sec:RVs_intro}

    Measuring RVs in a binary, where both stars contribute to most of the diagnostic lines, is a non-trivial task. We employed a Markov chain Monte Carlo (MCMC) method combined with a least-squares fitting method to derive the RVs of each component. In each MCMC step, the individual synthetic spectra, which were obtained from the spectroscopic analysis (see Sect.~\ref{sec:spec_analysis}), are shifted   by a different RV value until the combined synthetic spectrum matches the observations. Thus, we can obtain RVs from complex absorption profiles. For a more detailed explanation we refer to \citet[][Section 3.1.2 and Appendix A]{2022A&A...659A...9P}.
    
    The MCMC method returns a final probability distribution around the true solution. The errors quoted are the larger error margin of the 68\% confidence interval of the final distribution:  commonly used codes for inferring binary parameters from observables are not able to include asymmetric uncertainties. To minimise uncertainties introduced by the intrinsic variability of the lines, we fitted several lines that show distinguishable contributions of the primary and secondary components and calculated their mean RV. 
        
    The mean RVs obtained for each spectrum, and binary component are listed in Table~\ref{table:mean_RVs}. The fitted RVs of all lines used in this process are listed in Tables~\ref{table:RVs_prim} and \ref{table:RVs_sec}. The change in sign of the RV of the primary (from $\sim\!\SI{100}{km \, s^{-1}}$ to $\sim\!\SI{-80}{km\,s^{-1}}$) within $\sim\!\SI{2}{d}$ calls in to question the reported period of $\SI{24.2}{d}$ and limits the orbital period to a few days.
    
    \begin{table}
        \footnotesize
        \center
        \caption{Mean RVs of the primary and secondary, sorted by MJD. The listed RVs are corrected for the velocity of NGC 346 complex and for the barycentric motion.}
        \begin{tabular}{ c c  r r } 
            \hline
            \hline
            \rule{0cm}{2.8ex}
            \centering
                    ID$^{(a)}$ & MJD &\multicolumn{1}{c}{RV$_\mathrm{prim}$} &  \multicolumn{1}{c}{RV$_\mathrm{sec}$}\\
                       & &\multicolumn{1}{c}{$[\si{km\,s^{-1}}]$}&\multicolumn{1}{c}{$[\si{km\,s^{-1}}]$}\\
            \hline \rule{0cm}{2.8ex}%
            \rule{0cm}{2.4ex} 11 & 52249.01198 &  $11 \pm 10$ &  $10 \pm 21$ \\
            \rule{0cm}{2.4ex} 15 & 53277.25739 &  $56 \pm \,\,\,7$ & $-34 \pm \,\,\,7$ \\
            \rule{0cm}{2.4ex} 14 & 53277.25939 &  $51 \pm \,\,\,7$ & $-32 \pm \,\,\,9$ \\
            \rule{0cm}{2.4ex} 13 & 53277.26022 &  $42 \pm \,\,\,6$ & $-39 \pm \,\,\,7$ \\
            \rule{0cm}{2.4ex} 12 & 53277.28112 &  $48 \pm \,\,\,6$ & $-39 \pm \,\,\,7$ \\
            \rule{0cm}{2.4ex} 18 & 53280.19821 &  $18 \pm 15$ & $-12 \pm 29$ \\
            \rule{0cm}{2.4ex} 17 & 53292.06195 & $-57 \pm 15$ &  $41 \pm 27$ \\
            \rule{0cm}{2.4ex} 20 & 57611.32930 &  $98 \pm 14$ & $-57 \pm 19$ \\
            \rule{0cm}{2.4ex} 21 & 57613.24630 & $-78 \pm 15$ &  $67 \pm 23$ \\
            \rule{0cm}{2.4ex} 22 & 57613.28829 & $-77 \pm 14$ &  $73 \pm 21$ \\
            \rule{0cm}{2.4ex} 23 & 57617.23463 &  $55 \pm 13$ & $-22 \pm 19$ \\
            \rule{0cm}{2.4ex} 24 & 57618.10327 &  $72 \pm 13$ & $-34 \pm 22$ \\
            \rule{0cm}{2.4ex} 25 & 57620.14865 &  $36 \pm 13$ & $-34 \pm 22$ \\
            \rule{0cm}{2.4ex} 26 & 57620.19042 &  $39 \pm 13$ & $-34 \pm 22$ \\
            \rule{0cm}{2.4ex} 27 & 57620.23262 &  $43 \pm 11$ & $-32 \pm 22$ \\
            \rule{0cm}{2.4ex} 28 & 57621.20640 &  $64 \pm 13$ & $-24 \pm 22$ \\
            \rule{0cm}{2.4ex} 29 & 57621.24834 &  $49 \pm 11$ & $-30 \pm 20$ \\
            \rule{0cm}{2.4ex} 30 & 57622.14718 & $-85 \pm 13$ &  $72 \pm 21$ \\
            \rule{0cm}{2.4ex} 31 & 59216.11133 &  $92 \pm 15$ & $-55 \pm 19$ \\
        \hline
        \end{tabular}
                \rule{0cm}{2.8ex}%
                \tablefoot{
                \ignorespaces
                $^{(a)}$ Spectral ID taken from Table~\ref{tab:spectra}. %
                }
        \label{table:mean_RVs}
    \end{table}

\subsection{Radial velocity curve modelling}
\label{sec:rv_curve_mod}

    When fitting RVs obtained from spectroscopic data with observations separated by several years, it is possible to find multiple orbital solutions consistent with the observations. To obtain a constraint on alternative ephemeris, a Monte Carlo sampler suitable for two-body systems with sparse and/or noisy radial velocity measurements, namely \textit{The Joker} \citep{2017ApJ...837...20P}, is applied to the measured RVs. As \textit{The Joker} is only able to model the RV curve of one component at a time, its output is used in a second step as input parameters for the Physics of Eclipsing Binaries (PHEOBE) code \citep{prs1:05,prs1:16,hor1:18,jon1:20,con1:20}, which can model the orbits of both binary components simultaneously.
    
\subsubsection{Obtaining ephemerides with \textit{The Joker}}
\label{sec:joker}

    \textit{The Joker} uses a rejection sampling analysis on a densely sampled prior probability density function (pdf) and produces a sample of multimodal pdfs which contain the most important information about the ephemerides. The default pdfs of \textit{The Joker} code \citep[][their section 2]{2017ApJ...837...20P} are used here as the priors. As outlined in Section~\ref{sec:RVs_intro}, the RVs obtained from the MUSE spectra suggest that the orbital period should be  of the order of a few days (see Table~\ref{table:mean_RVs}). Hence, a log-uniform prior for the orbital period in the range of $\SIrange{0.1}{40}{d}$ is adopted for the fitting procedure. The upper limit was chosen as roughly two times the orbital period obtained by \citet{2004NewAR..48..727N} to enable their solution to be considered, while the lower limit was set to a reasonably small value. For the eccentricity we used the Beta distribution from \citet{10.1093/mnrasl/slt075} with $a=0.867$ and $b=3.03$. For the argument of the pericentre and the orbital phase a uniform prior without any restrictions is used. The RVs of the primary, which are more reliable since they have lower error margins, are employed as input data.
    
    From the $\num{1e7}$ samples created, $~340$ pass the rejection step, equivalent to an acceptance rate of $0.0034\%$. Afterwards, a standard MCMC method is used to sample around the most likely solution found from the rejection step in order to fully explore the posterior pdf. We find that the ephemerides are described best with a circular orbit ($e=0$) and an orbital period of $P_\mathrm{Joker}=\SI{3.07438\pm0.00002}{d}$. The complete set of emphemerides obtained with \textit{The Joker} are listed in Table~\ref{table:ephemeris_joker}.

    \begin{table}
        \footnotesize
        \center
        \caption{Ephemerides obtained using the primary's RVs and \textit{The Joker}.}
        \begin{tabular}{ c c c } 
            \hline
            \hline
            \rule{0cm}{2.8ex}
            \centering
                    parameter & value & unit\\
            \hline \rule{0cm}{2.8ex}%
            \rule{0cm}{2.8ex} $P$ &  $3.07438^{+0.00002}_{-0.00002}$ & $[\mathrm{d}]$\\
            \rule{0cm}{2.8ex} $e$ &  $0.00^{+0.09}_{-0.00}$ &\\
            \rule{0cm}{2.8ex} $\omega$ &  $-122.6^{+244.1}_{-38.4}$ & $[\mathrm{^\circ}]$\\
            \rule{0cm}{2.8ex} $M_0$ &  $-0.80^{+1.05}_{-0.81}$ & $[\mathrm{d}]$\\
            \rule{0cm}{2.8ex} $K$ &  $81.25^{+6.15}_{-6.05}$ & $[\mathrm{km\,s^{-1}}]$\\
            \rule{0cm}{2.8ex} $\varv_0$ &  $13.40^{+3.26}_{-3.27}$ & $[\mathrm{km\,s^{-1}}]$\vspace{0.15cm}\\
        \hline
        \end{tabular}
        \label{table:ephemeris_joker}
    \end{table}

\subsubsection{Obtaining orbital parameters with PHOEBE}
\label{sec:PHOEBE}

    The ephermerides obtained in Section~\ref{sec:joker} were determined by only using the RVs of the primary. This was done to get a first constraint on the orbital parameters; however, valuable information such as the mass ratio ($q$), the projected orbital separation ($a\,\sin\,i$), and the projected Keplerian masses ($M\,\sin\,i$) can only be obtained when fitting the RVs of the two components simultaneously. Therefore, the PHOEBE light and RV curve modelling software (version 2.3) was employed to model the observed RVs, with the built-in option of the MCMC sampler \verb+emcee+ \citep{for1:13}.
    
    The initial probability distributions of the period $P$, the argument and phase of the pericentre $\omega$ and $M_0$, and the barycentre velocity $\varv_0$ are assumed to be of Gaussian shape. The Gaussians are centred at the  best-fitting value as obtained by \textit{The Joker} (see Table~\ref{table:ephemeris_joker}) and have a standard deviation similar to their respective largest error margin. The ephemerides  obtained with \textit{The Joker} favour a circular orbit, and hence the eccentricity is fixed at $e=0$. To  constrain   the current mass ratio we employed the method of \citet{1941ApJ....93...29W} yielding $q_\mathrm{wilson}=1.34\pm0.07$. Because of a lack of information about other parameters, we used a uniform distribution in the range of $q_\mathrm{orb,\,i}=\SIrange{0.33}{3.0}{}$, where a starting value $q_\mathrm{orb,\,i}=q_\mathrm{wilson}$ is assumed. The initial projected orbital separation is approximated to be of the order of $a\,\sin\,i_\mathrm{i}=\SI{10}{\rsun}$ and that it is uniformly distributed over the range of $a\,\sin\,i_\mathrm{i}=\SIrange{0.1}{100}{\rsun}$.
    
\begin{figure}[tb]
    \footnotesize
        \centering
        \includegraphics[width=\hsize]{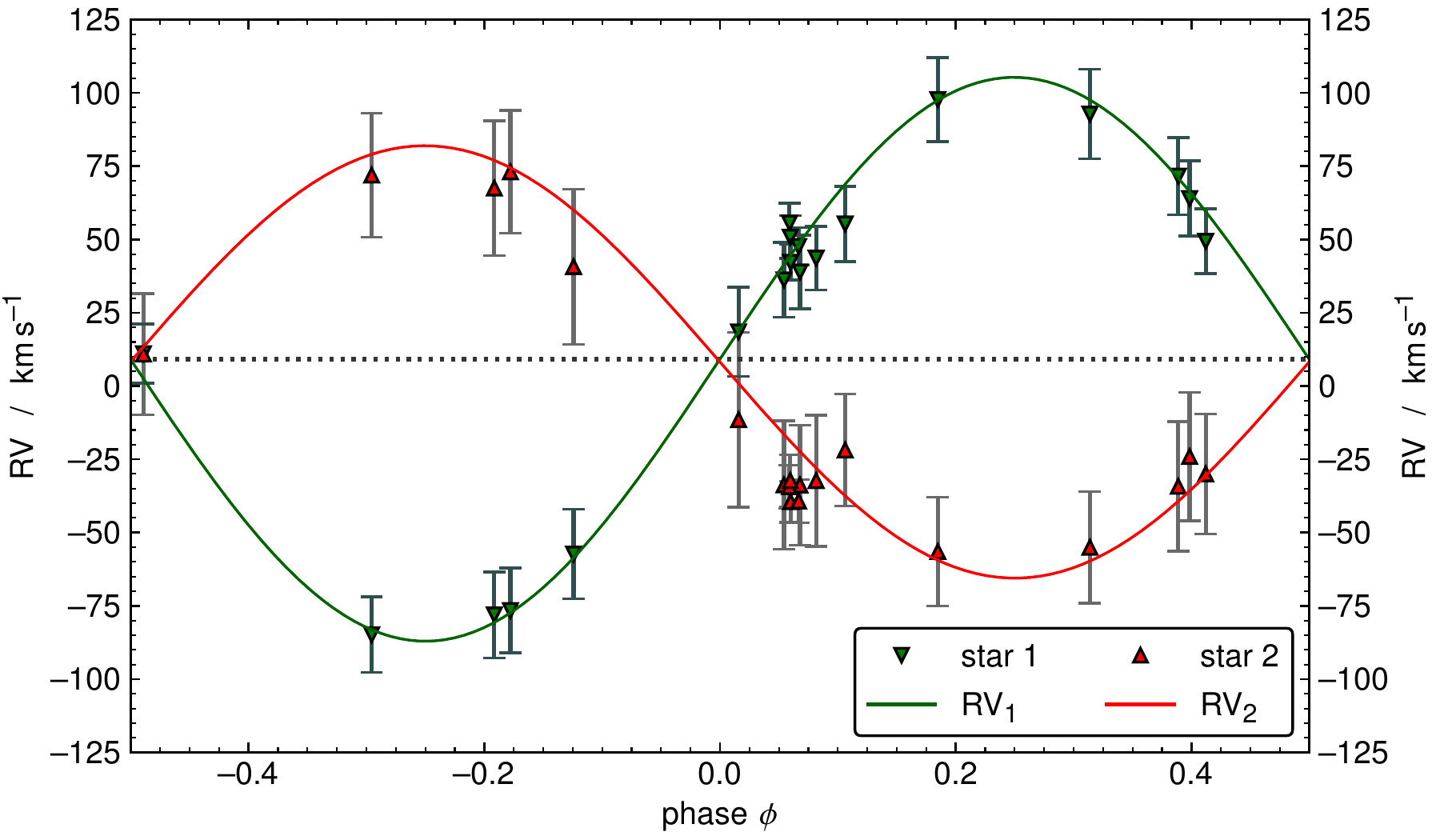}
        \caption{Observed (triangles) and synthetic (solid lines) RV curves of the primary (green) and secondary (red) component, as obtained by the PHOEBE code. The dashed black line indicates the barycentric velocity offset $\varv_0$.}
        \label{fig:phoebe_RVs}
    \end{figure}

    The best fit of the RV curve obtained with the PHOEBE code is shown in Fig.~\ref{fig:phoebe_RVs}. The final orbital parameters are listed in Tables~\ref{table:phoebe_results1} and \ref{table:phoebe_results2}. The RV fit yields masses without information about the inclination. By using the results from the spectroscopic analysis (see Sect.~\ref{sec:spec_analysis}) we estimate the inclination to be $i_\mathrm{orb}=\SI{16\pm1}{^\circ}$.
    
        \begin{table}[tb]
        \footnotesize
        \centering
        \caption{Ephemerides obtained using the RVs of both binary components and the PHOEBE code.}
        \begin{tabular}{ccc}
                \hline\hline \rule{0cm}{2.8ex}
                \rule{0cm}{2.8ex}parameter & value & unit\\ 
                \hline
                \rule{0cm}{2.8ex}$P$ & $3.07359^{+0.02481}_{-0.00001}$ & $[\mathrm{d}]$\\ 
                \rule{0cm}{2.6ex}$\omega_0$ & $242^{+9}_{-30}$ & $[\mathrm{^\circ}]$\\ 
                \rule{0cm}{2.6ex}$M_0$ & $1.5^{+7.8}_{-1.4}$ & $[\mathrm{d}]$\\ 
                \rule{0cm}{2.6ex}$\varv_0$ & $9.1^{+0.2}_{-0.5}$ & $[\mathrm{km\,s^{-1}}]$\\ 
                %\rule{0cm}{2.4ex}$e$ & \multicolumn{2}{c}{$0^{+0}_{-0}$}\\ 
                \rule{0cm}{2.6ex}$q_\mathrm{orb}$ & $1.47^{+0.13}_{-0.09}$ &\\ 
                \rule{0cm}{2.6ex}$a\,\sin\,i$ & $11.0^{+0.7}_{-0.3}$ & $[R_\odot]$\vspace{1ex}\\ 
                \hline
                \rule{0cm}{2.8ex}$i_\mathrm{orb}\,^{(a)}$ & $16^{+1}_{-1}$ & $[\mathrm{^\circ}]$\\ 
                \rule{0cm}{2.6ex}$a\,^{(b)}$ & $39.9^{+1.2}_{-2.5}$ & $[{R_\odot}]$\vspace{1ex}\\ 
                \hline %\vspace{0.01ex}
        \end{tabular}
                \rule{0cm}{2.8ex}%
                \tablefoot{
                \ignorespaces
                $^{(a)}$ Calibrated such that the orbital and spectroscopic masses agree within their uncertainties. $^{(b)}$ Calculated using $i_\mathrm{orb}$. %
                }
        \label{table:phoebe_results1}
    \end{table}
    
    \begin{table}[tb]
    \footnotesize
        \centering
        \caption{Orbital parameters of the individual binary components obtained using the PHOEBE code.}
        \begin{tabular}{cccc}
                \hline\hline \rule{0cm}{2.4ex}
                \rule{0cm}{2.8ex}parameter & primary & secondary & unit\\ 
                \hline
                \rule{0cm}{2.8ex}$K$ & $107.9^{+3.9}_{-3.9}$ & $73.3^{+2.7}_{-2.7}$ & $[\mathrm{km\,s^{-1}}]$\\ 
                \rule{0cm}{2.6ex}$M_\mathrm{orb}\,\sin\,i\,^3$ & $0.77^{+0.15}_{-0.07}$ & $1.13^{+0.21}_{-0.10}$& $[{M_\odot}]$\vspace{1ex}\\ 
                \hline
                \rule{0cm}{2.8ex}$M_\mathrm{orb}\,^{(a)}$ & $36.5^{+7.04}_{-3.5}$ & $53.8^{+10.1}_{-4.9}$ & $[{M_\odot}]$\\ 
                \rule{0cm}{2.6ex}$R_\mathrm{RL}^{\,\,\,\,\,\,\,\,\,(a)}$ & $13.8^{+0.8}_{-0.4}$ & $16.5^{+0.9}_{-0.4}$&$[{R_\odot}]$\vspace{1ex}\\ 
                \hline %\vspace{0.01ex}
        \end{tabular}
        \rule{0cm}{2.8ex}%
                \tablefoot{
                \ignorespaces
                $^{(a)}$ Calculated using $i_\mathrm{orb}$ (see Table~\ref{table:phoebe_results1}).  %
                }
        \label{table:phoebe_results2}
    \end{table}

    % (see Appendix~\ref{app:light_curve})
    To date, there is no observed light curve of NGC~346 SSN~7 reported in the literature. Since the PHOEBE code is capable of modelling a light curve, it can be used to make a prediction on the shape of a possible light curve (see Appendix~\ref{app:light_curve} for more details). From the predicted synthetic light curve, we conclude that a detector with a precision below $<\SI{18}{mmag}$ in the V band is needed to see any kind of variability. This precision is hardly achievable for stars in the SMC with current ground-based telescopes. Hence, it is not surprising that no light curve has been detected so far.

\section{Spectral analysis} % Matthew
\label{sec:spec_analysis}

    \begin{figure*}
        \centering
        \includegraphics[width=\hsize]{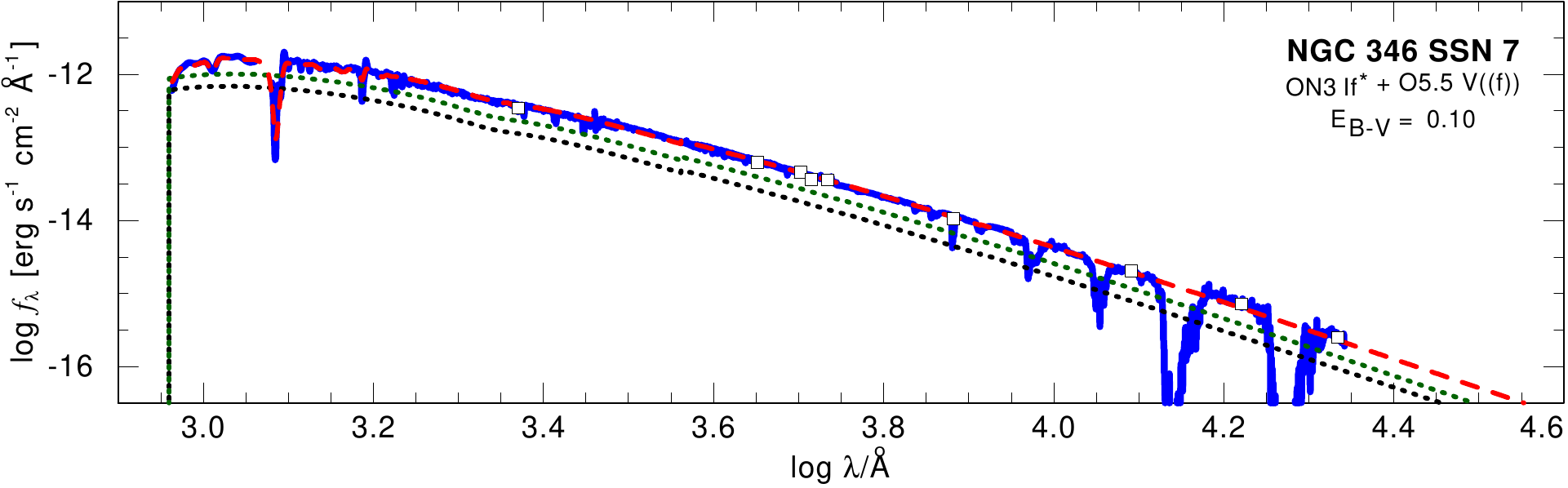}
        \caption{Observed (blue) and combined synthetic (red) SED. The individual synthetic SEDs of the primary and secondary are shown as the dotted black and green line, respectively. The photometry, listed in Table~\ref{tab:photometry}, is shown as open squares.
        }
        \label{fig:SED}
    \end{figure*}

\subsection{Stellar atmosphere modelling}% Matthew
\label{sec:analysis_PoWR}

    Expanding stellar atmospheres of hot stars (${T>\SI{15}{kK}}$) are in non-local thermal equilibrium (non-LTE) and can only be calculated using state of the art stellar atmosphere codes such as the Potsdam Wolf-Rayet \citep[PoWR, ][]{2002A&A...387..244G, 2011MNRAS.416.1456O, 2014A&A...565A..27H, 2015A&A...581A..21H,  2015ApJ...809..135S, 2004A&A...427..697H, 2015A&A...577A..13S, 2015A&A...581A..21H} code. Within PoWR, the radiative transfer equations and the rate equations are solved iteratively in the co-moving frame. The code is able to produce a synthetic spectrum that accounts for mass-loss, line blanketing, and wind clumping.

    For our stellar atmosphere models, we used tailored surface abundances that are representative of the chemical composition of the SMC. For the modelling of SSN~7 within NGC~346, the initial H abundance is adopted from \citet{asp1:05}, while the initial abundances of C, N, O, Mg, and Si are based on the works of \citet{hun1:07} and \citet{tru1:07}. The abundances of P and S are taken from \citet{sco1:15} and are scaled down to the metallicity of the SMC (${Z_\mathrm{SMC}=1/7\,\zsun}$). The iron group elements (Fe, Sc, Ti, V, Cr, Mn, Co, and Ni) are either taken from \citet{tru1:07} or, when not available, taken from solar values \citep{sco1:15} and multiplied by a factor $1/7$ to account for the lower metallicity of the SMC (Table~\ref{table:ion_levels}). The numerous complex iron lines are handled with a `superlevel' approach \citep{2002A&A...387..244G}.

    Microturbulence within a stellar atmosphere leads to a Doppler-broadening of spectral lines. The PoWR models used assume a depth-dependent microturbulence starting in the photosphere at ${\xi_\mathrm{ph}=10\,\mathrm{km\,s}^{-1}}$ and increases outwards linearly with the wind velocity $\xi(r)=0.1\cdot\varv(r)$. 
    
    Our PoWR models account for wind inhomogeneities and introduced optically thin clumps (microclumping) within the stellar atmosphere. The density contrast of the clumps within the wind is described by the clumping factor $D$. For NGC~346 SSN~7, a depth-dependent clumping factor is used, starting at the sonic radius and reaching $D=10$ at ten times the stellar radius.

    For the velocity stratification within the wind, a double $\beta${-}law \citep{hil1:99} of the form 
    \begin{equation}
      \label{eq:beta_law_2}
        \varv(r) = \varv_\infty \left[f\left(1-\frac{r_0}{r}\right)^{\beta_1}+(1-f)\left(1-\frac{r_1}{r}\right)^{\beta_2}\right]
    \end{equation}
    is adopted. The inner part of the velocity stratification is described by an exponent  $\beta_1=0.8$, which is typical for O stars \citep{pau1:86}. The outer part is described with an exponent $\beta_2=4$ and only contributes with a fraction of $f=0.4$. The parameters $r_0$ and $r_1$ are set within the code, such that the wind is smoothly connected.

    The effective temperatures quoted in this work are related via the Stefan-Boltzmann law to the radius at which the Rosseland mean continuum optical depth is $\tau=2/3$. Stellar and wind parameters can be obtained by adjusting the synthetic spectrum until it matches the observed spectra simultaneously. This also means that the abundance of a specific element  (e.g. H, C, N, O) can be adjusted if needed to match the strength of observed lines.
    
\subsection{Resulting spectroscopic parameters}

    Deriving stellar and wind parameters from stellar spectra in which lines are blended by two stars of a binary is a difficult task. The following subsections describe the strategy employed and the final parameters that are listed in Table~\ref{table:SSN7_binary_fit_parameters}.
    
\subsubsection{Rotation rates}

    Before starting the spectral fitting process, accurate values for the projected rotational velocities ($\varv\sin i$) of each component are needed. These impact the depth and shape of all synthetic lines and hence conclusions drawn from the spectroscopic fit. We employ the \textsc{iacob-broad} tool \citep{2014A&A...562A.135S} which uses a combination of a Fourier transform (FT) and a goodness of fit (GOF) method to obtain accurate measurements of $\varv\sin i$. 
    
    When obtaining projected rotation rates, the preference is to use metal lines over helium lines as the latter are, in addition to the rotational broadening, also pressure broadened. For the primary, we use the 
    \NIV{4057} as well as the \NIV{6381} line. As there are no metal lines that solely can be attributed to the secondary, we use the prominent \HeI{6875} and the \HeI{7065} line.
    The projected rotational velocities of the primary and secondary component are best described with ${\varv_1 \sin i = 135 \pm 10 \ \mathrm{km \ s^{-1}}}$ and ${\varv_2 \sin i = 185 \pm 10 \ \mathrm{km \  s^{-1}}}$, respectively.
    
    Most of the observed lines are formed in the rotating photosphere of a star, while fewer but still important lines in the UV are formed within the static (non-rotating) wind. To account for this, the synthetic PoWR models used in this work are calculated with a rigidly rotating photosphere and a non-rotating wind, as described in \cite{2014A&A...562A.118S}.

\subsubsection{Luminosity, temperature, and surface gravity}

\begin{figure*}
        \centering
        \includegraphics[width=0.85\hsize]{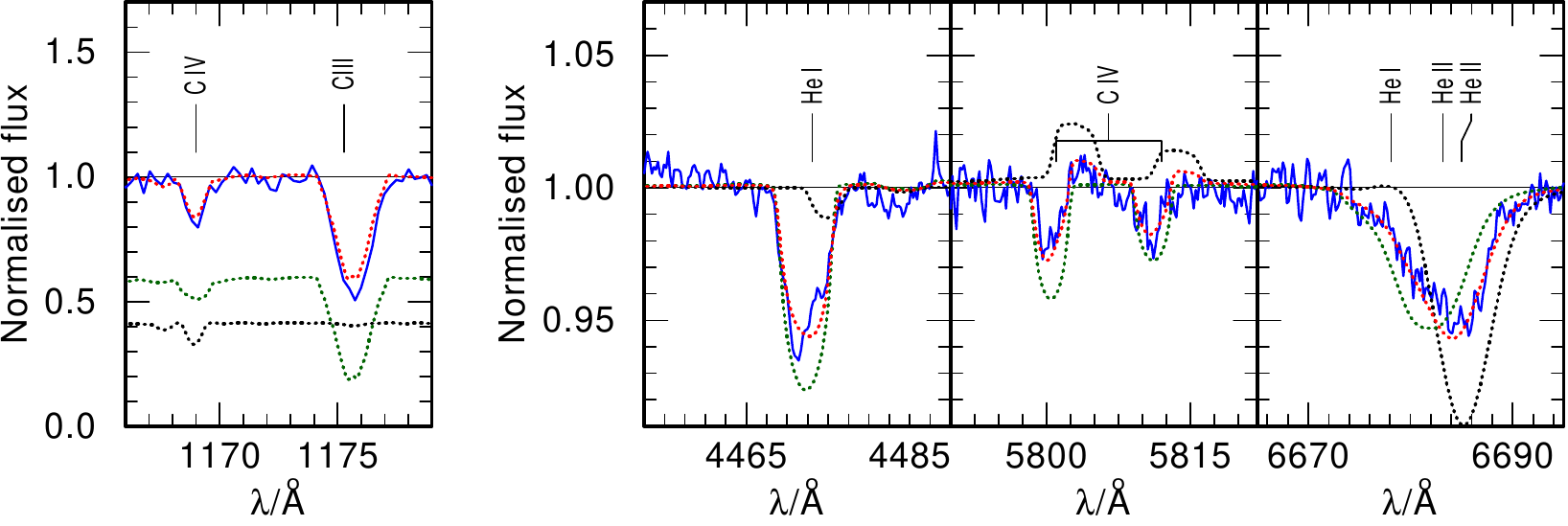}
        \caption{Close-ups on the carbon and helium lines  used to constrain the luminosity ratio (solid blue) compared to the combined synthetic model (dotted red). \textit{Left:} Observed FUSE spectra (ID 01). The weighted synthetic spectra of the primary and secondary components are shown as a dotted black and green line. \textit{Right:} Observed X-Shooter spectra (ID 09 and 31). The unweighted synthetic spectra of the primary and secondary components are shown as a dotted black and green line.}
        \label{fig:CIII_LUMSET}
    \end{figure*}

    To  constrain the luminosity of each binary component, we carefully looked at the ratios of isolated and blended helium and metal lines. One of the key diagnostic lines was the \CIII{1175} in the UV, which  originates purely from the secondary, giving an extra constraint on the luminosity ratio. A selected set of the lines used is illustrated in Fig.~\ref{fig:CIII_LUMSET}. Determining the luminosity ratio is an iterative process and is revisited after each modelling change, as changes in temperature and surface gravity also affect the shape of key diagnostic lines. The luminosity of each component is ${\log (L_1 \ / \ L_\odot) = 5.75}$ and ${\log (L_1 \ / \ L_\odot) = 5.78}$ for the primary and secondary, respectively. 

\begin{figure*}
        \centering
        \includegraphics[width=0.97\textwidth]{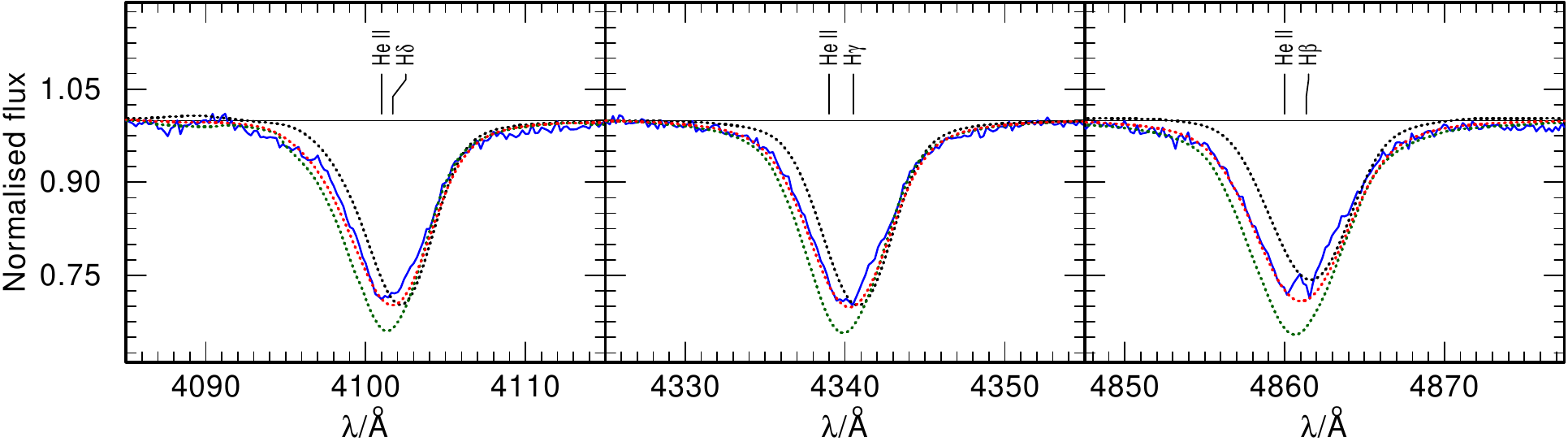}
        \caption{Close-ups on the Balmer lines in the X-Shooter spectrum (ID 09; solid blue) compared to the combined synthetic model (dotted red). The unweighted synthetic spectra of the primary and secondary components are shown as dotted black and green line.
        }
        \label{fig:spec_balmerlines}
    \end{figure*}

    The surface gravity of each of the component stars is determined by fitting the wings of the Balmer lines in the spectra with the biggest RV shifts. We find that the primary and the secondary have similar surface gravities of $\log g = 3.7\pm0.1$ (see Fig.~\ref{fig:spec_balmerlines}).

    The temperatures of the stars are derived using the ratio between the \ion{He}{I} and \ion{He}{II} lines associated with each component (see Table~\ref{table:lines_considered}). To get a more accurate constraint on the primary's temperature, the ratio between the different observed nitrogen lines are used, namely \NIVt{3479,\,3483,\,3485}, \NIV{4058}, \NVd{4604,\,4620}, \NIIId{4634,\,4641}, and \NIVt{6212,\,6216,\,6220}. We find for the primary $T_\mathrm{eff,\,1}=\SI{43.6\pm2}{kK}$, while the secondary is best fit with $T_\mathrm{eff,\,2}=\SI{38.7\pm2}{kK}$.

    \begin{table}
        \small
        \center
        \caption{Abundances and ionisation stages employed in the PoWR models.}
        \begin{tabular}{ c | c c | c c } 
            \hline
            \hline
            \centering\rule{0cm}{2.8ex}
            Element & \multicolumn{2}{c|}{Primary}  & \multicolumn{2}{c}{Secondary} \\ 
            \rule{0cm}{2.4ex}& mass fr. & ion. stages & mass fr. & ion. stages\\
            \hline\rule{0cm}{2.8ex}
            \rule{0cm}{2.4ex}H & 0.60 & \textsc{i}, \textsc{ii} & 0.7375 &   \textsc{i}, \textsc{ii}\\
            \rule{0cm}{2.4ex}He & 0.39 &  \textsc{i}, \textsc{ii}, \textsc{iii}&  0.26 & \textsc{i}, \textsc{ii}, \textsc{iii}\\
            \rule{0cm}{2.4ex}C & $\num{1e-5}$ & \textsc{ii} -- \textsc{v}& $\num{21e-5}$ &\textsc{ii} -- \textsc{v}\\
            \rule{0cm}{2.4ex} N  & $\num{138e-5}$ &\textsc{ii} -- \textsc{vi} &$\num{40e-5}$ & \textsc{ii} -- \textsc{vi}\\
            \rule{0cm}{2.4ex}O & $\num{4e-5}$ &\textsc{ii} -- \textsc{vii} & $\num{110e-5}$ &\textsc{ii} -- \textsc{vii}\\
            \rule{0cm}{2.4ex}Mg & $\num{10e-5}$ & \textsc{ii}, \textsc{iii}, \textsc{iv} & $\num{10e-5}$ &  \textsc{ii}, \textsc{iii}, \textsc{iv}\\
            \rule{0cm}{2.4ex}Si &$\num{1.3e-5}$ & \textsc{iii} -- \textsc{viii}& $\num{1.3e-5}$ & \textsc{iii} -- \textsc{vi}\\
            \rule{0cm}{2.4ex}P & $\num{8.3e-7}$ & \textsc{iv}, \textsc{v}, \textsc{vi}&  $\num{8.3e-7}$ & \textsc{iv}, \textsc{v}, \textsc{vi}\\
            \rule{0cm}{2.4ex}S & $\num{4.4e-5}$ &  \textsc{iii} -- \textsc{vi}& $\num{4.4e-5}$ & \textsc{iii} -- \textsc{vi}\\
            \rule{0cm}{2.4ex}Fe & $\num{3.5e-4}$ &  \textsc{ii} -- \textsc{ix}& $\num{3.5e-4}$ & \textsc{ii} -- \textsc{ix}\vspace{0.0cm}\\
            \hline
        \end{tabular}
        \label{table:ion_levels}
    \end{table}

\subsubsection{Surface abundances}
\label{sec:surface_abun}

    During the determination of the temperature of the primary, we find that it is possible to match the ratio of the \ion{He}{I} and \ion{He}{II} lines, but these lines are systematically too weak. Hence, we lowered the primary's surface hydrogen abundance to $X_\mathrm{H,\,1} = 0.60\pm0.1$. Furthermore, to fit the observed nitrogen lines, which are also crucial for the temperature determination, we find that the nitrogen abundance of the primary is enhanced to $X_\mathrm{N,\,1}=138^{+10}_{-30}\times10^{-5}$ (i.e. $2\,N_\odot$).
    
    In addition, the spectrum contains carbon and oxygen lines that show clear contributions from the primary. To constrain its surface oxygen abundance, we used the \OIVt{1337{-}1343{-}1344} 
    triplet in the UV, the \ion{O}{iv}
    multiplet around $\sim 3000$\,\AA, and the 
    \ion{O}{iii} lines in the range from $\SIrange{1409}{1412}{\AA}$.
    We find that oxygen in the primary's atmosphere is depleted with $X_\mathrm{O,\,1}=4^{+6}_{-3}\times10^{-5}$. 
    Regarding the carbon abundance, we used \CIV{1169} in the UV and the \CIVd{5801,\,5812} doublet in the optical, yielding a surface carbon abundance of $X_\mathrm{C,\,1}=1^{+9}_{-0}\times10^{-5}$ in the primary.
    
    Determining the surface abundance of the secondary is more complicated as it shows little contribution to the observed metal lines. We can see in some spectra that the secondary contributes to the \NIIId{4634,\,4641} doublet. Hence, we tried to use this as a constraint on its surface nitrogen abundance and find the best fit with $X_\mathrm{N,\,2}=60^{+20}_{-10}\times10^{-5}$. We can also see a contribution to the \CIVd{5801,\,5812} doublet in the optical, and    the \CIII{1175} in the UV. These lines can be well reproduced with the initial carbon abundance of $X_\mathrm{C,\,2}=21^{+0}_{-5}\times10^{-5}$. Unfortunately, we cannot see a clear contribution of the secondary to the oxygen lines. As the nitrogen abundance is increased and theory predicts that the amount of CNO material should be $X_\mathrm{CNO} \approx {137 \times 10^{-5}}$ for stars in the SMC \citep{1998RMxAC...7..202K}, we scaled down the oxygen abundance of the secondary to $X_\mathrm{O,\,2}=\num{80e-5}$. We note that this does not have any impact on the quality of our fit.

    \begin{figure*}
        \centering
        \includegraphics[width=\hsize]{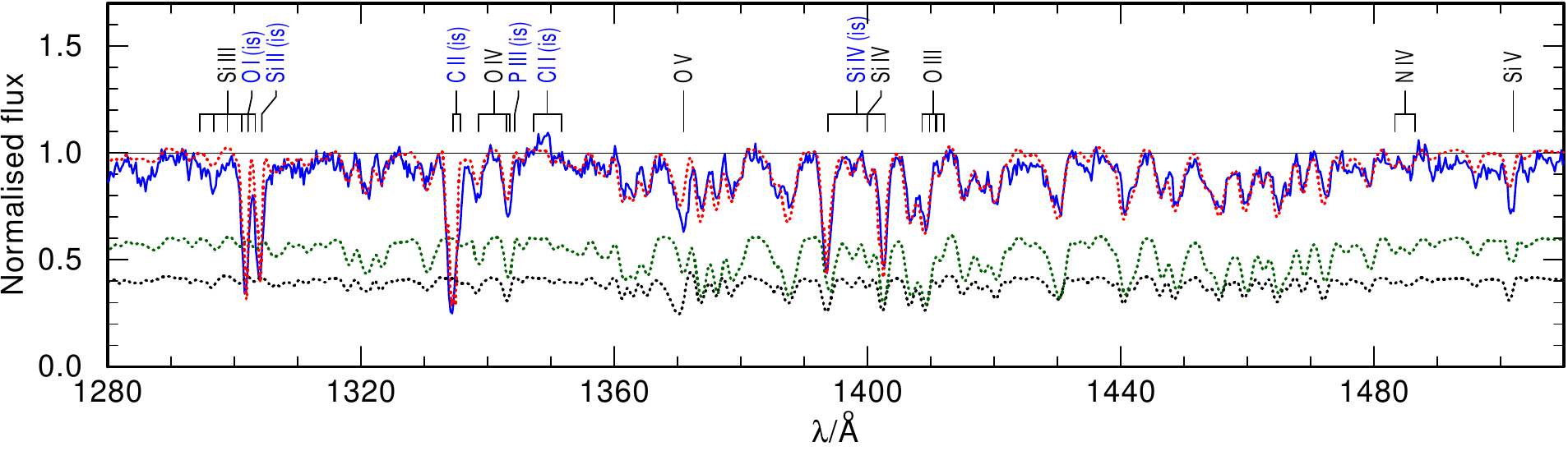}
        \caption{Close-up on the iron forest of the HST FOS (ID 03) spectrum (solid blue) compared to the combined synthetic model flux (dashed red). The weighted synthetic model spectra of the primary and secondary are shown as a dotted black and green line, respectively. The iron forest is used to check the iron abundance and      to constrain the microturbulence within the stellar atmosphere.}
        \label{fig:ironforest}
    \end{figure*}
    
    \begin{figure*}
        \centering
        \includegraphics[width=1\hsize]{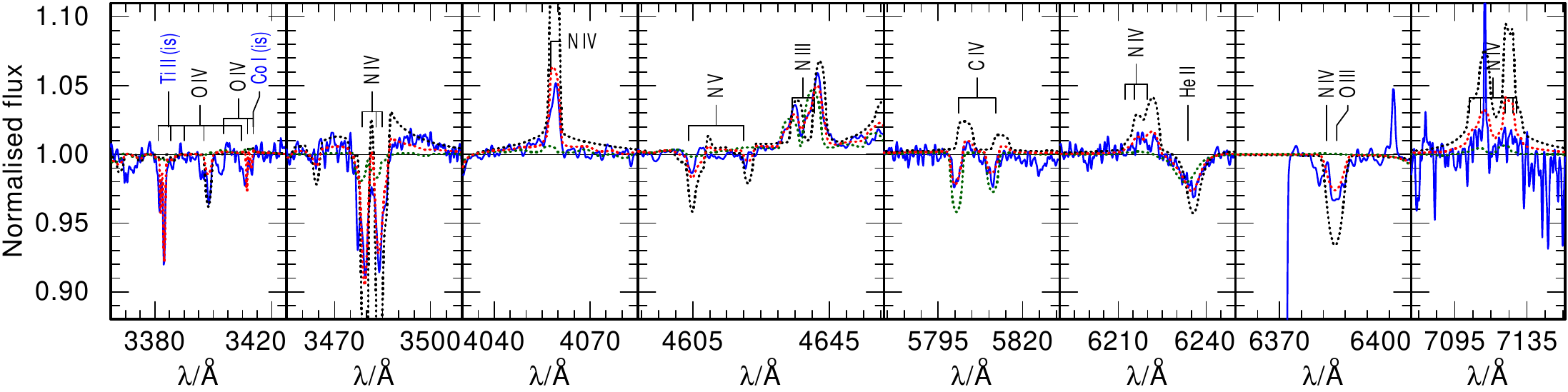}
        \caption{Close-ups on the metal lines in the X-Shooter spectrum (ID 09 and 31; solid blue) compared to the combined synthetic spectrum (dashed red). The unweighted synthetic spectra of the primary and the secondary component are shown as a dotted black and green line, respectively.}
        \label{fig:spec_xshooter_metal}
    \end{figure*}

   Figure~\ref{fig:spec_xshooter_metal} shows the all-important metal lines used to fit the abundances of the primary and secondary. The changed surface abundances of the primary, especially the lower surface hydrogen abundance and the increased nitrogen, hint at previous binary interaction that stripped off parts of the envelope (see Sect.~\ref{sec:discussion}).

\subsubsection{Spectroscopic stellar masses}
    
    Given the fundamental stellar parameters ($T_\mathrm{eff}$, $\log g$, $L$) of each binary component, we can calculate the spectroscopic masses. We find that the mass of the primary component is ${M_\mathrm{spec,\,1}=32^{+19}_{-13}\,\msun}$, surprisingly low for its luminosity. The mass of the secondary is ${M_\mathrm{spec,\,2}=55^{+32}_{-21}\,\msun}$. The low mass of the primary strengthens our argument that the binary components must have interacted in the past.

    To further strengthen this hypothesis, we calculated the Roche radius, using the results from the orbital analysis in combination with the spectroscopic masses and find that both stars are currently slightly overfilling their Roche lobe with ${R_1/R_\mathrm{RL}=1.01}$ and ${R_2/R_\mathrm{RL}=1.03}$. This strongly suggests that the system   interacted in the past, and that both stars are still currently in contact.
    
\subsubsection{Wind velocities and mass-loss rates}
\label{sec:mass_loss_rates_fitting}
    So far only photospheric lines were considered in the spectroscopic analysis. However, in the spectra there are several lines that are sensitive to wind parameters, like the \NVd{1239,\,1243} and \CIVd{1548,\,1551} doublets in the UV, and  \textsc{H}$\alpha$ and \HeII{4686} in the optical. Both stars are very bright, massive, and hot, meaning that     both of them are expected to contribute to the wind diagnostic lines. Within the FUV FUSE observation, no measurable \ion{P}{v} or \ion{S}{IV} wind resonance lines are present. This is consistent with our composite model.
    
\begin{figure*}
        \centering
        \includegraphics{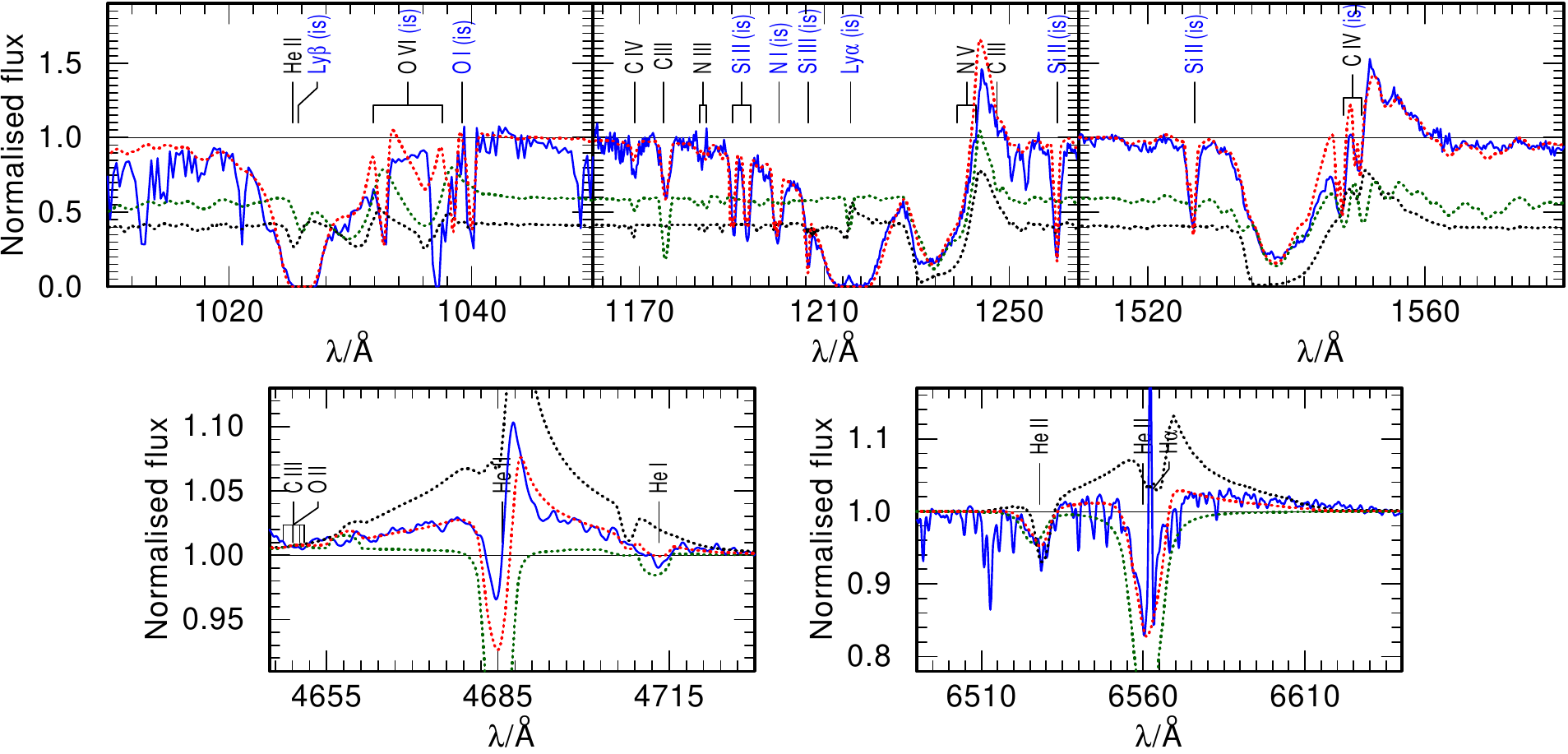}
        \caption{Close-ups on the key wind diagnostic lines (solid blue) in the FUV, UV and optical spectra overplotted by the combined synthetic model (dashed red). In the upper panels the weighted and in the lower panels the unweighted synthetic spectra of the primary and secondary components are shown as dotted black and green line, respectively. \textit{Upper left:} The \ion{O}{VI} line of the FUSE (ID 1) spectrum. This line is sensitive to X-rays. \textit{Upper middle:} The \ion{N}{V} resonance doublet in the HST FOS (ID 3) spectrum. \textit{Upper right: } The \ion{C}{IV} resonance doublet in the HST FOS (ID 3) spectrum. \textit{Lower left:} The \HeII{4686} line in the X-Shooter spectrum (ID 9), showing strong wind emission features from the primary. \textit{Lower right:} $\mathrm{H}\alpha$ of the X-Shooter spectrum (ID 31), again, the primary dominates the wind emission of this line.}
        \label{fig:wind_lines}
    \end{figure*}

    The primary motion is reflected by the emission parts of \textsc{H}$\alpha$ and \HeII{4686}, while the secondary's motion is evident from the absorption components of the same lines. We conclude that the primary must have a stronger mass-loss rate compared to the secondary. We used the \textsc{H}$\alpha$ and \HeII{4686}, in combination with \NVd{1239,\,1243} and \CIVd{1548,\,1551}, to calibrate the primary's mass-loss rate to ${\log (\dot{M}_1 / (\msunpyr)) = -5.4 \pm 0.1}$. With this mass-loss rate, the resonance doublets in the UV were fully saturated in the synthetic spectrum of the primary, yet the observed profiles were not fully saturated. The secondary contributes roughly 60\% of the UV flux, so we fit the combined spectra with a partially saturated secondary wind. The secondary is fit with a mass-loss rate of ${\log (\dot{M}_2 \ / \ (\msunpyr)) = -7.3 \pm 0.1}$, which results in the combined synthetic spectrum matching the observed spectrum (Fig.~\ref{fig:wind_lines}). 
    
    The terminal wind velocity of the primary is adjusted to match the blue edge of the \NVd{1239,\,1243} absorption trough, giving ${\varv_{\infty,\,1} = \SI{2500\pm100}{km\,s^{-1}}}$. For the secondary, we experimented with different values of $\varv_\infty$ to see how the slope of the \CIV{1548,1551} resonance doublet changes. The best fit that matches the profile in the STIS and FOS spectrum was obtained when using the same terminal wind velocity for the secondary as for the primary (${\varv_{\infty,\,2} = \SI{2500\pm300}{km\,s^{-1}}}$).

\subsubsection{Required X-ray flux}

    The \OVId{1032,\,1038} resonance doublet in the FUV shows weak indications of a P\,Cygni profile, which cannot be modelled by default in a stellar atmosphere model. \citet{cas1:79} were the first to suggest that these high ionisation stages must be `super-ionised'  by a hot X-ray plasma that originates from within the wind.
    
    We include, in our primary and secondary, a hot plasma that is smoothly distributed throughout the wind, starting at $1.1\,R_\ast$ and having a temperature of $T_\mathrm{X}=\SI{3}{MK}$. We find that the  \OVId{1032,\,1038} resonance doublet (Fig.~\ref{fig:wind_lines}) can be explained with an X-ray luminosity of $\log (L_\mathrm{X,\,1}/(\mathrm{erg\,s}^{-1}))=32.4$ and $\log (L_\mathrm{X,\,2}/(\mathrm{erg\,s}^{-1}))=32.1$ for the primary and secondary, respectively. This is much lower than the X-Ray luminosity observed from the core of NGC~346 and comparable to X-rays needed for other O stars \citep{2018A&A...620A..89N}.

        \begin{table*}[t]
        \centering
        \caption{Stellar  parameters derived from the spectroscopic analysis, compared to those of the best-fitting MESA model}
        \begin{tabular}{c|cc|cc | c } 
                \hline\hline \rule{0cm}{2.4ex}
                \rule{0cm}{2.8ex} & \multicolumn{2}{c|}{Spectroscopic fit} & \multicolumn{2}{c|}{MESA model}& Unit\\
                \rule{0cm}{2.4ex}& SSN~7a & SSN~7b & mass donor & mass gainer & \\
                \hline
            \rule{0cm}{2.8ex} $\log\,L$& $5.75\pm{0.15}$ & $5.78\pm{0.15}$ & $5.76\pm{0.1}$ &  $5.74\pm{0.1}$ & $[L_\odot]$\\
            \rule{0cm}{2.4ex} $T_\mathrm{eff}$ & $43.6\pm{1.0}$ & $38.7\pm{1.0}$ & $43.6\pm{1.5}$ &  $38.8\pm{1.5}$ & $[\mathrm{kK}]$\\
            \rule{0cm}{2.4ex} $\log\,g$ &$3.7\pm{0.1}$ & $3.7\pm{0.1}$ & $3.68\pm{0.05}$& $3.7\pm{0.05}$ & $[\mathrm{cm \ s}^{-2}]$\\
            \rule{0cm}{2.4ex} $\varv \sin i$ & $135\pm10$ & $150\pm10$ & - & - & $[\mathrm{km\,s^{-1}}]$\\
            \rule{0cm}{2.4ex} $\varv_\mathrm{rot}$ & - & - & $222\pm{10}$ & $274\pm{10}$ & $[\mathrm{km\,s^{-1}}]$\\
            \rule{0cm}{2.4ex} $\log \dot{M}$  & $-5.4\pm0.1$ & $-7.3\pm0.4$ & $-5.87\pm{0.1}\,^{(a)}$& $-5.75\pm{0.1}\,^{(a)}$ & $[M_\sun\,\textrm{yr}^{-1}]$ \\
            \rule{0cm}{2.4ex} $\varv_\infty$& $\num{2500\pm100}$ & $\num{2500\pm300}$ & - & - & $[\mathrm{km\,s^{-1}}]$\\
            \rule{0cm}{2.4ex} $R$ & $13.2\pm1.7$ & $17.3\pm2.1$ &  $13.5\pm{0.3}$& $16.3\pm{0.5}$ &$[R_\odot]$ \\
            \rule{0cm}{2.4ex} $R_\mathrm{RL}\,^{(b)}$ & $13.1^{+3.8}_{-2.3}$ & $16.8^{+4.8}_{-3.3}$ &  $13.5\pm{0.3}$& $16.3\pm{0.5}$ &$[R_\odot]$ \\
            \rule{0cm}{2.4ex} $R/R_\mathrm{RL}$ & $1.01$ & $1.03$ &  $1$& $1$ & \\
            \rule{0cm}{2.4ex} $M$ & $32^{+19}_{-13}$ & $55^{+32}_{-21}$ & $33\pm{2}$ & $50\pm{5}$& $[M_\odot]$\\
            \rule{0cm}{2.4ex}$X_\mathrm{H}$ & $0.60\pm0.05$ & 0.7375 & $0.51\pm0.1$ & $0.71\pm0.02$ & $[\mathrm{mass\,fr.}]$\\
                \rule{0cm}{2.4ex}$X_\mathrm{He}$ & $0.39\pm0.05$ & 0.26 &  $0.50\pm0.1$ & $0.28\pm0.02$ & $[\mathrm{mass\,fr.}]$\\
                \rule{0cm}{2.4ex}$X_\mathrm{C}\times 10^5$ & $1^{+1}_{-0}$ & $21\pm5$ & $2\pm2$ & $26\pm2$ & $[\mathrm{mass\,fr.}]$\\
                \rule{0cm}{2.4ex}$X_\mathrm{N}\times 10^5$ & $138^{+10}_{-30}$ & $40^{+20}_{-10}$ & $138\pm2$ & $43\pm5$ &  $[\mathrm{mass\,fr.}]$\\
                \rule{0cm}{2.4ex}$X_\mathrm{O}\times 10^5$ & $4^{+6}_{-1}$ & 80$^{(c)}$& $4\pm2$ & $80\pm5$ & $[\mathrm{mass\,fr.}]$\\ 
            \hline
            
            \hline
        \end{tabular}
        \rule{0cm}{2.8ex}%
                \tablefoot{
                \ignorespaces
                $^{(a)}$ Only wind mass-loss rates, according to the wind recipe used in our calculations. $^{(b)}$ Calculated using the period $P=\SI{3.07}{d}$, as from the orbital analysis (see Sect.~\ref{sec:orb_analysis}), and using the spectroscopic mass ratio.  $^{(c)}$ Set to fulfil ${\Sigma\,\mathrm{CNO}_\mathrm{i}\approx \mathrm{const.} = \num{137e-5}}$.
                }
        \label{table:SSN7_binary_fit_parameters}
    \end{table*}

\section{Stellar evolution modelling}
\label{sec:stellar_evo_mod}

    In the previous sections we determine the stellar and wind parameters of the binary components, and refine the binary period, and we establish that the system is most likely in contact. With this information at hand, we proceed to model the past, present, and future evolution of this remarkable system.

    The evolution of SSN~7 in this work is modelled using the Modules for Experiments in Stellar Astrophysics \citep[MESA;][]{Paxton2011,Paxton2013,Paxton2015,Paxton2018,Paxton2019} code (v.15140). Our  aim is to understand, from the predictions of the stellar evolutionary models, the current evolutionary stage of our target rather than finding a perfect fine-tuned fit for our empirically derived parameters. The orbital and spectroscopic parameters (Sect.~\ref{sec:analysis_PoWR} and Sect.~\ref{sec:analysis_components}) limit the possible previous evolutionary paths. To find a fitting model, we explore a parameter space in the following ranges: donor masses in the range  ${M_{1,\,\mathrm{i}} = \SIrange{45}{65}{\msun}}$, initial accretor masses with ${M_{2,\,\mathrm{i}} = \SIrange{30}{55}{\msun}}$, and initial orbital periods in the range  ${P_{\mathrm{i}} = \SIrange{2}{7}{d}}$.

\subsection{Stellar and binary input physics}
    
    The employed binary evolutionary models are constructed using the input files provided by \citet{mar1:16}.\footnote{\url{https://github.com/orlox/mesa_input_data/tree/master/2016_binary_models}} A brief summary of the assumed input physics follows.
   
    In the binary models, following the work of \citet{bro1:11}, tailored initial abundances are used. The initial abundances  are $X_\mathrm{H}= 0.7460$, $X_\mathrm{He}=0.2518$, and $Z=0.0022$, with $Z$ being the total metal fraction. The individual initial metal abundances are listed in Table~\ref{table:initial_Z_MESA}. The abundances are comparable to fit in our atmospheric model (see Table~\ref{table:ion_levels}).

    \begin{table}
        \footnotesize
        \center
        \caption{Initial chemical abundances of our stellar evolutionary models.}
        \begin{tabular}{ c r c } 
            \hline
            \hline
            \rule{0cm}{2.8ex}
            \centering
                    element & \multicolumn{1}{c}{mass fraction} & reference\\
            \hline \rule{0cm}{2.8ex}%
            \rule{0cm}{2.4ex} $\mathrm{C}$  & $20.82\times10^{-5}$  & \citet{1998RMxAC...7..202K}\\
            \rule{0cm}{2.4ex} $\mathrm{N}$  & $3.28\times10^{-5}$   & \citet{1998RMxAC...7..202K}\\
            \rule{0cm}{2.4ex} $\mathrm{O}$  & $113.07\times10^{-5}$ & \citet{1998RMxAC...7..202K}\\
            \rule{0cm}{2.4ex} $\mathrm{Mg}$ & $9.32\times10^{-5}$   & \citet{tru1:07}\\
                                            &                       & \citet{hun1:07}\\
            \rule{0cm}{2.4ex} $\mathrm{Si}$ & $1.30\times10^{-5}$   & \citet{tru1:07}\\
                                            &                       & \citet{hun1:07}\\
            \rule{0cm}{2.4ex} $\mathrm{Fe}$ & $33.74\times10^{-5}$  & \citet{1999ApJ...518..405V} \vspace{0.15cm}\\
        \hline
        \end{tabular}
         \tablefoot{All elements that are not listed here correspond to the solar abundances from \citet{asp1:05} scaled down by a factor of $1/5$.}
        \label{table:initial_Z_MESA}
    \end{table}

    \begin{figure*}[t]
        \centering
        \includegraphics[width=\hsize]{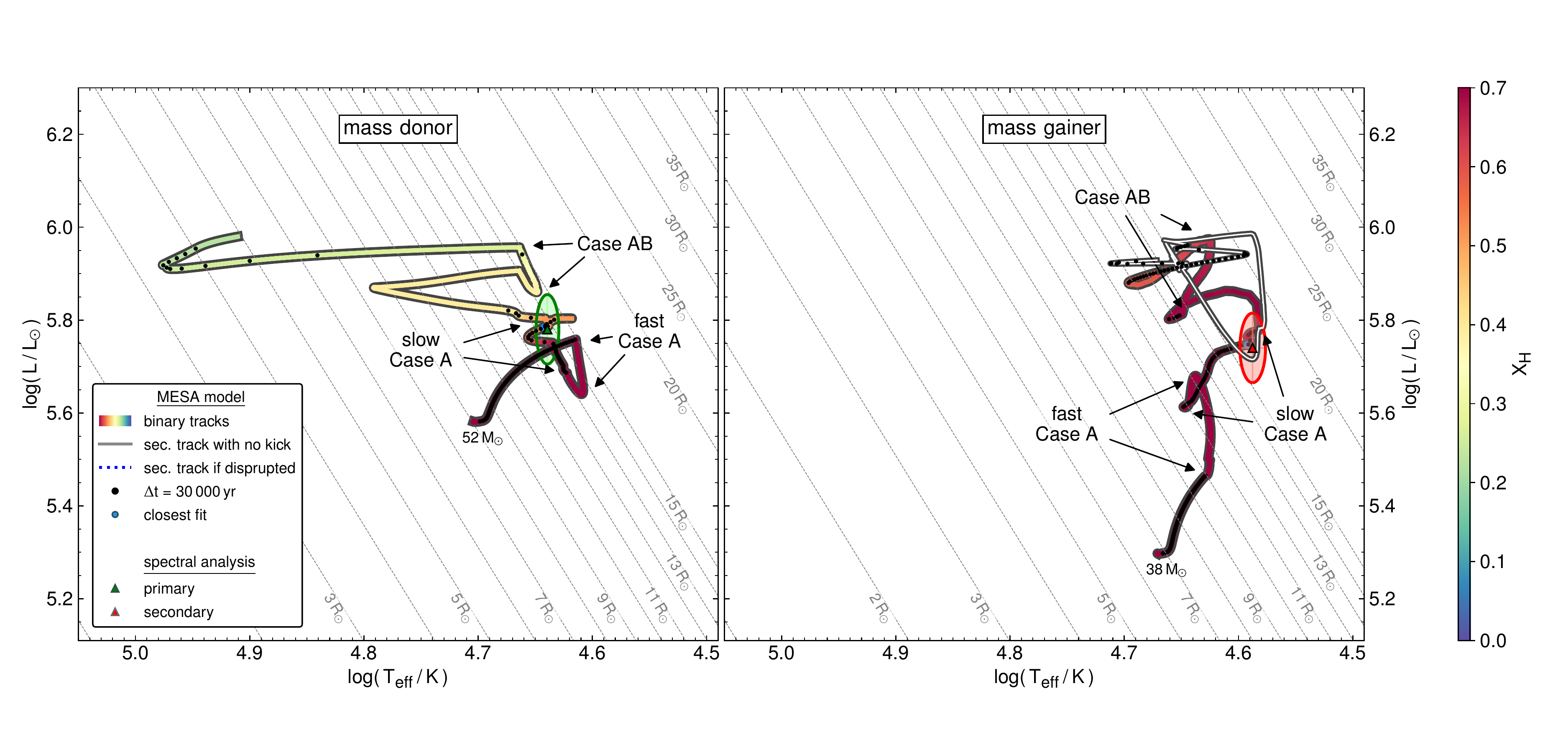}
        \caption{Evolutionary tracks of the donor (left) and the accretor (right) colour-coded by their surface H-abundance. The track of the mass gainer is only colour-coded until the primary dies and then continues as a solid white line,  assuming that the binary is not disrupted by a supernova explosion. The tracks are overlayed by black dots that show equidistant time steps of $\SI{30\,000}{yr}$ to highlight phases in which the stars spend most of their time. The spectroscopic results are shown as green and red triangles surrounded by error ellipses. Start and end phases of fast and slow Case A, as well as Case AB mass-transfer are indicated by arrows. The iso-contours of equal radii are indicated in
the background as dashed grey lines.}
        \label{fig:MESA_tracks}
    \end{figure*}
    
    Mixing is included as follows. Convection is modelled as standard mixing length theory \citep[MLT,][]{boe1:58} using the Ledoux criterion and $\alpha_\mathrm{mlt}=1.5$. Semiconvective mixing is assumed to be efficient with $\alpha_\mathrm{sc}=1$ \citep{lan1:83,sch1:19}. For core hydrogen-burning stars, the step overshooting is assumed to be up to $0.335H_\mathrm{P}$ \citep{bro1:11,sch1:19}. Furthermore, rotational mixing is included as a diffuse process, including dynamical shear, secular shear, and  Goldreich-Schubert-Fricke instabilities, as well as Eddingtion-Sweet circulations \citet{Heg1:00}. The efficiency parameters are set to  $f_c = 1/30$ and $f_\mu = 0.1$. Thermohaline mixing is modelled with $\alpha_\mathrm{th}=1$ \citep{kip1:80}. Lastly, the models take into account momentum transport originating from a magnetic field as expected from a Taylor-Spruit dynamo \citep{spr1:02}.
    
    To avoid numerical complications in the late evolutionary stages of the star, the evolution of both components is only modelled until core helium depletion. Furthermore, for core helium burning stars with core masses above $>14\,\msun$ a more efficient mixing theory, MLT++ \citep[section 7.2 of][]{Paxton2013}, is employed.

    Mass-loss via a stellar wind is included in the models in the following way. For hydrogen-rich ($X_\mathrm{H}\geq0.7$) OB stars, the recipe of \citet{2001A&A...369..574V} is used. As soon as the temperature of any model drops below the bi-stability jump \citep[][their equations 14 and 15]{2001A&A...369..574V}, the maximum of the mass-loss rate of \citet{2001A&A...369..574V} or \citet{nie1:90} is taken. During the WR stage, when the surface hydrogen abundance drops below $X_\mathrm{H}\leq0.4$, the mass loss rates of \citet{she1:19,she2:19} are used. For surface abundances between ${0.7>X_\mathrm{H}>0.4}$, the mass-loss rate is linearly interpolated between the mass-loss rates of \citet{2001A&A...369..574V} and \citet{she1:19,she2:19}.

    To reduce the free parameter space, it is presumed that the components of SSN~7 are initially tidally locked and that the orbit is circularised. This is a reasonable approach for a binary with a period as short as $\SI{3.1}{d}$. 
    
    Mass transfer by Roche-lobe overflow is modelled implicitly using the contact scheme of the MESA code. The mass transfer is modelled   conservatively as long as the mass-gainer is below critical rotation. Whenever the mass-gainer spins up to critical rotation, it will stop the accretion, and we assume that the material is directly lost from the system. In addition to the mass transfer by Roche-lobe overflow, mass accretion from the stellar wind is taken into account.
    
    At the instance of time when the donor star depletes helium in its core, its model is stopped. It is assumed that the primary directly collapses into a BH without a supernova explosion and without a kick. Further mass transfer via Roche-lobe overflow and wind from the companion  onto the BH accretion onto the BH is limited by the Eddington accretion rate.
    
\subsection{Resulting binary models}
\label{sec:MESA_results}

    The best fit of the empirically derived fundamental stellar parameters of both binary components is achieved with an initial primary mass of $52\,\msun$, an initial secondary mass of $38\,\msun$, and an initial orbital period of $\SI{2.7}{d}$. The corresponding tracks of the mass donor and mass gainer in the Hertzsprung--Russell diagram (HRD) are shown in Fig.~\ref{fig:MESA_tracks}. The binary evolutionary model that matches the spectral fit parameters best predicts that the observed stars are currently in contact in a slow Case A mass-transfer phase. This prediction would make SSN~7 the most massive Algol-like system known so far, making it an ideal laboratory to study the ongoing effects and efficiencies of mass transfer in binaries hosting massive stars.

    The parameters of the closest fitting binary evolution model are listed in Table~\ref{table:SSN7_binary_fit_parameters}. Errors quoted are sophisticated guesses on the uncertainties of the binary evolutionary models from the experiences we gained when tailoring the evolutionary models to match the empirically derived stellar parameters. For more accurate error margins, a fine grid of detailed stellar evolution models is needed.
    
    The empirically derived fundamental stellar parameters as well as the observed period of $\SI{3.1}{d}$ can be explained by our binary evolutionary model. Our favourite model predicts that the binary has an age of $\SI{4.2}{Myr}$. The only discrepancy found is between the predicted and observed rotation rates. We discuss this in detail in Sect.~\ref{sec:vrot}.

    It is predicted that the primary, after it depletes the hydrogen in its core,  expands and initiates another more rapid mass-transfer phase, also called Case AB mass-transfer. During this phase it  strips off large fractions of its H-rich envelope and evolves bluewards in the HRD towards the regime where WR stars can be found. During its WR phase, our model is predicted to have some H remaining in the envelope (${X_\mathrm{H}\sim0.2}$). During this time, the secondary, which will be rejuvenated from mass accretion, is still core hydrogen burning.
    
    After the primary, which is the current mass donor, depletes the  helium in its core, we assume that it directly collapses into a BH. The formed BH has a mass of $25.5\,\msun$, which should be considered as an upper limit since we assumed a direct collapse. However, this assumption allows us to make predictions about the future evolution of the secondary. Our model predicts that the secondary will evolve off the main sequence  and will initiate mass transfer on the BH, stripping parts of its H-rich envelope (${X_\mathrm{H}\sim0.4}$). This star does not evolve bluewards to the WR regime and the Of/WNh population. This is linked to the increased envelope-to-core mass ratio, leading to the formation of a steep chemical gradient \citep[][]{pauli2023,2018A&A...611A..75S}. After a short time (of the order of a few hundred thousand years), the remaining star will also collapse to a BH with a mass of $25\,\msun$. The two black holes orbit each other every $\SI{4.85}{d}$. Following \citet{peters1964} we calculated the merger  timescale of the binary to be $\tau_\mathrm{merger} = \SI{18.5}{Gyr}$. This is longer than the age of the Universe. We note that our estimated value should be considered a lower limit as the BH masses of the individual components could be lower, which would result in even longer merger timescales.
    
\section{Discussion}
\label{sec:discussion}
    
\subsection{Orbital, spectroscopic, and stellar evolutionary masses}
    
    The most reliable mass estimate is the projected orbital mass of the system (see Sect.~\ref{sec:orb_analysis}). Since we do not have a light curve available, it is only possible to compare it with the spectroscopic result by choosing an inclination. However, this introduces additional uncertainties and makes the orbital masses inaccurate. A better way to compare the results is by using the derived mass ratios. 
    
    In Section~\ref{sec:orb_analysis} we obtained $q_\mathrm{wilson}=1.34\pm0.07$ using the fitting method from \citet{1941ApJ....93...29W}  and $q_\mathrm{PHOEBE}=1.47^{+0.13}_{-0.09}$ when modelling the RV curve in detail with the PHOEBE code. From our spectroscopic analysis, we derived a somewhat higher mass ratio of $q_\mathrm{spec}=1.7^{+2.9}_{-1.0}$. The large error margins arise from the uncertainties from fitting the surface gravity and from the uncertainties introduced in the multiple solutions that can be found for the luminosity ratio (i.e. ${M_\mathrm{spec}\propto R^2\propto L}$). Hence, it could be that the spectroscopic mass ratio is overestimated. However, the conclusion that the primary component is more massive than the secondary component remains.

    During the tailoring process of our evolutionary model towards the empirically derived stellar parameters, we exploited various combinations of the mass ratio of the two components. In these models, the mass and the position of the primary were always only explainable when it was in a slow Case A mass-transfer phase. The secondary in these models had different initial masses, and thus the mass during the Case A mass transfer was also different (i.e. lower) compared to our best-fitting model.

    \begin{figure}[t]
        \centering
        \includegraphics[trim={1cm 0.5cm 1.5cm 1cm},clip, width=\hsize]{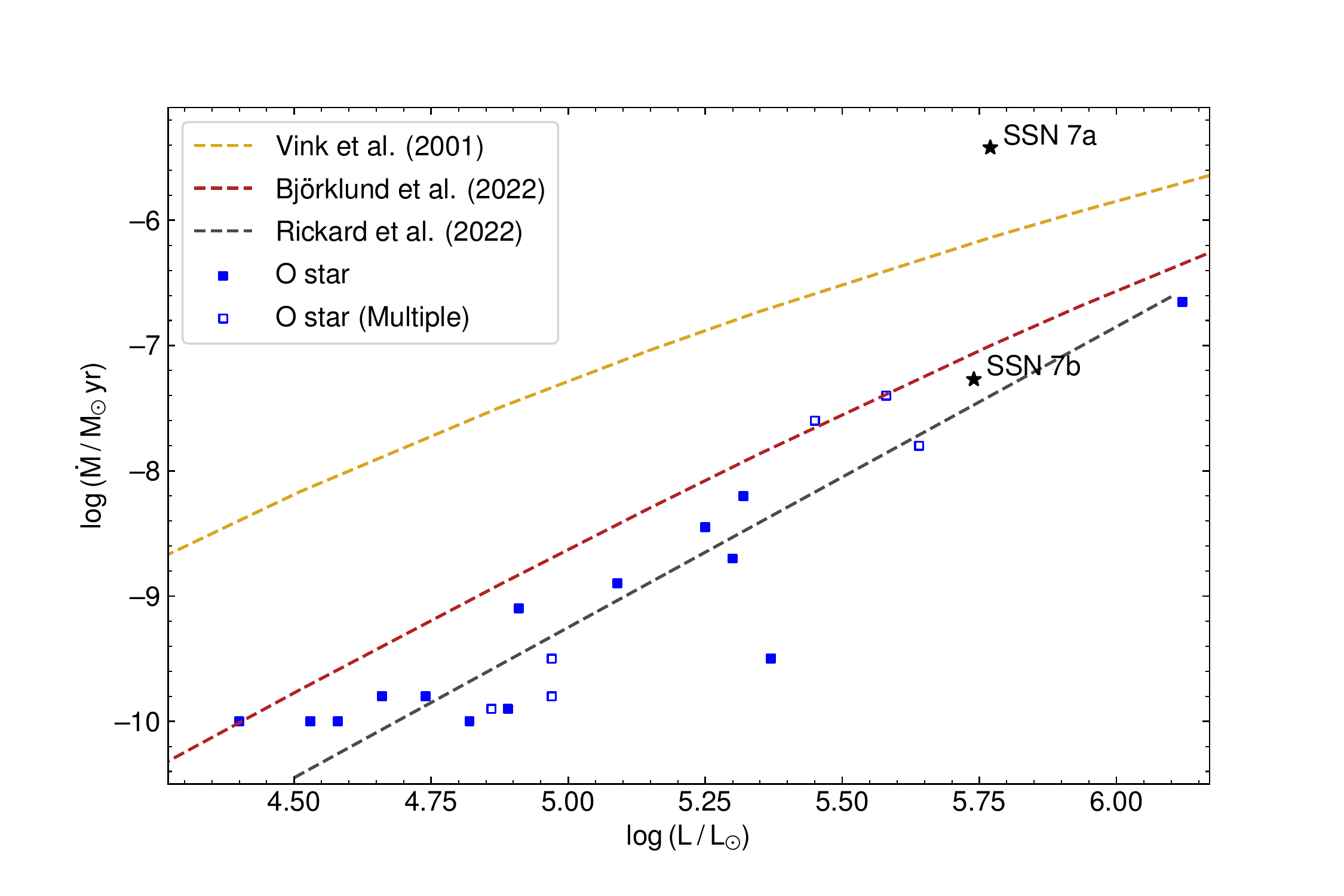}
        \caption{Observed mass-loss rates of O stars in NGC\,346 (including SSN~7) as a function of stellar luminosity compared to theoretical and empirical relations. The position of both binary components of SSN~7 are marked by black stars. The data is complemented by O stars from \citet{2022RICKARDa}. Apparently single stars are shown as filled blue squares and known muliple systems as open blue squares. In the background the empirical relation of main-sequence O-type stars from \citet{2022RICKARDa} is included as dashed black line. In addition, we show the theoretical relations of \citet{2001A&A...369..574V} and \citet{2022arXiv220308218B} for main-sequence stars as dashed yellow and dashed red line, respectively.}
        \label{fig:Mdot_LogL}
    \end{figure}

\subsection{Effects of the inclination on the final parameters}

\subsubsection{Fundamental stellar and wind parameters}
    
    As discussed in the previous section, the orbital mass ratio and the spectroscopic mass ratio agree within their respective uncertainties. This agreement enabled us to estimate  the inclination of the system to be ${i_\mathrm{orb}=\SI{16\pm1}{^\circ}}$ (see Sect.~\ref{sec:PHOEBE}). This implies that we must be looking at the system pole-on. Given our estimates on the rotational velocities and the fact that  the two stars are in contact, they must be oblate and deformed. Hence, the stellar parameters of these stars might no longer be uniform over the surface of the star.
    
    Under the assumption of a star rotating close to break-up velocity, the effective surface gravity at the equator is ${g_\mathrm{equator}\approx0}$ as the gravitational and centrifugal forces cancel each other out. On the other hand, the effective surface gravity on the pole must be rather unperturbed and is mostly affected by gravity. Since we are looking at the pole region of the star, we conclude that our estimate on the surface gravity is reliable within its uncertainties, and hence also our estimate on the inclination.
    
    Due to the effect of gravity darkening, an ellipsoidal star is cooler at the equator and hotter at the poles. \citet{Abdul-Masih2023} studied the effect of rotation and inclination on the determined effective temperature by using synthetic models created with the {\sc spamms} code on a three-dimensional surface. He concluded that in the case of rapid rotation and low inclination, the star’s temperature can be overestimated by $10\%$. This means that both of our stars actually might be $\SIrange{3}{4}{kK}$ cooler. Furthermore, \citet{Abdul-Masih2023} finds that the surface helium abundance of a rapidly rotating pole-on star can be underestimated by as much as $60\%$, which would account for the discrepancy between the observed helium abundance and the helium abundance from the binary evolutionary models.

    Given the ellipsoidal shape and hence the different radii and temperatures, it might be evident that the star’s luminosity also depends on the inclination angle. We calculated the expected luminosity of our target at the pole and the equator and find a difference ${\Delta \log L/\lsun \approx 0.15}$, which is within our given uncertainties. The impact of a somewhat lower luminosity on our final results is too small to impact our final conclusions.

    Due to the different effective surface gravities between the poles and the equator, wind parameters vary with inclination. The wind is slower and weaker in the equatorial regions, while it is faster and more powerful at the poles. An average mass-loss rate for the whole star would thus be lower than any measured from a low inclination angle, such as in our case.  We would therefore treat our stated mass-loss rates as upper limits.

\subsubsection{Discrepancy between observed and predicted rotation velocity}
\label{sec:vrot}
    The binary evolutionary models agree with almost all spectroscopically derived stellar parameters, except the rotational velocity. Under the assumption that the inclination determined from the orbital analysis (see Sect.~\ref{sec:orb_analysis}) is reliable, the observed rotation velocities of the binary components  (${\varv_1=\SI{490\pm47}{km\,s^{-1}}}$ and ${\varv_2=\SI{544\pm49}{km\,s^{-1}}}$) are a factor of two higher than those predicted by the binary evolutionary models.

    Given that we are looking at the binary pole-on, and that both stars are deformed due to their fast rotation, this means that the assumption of spherical symmetry breaks down in the evolutionary models. 
    \citet{zahn2010} have shown this for a uniformly rapidly rotating star ${R_\mathrm{equator}\approx1.5 R_\mathrm{pole}}$. Hence, under the assumption that ${R\approx R_\mathrm{pole}}$, the rotation rates of the binary model need to be corrected to ${\varv_\mathrm{mod,\,\,1}\approx\SI{325\pm45}{km\,s^{-1}}}$ and ${\varv_\mathrm{mod,\,\,2}\approx\SI{425\pm50}{km\,s^{-1}}}$. This brings observations and predictions closer together, but still shows some discrepancy. We note that our target is a contact binary, meaning that the stars are tear-drop shaped rather than ellipsoidal. This makes an estimate of the rotation rates even more complicated.

    On the other hand, it is worth mentioning that in our stellar atmosphere calculations we neglect additional broadening mechanisms such as macroturbulence and asynchronous rotation (e.g. introduced by differential rotation). These effects are compensated for in our estimate of the projected rotational velocity $\varv \sin i$, and hence might lead to an overestimation, which can explain the observed discrepancy.

\subsection{Empirical mass-loss rates}

    From our spectroscopic analysis (Sect.~\ref{sec:mass_loss_rates_fitting}) we derived that the primary dominates most of the wind lines (${\log\,( \dot{M}_1 \, / \, (\msunpyr) )= -5.4}$), while the secondary still has a significant and noticeable contribution in the wind UV lines (${\log\,( \dot{M}_2 \, / \, (\msunpyr) )=-7.3}$). In the following we want to determine whether the chosen mass-loss rates agree with observations and theoretical expectations.

    Following \citet{2022RICKARDa}, we would expect from a star that has evolved in isolation and has a current luminosity of $\log(L/\lsun)=5.75$ a mass-loss rate of only ${\log\,( \dot{M}_\mathrm{Rickard} \, / \, (\msunpyr) )=-7.35}$. This is two orders of magnitude lower than that derived for the primary from empirically spectroscopic analysis. However, the primary has lost a huge amount of mass and has an unusual $L /\/ M$ ratio that is not considered in their relation. On the other hand, theoretical mass-loss recipes take into account several parameters, such as luminosity, mass, temperature and metallicity. For comparison, we calculated the mass-loss rate predicted with the frequently used \citet{2001A&A...369..574V} to be ${\log\, (\dot{M}_\mathrm{Vink, \ 1} \, / \, (\msunpyr)) = -5.9}$. A more updated recipe for O-type stars is from \citet{2022arXiv220308218B}. Their mass-loss recipe predicts ${\log\,( \dot{M}_{\mathrm{Bj\"orklund \ 2022,\ 1 }}\, / \, (\msunpyr) )=-6.6}$, which is one order of magnitude lower than the observed mass-loss rate. However, these mass-loss rates are calculated for H-rich ($X_\mathrm{H}>0.7$). The primary has already lost a significant amount of its envelope, and its surface H-abundance already dropped to $X_\mathrm{H}\approx0.6$ (see Table~\ref{table:SSN7_binary_fit_parameters}). For completeness, we calculated the corresponding WR mass-loss rates according to \citet{she1:19,she2:19} for stars with $X_\mathrm{H}>0.4$, yielding ${\log\, (\dot{M}_\mathrm{Shenar, \ 1} \, / \, (\msunpyr)) = -5.5}$ and being in agreement with the empirically derived values. 

    The secondary seems to be more well behaved. Its empirically derived mass-loss rates agrees well with the expectations from other O stars in the SMC  \citep{2022RICKARDa}. The theoretical predicted rates from \citet{2001A&A...369..574V} and \citet{2022arXiv220308218B} are ${\log\, (\dot{M}_\mathrm{Vink, \ 2} \, / \, (\msunpyr)) = -6.2}$ and ${\log\,( \dot{M}_{{\mathrm{Bj\"orklund \ 2022}},\ 1} \, / \, (\msunpyr) )=-7.1}$, respectively. Only the latter   agrees within the error margins of the observed mass-loss rates. For comparison, we illustrate in Fig.~\ref{fig:Mdot_LogL} the expected mass-loss rates from other O-type stars in NGC~346 compared to our spectroscopically fit values, alongside theoretical prescriptions.

\subsection{SSN 7 and its role in NGC346}
\label{sec:ionisation}
    In the recent paper of \citet{2022RICKARDa}, it is noted that the massive giant NGC~346 SSN~9 (MPG~355, O2III(f *), \citealt{2000PASP..112.1243W}) is the major ionising source of the core region. In their work it is stated that the star outshines the remaining O-star population and is responsible for 50\% of the hydrogen ionising flux (($\log{Q}_{\mathrm{H,\,SSN\,9}} = 49.98$). However, in their analysis SSN~7 was neglected  because of its binary nature. According to our stellar atmosphere models, we find that the two early-type stars in the binary have a combined hydrogen ionising flux of $\log{Q}_{H,\,\mathrm{SSN7\,a+b}} = 49.88$. This makes SSN~7 one of the two major contributors to the H ionising flux of the core of NGC~346. We stated in \citet{2022RICKARDa} that the morphology of the ionising front within images of NGC~346 centre upon SSN~9. The location of SSN~7 is directly `behind' SSN~9 from the position of the ionising front (Fig.~\ref{fig:NGC_346_Obs}).

\begin{figure}[t]
        \centering
        \includegraphics[width=\hsize]{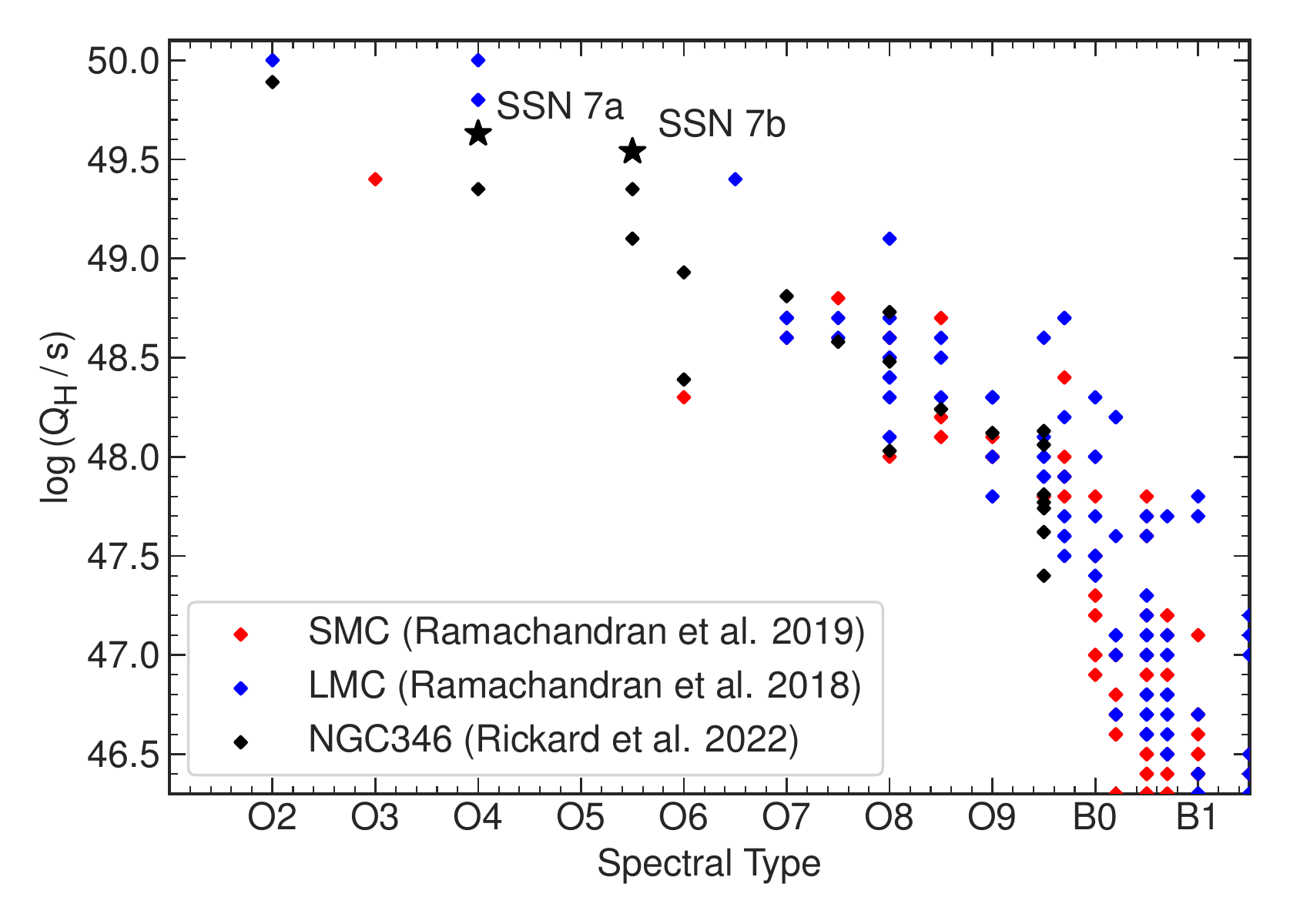}
        \caption{$\log Q_H$ to spectral type. SMC sample stars analysed by \citet{2019A&A...625A.104R} are included (red diamonds) as well as LMC stars from \citet{2018A&A...615A..40R} (blue diamonds) and O stars from the core of NGC~346 from  \citet{2022RICKARDa} (black diamonds). The two components of the SSN~7 binary are shown separately.}
        \label{fig:logQ_to_spt}
    \end{figure}

    We compared the calculated H ionising photon count of both components of SSN~7 to other massive stars within both the SMC and LMC (Fig.~\ref{fig:logQ_to_spt}). The predicted ionising flux is consistent with the spectral types assigned to each component by \citet{2000PASP..112.1243W}, even though the primary has an outstanding L/M ratio and the secondary that is suspected to have accreted several solar masses.
    
    For  helium ionising photons, the model of the primary gives $\log{Q}_{\mathrm{He\,I,\,SSN\,7a}} = 48.98$ and $\log{Q}_{\mathrm{HeII,\,SSN\,7a}} = 40.81$, while the secondary model gives $\log{Q}_{\mathrm{He\,I,\,SSN\,7b}} = 48.58$ and $\log{Q}_{\mathrm{HeII,\,SSN\,7b}} = 43.14$. Combined, the helium ionising fluxes of the two components of the binary combined are $\log{Q}_{\mathrm{\mathrm{HeI,\,SSN7\,a+b}}} = 49.12$ and $\log{Q}_{\mathrm{\mathrm{HeI,\,SSN7\,a+b}}} = 43.14$, or 25\% and 0.04\% of the total of the O stars analysed in the core of NGC~346 between this paper and \citet{2022RICKARDa}. This is compared to $\log{Q}_{\mathrm{HeI,\,SSN~9}} = 49.41$ and $\log{Q}_{\mathrm{HeII,\,SSN~9}} = 46.50$, or 31\% and 99\%. This is notable because the combined \ion{He\,\textsc{i}} ionising flux for the binary is comparable to that of SSN~9, but not for the \ion{He\,\textsc{ii}} ionising flux, where SSN~9 dominates.

\subsection{Age of the NGC~346 O-star population}

\begin{figure}[t]
        \centering
        \includegraphics[width=\hsize]{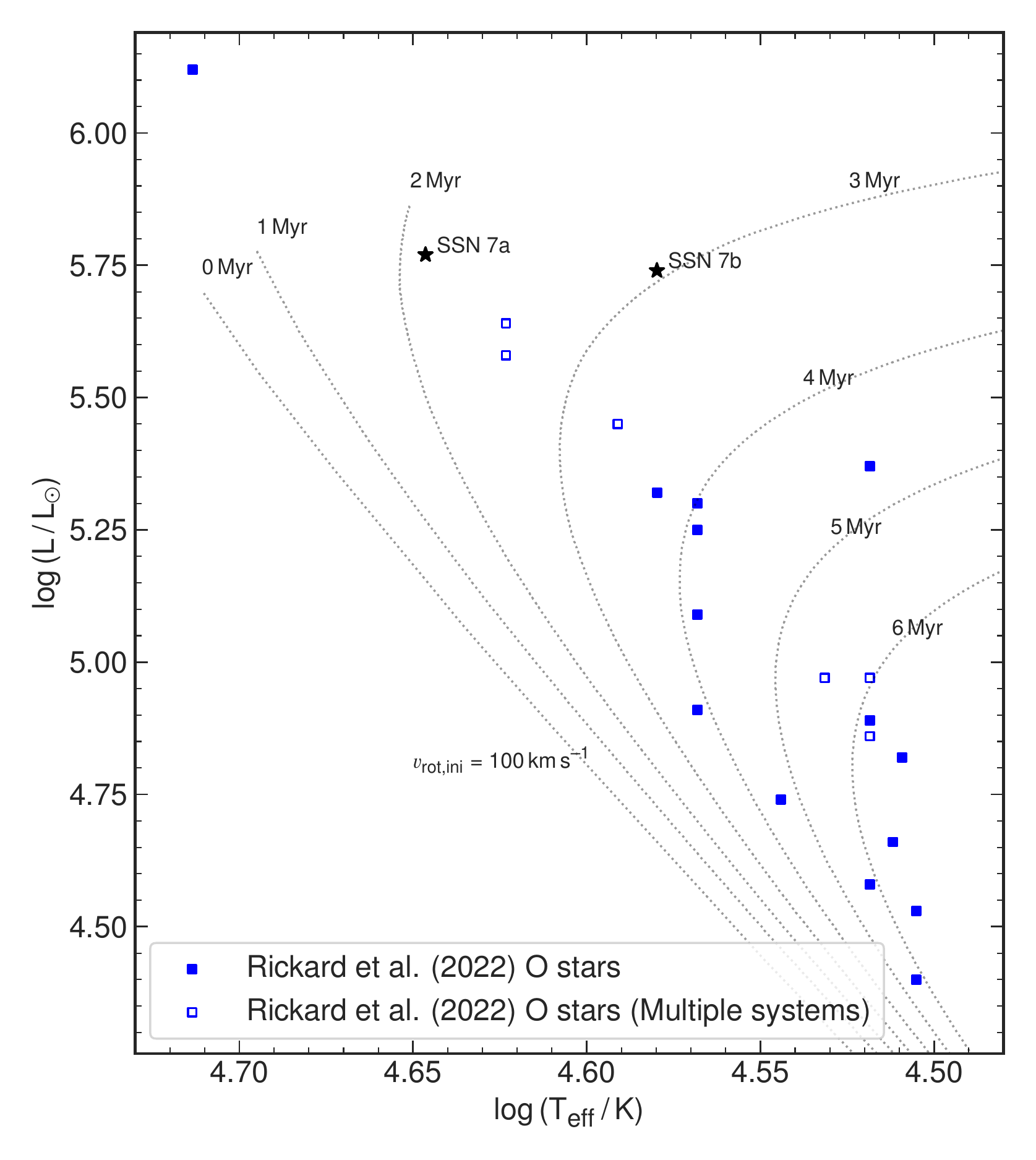}
        \caption{HR diagram showing the position of the two components of SSN 7 as determined by our spectrographic analysis. Black stars are the two SSN 7 components. Blue symbols are the other O stars within the central core of NGC 346 \citep{2022RICKARDa}. Open symbols  are noted spectral binaries \citep{2019AA...626A..50D}. Dotted lines are isochrones with $Z=0.002$ \citep{2013A&A...558A.103G} and an initial rotation of ${100 \ \mathrm{km \ s^{-1}}}$.}
        \label{fig:HR}
    \end{figure}

    During our evolutionary analysis it was revealed that the system of SSN 7 has an age of $\SI{4.2}{Myr}$ (see Sect.~\ref{sec:MESA_results}). Meanwhile, much younger ages for NGC346 have been estimated in previous works, such as $1~-~2.6$~Myr \citep{2019AA...626A..50D} and $1~-~2$~Myr \citep{2000PASP..112.1243W}. In Fig.~\ref{fig:HR} we compare the empirically derived stellar parameters of SSN~7 and other O stars \citep{2022RICKARDa} to isochrones from \citet{2013A&A...558A.103G}. According to these isochrones, the age of the primary should be $\SI{2}{Myr}$, while the secondary would be $\SI{3}{Myr}$ old. This mismatch between the age predicted through evolutionary modelling and the age compared to single-star isochrones is not an isolated example and could lead to false conclusions on the age of the cluster.

    When inspecting Fig.~\ref{fig:HR} in more detail we can see that most of the luminous O stars in NGC~346 are known binaries, or higher order multiple systems, with the one exception of SSN~9. However, this star is unusually luminous and hot and undermassive for its position in the HRD, hinting towards a binary nature. Taking into account only the apparent single O stars in this cluster, we conlcude the age of the O-star population, according to single-star evolution isochromes, to be $\SI{4}{Myr}$, which is consistent with the age derived from our binary models. 

    With evidence that the core of NGC~346 has not yet been affected by any supernova explosion \citep{2003ApJ...586.1179D}, the fate of more massive stars within NGC~346 comes into question. Either none had formed yet, or those that did have imploded into BHs directly.

\subsection{Luminosity class and spectral type}

    In the previous work of \citet{1978ApJ...224L.133W}, SSN~7 was classified as O4 III(n)(f), but it was not known to be a binary. In a recent work,  \citet{2019AA...626A..50D}   propose  that the presence of \HeI{}{4471} absorption and weak \HeII{4686} emission suggest that the primary should be a O4 If star. With the new insights gained from the detailed multi-epoch observations, we assign the \HeI{}{4471} absorption mainly to the secondary component, meaning that the primary has an earlier type.

    Here we want to use the newly gained insights from the spectral analysis to reclassify the primary and secondary star. It is evident that the primary in our model shows only a very weak contribution to the \HeI{4471} line (see Fig.~\ref{fig:CIII_LUMSET}), meaning that it must be of very early type (O2-3.5). Following \citet{2002AJ....123.2754W} it is suggested to use for such early-type stars the ratio of   \NIV{4058} to \NIII{4640} emission. The multi-epoch observations have shown that the \NIII{4640} emission line is a blend of the primary and secondary components, while the  \NIV{4058} purely originates from the primary (see Fig.~\ref{fig:spec_xshooter_metal} and Table~\ref{table:lines_considered}). Guided by our stellar atmosphere models we find that \NIV{4058} $>$ \NIII{4640}, while we see very weak or no absorption in \HeI{4471}, implying a spectral type of O3. 

    According to our spectral analysis we find that in the primary nitrogen is enhanced, while carbon and oxygen are deficient. Therefore, we add the qualifier `N' to the spectral type. Following the classification criteria of \citet{2011ApJS..193...24S} and taking into account that the \NIV{4058} emission feature is greater than the \NIII{4640} emission and that the \HeII{4686} is also in emission we suggest   adding the suffix `f$^\ast$'. In \citet{2014ApJS..211...10S} the luminosity classes are linked to the spectral type as a function of the f phenomenon, suggesting a luminosity class I for the primary. Because of the high temperature of the star, combined with the low \ion{Si}{iv} abundance ($\sim\,2\%$ solar), no \SiIVd{1394,\,1403} P\,Cygni lines are observed \citep{2002ApJ...579..774C}, so the presence or absence of this line cannot contribute to our luminosity classification. The final spectral type we  assign to the primary is ON3\,If$^\ast$. 
    
    Observational counterparts with similar spectral type at low metallicity are sparse. Galactic examples are Cyg~OB2-7 and Cyg~OB2-22A. By comparison of their observed spectra to the synthetic spectrum of our primary we find a good agreement between the morphology of the lines, supporting our choice on the spectral type.

    For the secondary, we limit its spectral type to O4-8 as it shows \HeI{4471} in absorption, while   \HeII{4542}$\,>\,$\HeI{4388} and   \HeII{4200}$\,>\,$\HeI{4144,} and it lacks \ion{Si}{III}\,$\lambda4552$, which can be seen in the earliest O-type stars. Hence, we follow \citet{1971ApJ...170..325C} and use the logarithmic ratio of the equivalent widths of \HeI{4471} and \HeII{4542}. Unfortunately, the \HeII{4542} line is blended by the companion. Therefore, we decided to use the equivalent widths of the synthetic spectrum of the secondary, yielding $\log(\mathrm{EW}(4471)/\mathrm{EW}(4542))=-0.45$. This corresponds to a spectral type of O5.5. As mentioned above, the \NIII{4640} emission line is a blend from the primary and secondary component. As the secondary does not contribute to the \NIV{4058} line and shows strong \HeII{4686} absorption we add the suffix `((f))' \citep{2011ApJS..193...24S}. Following \citet{2014ApJS..211...10S}, given the spectral type and the f phenomenon, the star should have a luminosity class V. We use the criterion for the luminosity class estimation introduced by \citet{2018A&A...616A.135M}. The criterion is based on the equivalent width of \HeII{4686} absorption line in the synthetic model of the secondary. With an equivalent with of ${\mathrm{EW}(4868))=0.74}$, this star is confirmed to have a luminosity class V. The final spectroscopic classification of the binary is ON3\,If$^\ast$+O5.5\,V((f)).

\section{Conclusions}

    We have conducted the first consistent multi-wavelength multi-epoch spectral analysis of the SB 2 contact binary SSN~7 located in the NGC~346 cluster in the SMC. The detailed analysis revealed that SSN~7 is a system with two binary components that have comparable luminosity ($\log(L/\lsun)\approx5.75$), while having divergent stellar and wind parameters. The primary is the hotter ($T_\mathrm{eff,\,1}=\SI{43.6}{kK}$) but less massive ($M_1=32\,\msun$) component. We revealed that the surface hydrogen abundance is $X_\mathrm{H}=0.6$, and that it has CNO values close to those of the CNO equilibrium. The mass-loss rate of the primary,   ${\log\,( \dot{M}_1 \, / \, (\msunpyr) )= -5.4}$, is exceptionally high for its luminosity. The secondary is slightly cooler ($T_\mathrm{eff,\,2}=\SI{38.7}{kK}$), but more massive  ($M_2=55\,\msun$). The secondary has a mass-loss rate of ${\log\,( \dot{M}_2 \, / \, (\msunpyr) )= -7.3}$, which is in agreement with other O stars within the SMC that have similar luminosity. The findings suggest that the two binary components exchange mass. Based on the newly gained insights we reclassified the system as ON3\,If$^\ast$+O5.5\,V((f)).

    The multi-epoch RV analysis of 19 separated observations, including 11 spaced over ten days, allowed us to constrain the orbital parameters. The best fit is achieved with a period of ${P=\SI{3.07}{d}}$ and a circular orbit. The orbital analysis confirms a mass ratio of ${q_\mathrm{orb}=1.5}$, favouring the secondary to be more massive and supporting the theory of a binary interaction. Furthermore, given the orbital and stellar parameters, we calculated the Roche radius and found that both components are filling their Roche lobes and must be in contact.

    From initial abundances consistent with NGC~346, we have modelled possible stellar evolution pathways that explain the empirically derived stellar and orbital parameters of SSN~7. Our favourite model predicts that the system has an age of $\sim 4$~Myr. In this context, we review the age of the O-star  population within the core of NGC~346 and see that no single star (except for the peculiar massive object SSN~9) is to the left of the 4 Myr isochrone. We postulate that the age of the O-star population within the core of NGC~346 is older than $\gtrsim4$~Myr.

    The binary evolutionary models are capable of explaining the currently observed contact phase and predict that both stars are still undergoing slow Case A mass transfer. This makes SSN~7 the most massive Algol-like system known to date, enabling the study of the physics of binary interactions in a way previously unavailable. This will shape our understanding of post-interaction massive binaries, binary BHs, and ultimately, the progenitors of GW events.

    \begin{acknowledgements}
        DP acknowledges financial support by the Deutsches Zentrum  f\"ur  Luft  und  Raumfahrt  (DLR)  grant  FKZ  50  OR  2005. MJR and DP  acknowledge Prof. Wolf-Rainer Hamann, Prof. Lida Oskinova, Dr. Rainer Hainich, and Dr. Heldge Todt of the Institut f\"{u}r Physik und Astronomie, Universit\"{a}t Potsdam. Dr. Andreas Sander and Dr. Varsha Ramachandran of Zentrum für Astronomie der Universit\"{a}t Heidelberg, Astronomisches Rechen-Institut, and Dr. Tomer Shenar of the Institut for Astronomy, University of Amsterdam, and all other contributors to the PoWR code. This publication has benefited from a discussion at a team meeting sponsored by the International Space Science Institute at Bern, Switzerland.
    \end{acknowledgements}

%-use BibTeX with the regular commands:
\bibliographystyle{aa} % style aa.bst
\bibliography{bib} % your references Yourfile.bib

\begin{thebibliography}{114}
\expandafter\ifx\csname natexlab\endcsname\relax\def\natexlab#1{#1}\fi

\bibitem[{{Abbott} {et~al.}(2016){Abbott}, {Abbott}, {Abbott}, {Abernathy},
  {Acernese}, {Ackley}, {Adams}, {Adams}, {Addesso}, {Adhikari}, {Adya},
  {Affeldt}, {Agathos}, {Agatsuma}, {Aggarwal}, {Aguiar}, {Aiello}, {Ain},
  {Ajith}, {Allen}, {Allocca}, {Altin}, {Anderson}, {Anderson}, {Arai},
  {Araya}, {Arceneaux}, {Areeda}, {Arnaud}, {Arun}, {Ascenzi}, {Ashton}, {Ast},
  {Aston}, {Astone}, {Aufmuth}, {Aulbert}, {Babak}, {Bacon}, {Bader}, {Baker},
  {Baldaccini}, {Ballardin}, {Ballmer}, {Barayoga}, {Barclay}, {Barish},
  {Barker}, {Barone}, {Barr}, {Barsotti}, {Barsuglia}, {Barta}, {Bartlett},
  {Bartos}, {Bassiri}, {Basti}, {Batch}, {Baune}, {Bavigadda}, {Bazzan},
  {Bejger}, {Bell}, {Berger}, {Bergmann}, {Berry}, {Bersanetti}, {Bertolini},
  {Betzwieser}, {Bhagwat}, {Bhandare}, {Bilenko}, {Billingsley}, {Birch},
  {Birney}, {Biscans}, {Bisht}, {Bitossi}, {Biwer}, {Bizouard}, {Blackburn},
  {Blair}, {Blair}, {Blair}, {Bloemen}, {Bock}, {Boer}, {Bogaert}, {Bogan},
  {Bohe}, {Bond}, {Bondu}, {Bonnand}, {Boom}, {Bork}, {Boschi}, {Bose},
  {Bouffanais}, {Bozzi}, {Bradaschia}, {Brady}, {Braginsky}, {Branchesi},
  {Brau}, {Briant}, {Brillet}, {Brinkmann}, {Brisson}, {Brockill}, {Broida},
  {Brooks}, {Brown}, {Brown}, {Brown}, {Brunett}, {Buchanan}, {Buikema},
  {Bulik}, {Bulten}, {Buonanno}, {Buskulic}, {Buy}, {Byer}, {Cabero},
  {Cadonati}, {Cagnoli}, {Cahillane}, {Calder{\'o}n Bustillo}, {Callister},
  {Calloni}, {Camp}, {Cannon}, {Cao}, {Capano}, {Capocasa}, {Carbognani},
  {Caride}, {Casanueva Diaz}, {Casentini}, {Caudill}, {Cavagli{\`a}},
  {Cavalier}, {Cavalieri}, {Cella}, {Cepeda}, {Cerboni Baiardi}, {Cerretani},
  {Cesarini}, {Chamberlin}, {Chan}, {Chao}, {Charlton}, {Chassande-Mottin},
  {Cheeseboro}, {Chen}, {Chen}, {Cheng}, {Chincarini}, {Chiummo}, {Cho}, {Cho},
  {Chow}, {Christensen}, {Chu}, {Chua}, {Chung}, {Ciani}, {Clara}, {Clark},
  {Cleva}, {Coccia}, {Cohadon}, {Colla}, {Collette}, {Cominsky}, {Constancio.},
  {Conte}, {Conti}, {Cook}, {Corbitt}, {Cornish}, {Corsi}, {Cortese}, {Costa},
  {Coughlin}, {Coughlin}, {Coulon}, {Countryman}, {Couvares}, {Cowan},
  {Coward}, {Cowart}, {Coyne}, {Coyne}, {Craig}, {Creighton}, {Cripe},
  {Crowder}, {Cumming}, {Cunningham}, {Cuoco}, {Dal Canton}, {Danilishin},
  {D'Antonio}, {Danzmann}, {Darman}, {Dasgupta}, {Da Silva Costa}, {Dattilo},
  {Dave}, {Davier}, {Davies}, {Daw}, {Day}, {De}, {DeBra}, {Debreczeni},
  {Degallaix}, {De Laurentis}, {Del{\'e}glise}, {Del Pozzo}, {Denker}, {Dent},
  {Dergachev}, {De Rosa}, {DeRosa}, {DeSalvo}, {Devine}, {Dhurandhar},
  {D{\'\i}az}, {Di Fiore}, {Di Giovanni}, {Di Girolamo}, {Di Lieto}, {Di Pace},
  {Di Palma}, {Di Virgilio}, {Dolique}, {Donovan}, {Dooley}, {Doravari},
  {Douglas}, {Downes}, {Drago}, {Drever}, {Driggers}, {Ducrot}, {Dwyer}, {Edo},
  {Edwards}, {Effler}, {Eggenstein}, {Ehrens}, {Eichholz}, {Eikenberry},
  {Engels}, {Essick}, {Etzel}, {Evans}, {Evans}, {Everett}, {Factourovich},
  {Fafone}, {Fair}, {Fairhurst}, {Fan}, {Fang}, {Farinon}, {Farr}, {Farr},
  {Favata}, {Fays}, {Fehrmann}, {Fejer}, {Fenyvesi}, {Ferrante}, {Ferreira},
  {Ferrini}, {Fidecaro}, {Fiori}, {Fiorucci}, {Fisher}, {Flaminio}, {Fletcher},
  {Fournier}, {Frasca}, {Frasconi}, {Frei}, {Freise}, {Frey}, {Frey},
  {Fritschel}, {Frolov}, {Fulda}, {Fyffe}, {Gabbard}, {Gair}, {Gammaitoni},
  {Gaonkar}, {Garufi}, {Gaur}, {Gehrels}, {Gemme}, {Geng}, {Genin}, {Gennai},
  {George}, {Gergely}, {Germain}, {Ghosh}, {Ghosh}, {Ghosh}, {Giaime},
  {Giardina}, {Giazotto}, {Gill}, {Glaefke}, {Goetz}, {Goetz}, {Gondan},
  {Gonz{\'a}lez}, {Gonzalez Castro}, {Gopakumar}, {Gordon}, {Gorodetsky},
  {Gossan}, {Gosselin}, {Gouaty}, {Grado}, {Graef}, {Graff}, {Granata},
  {Grant}, {Gras}, {Gray}, {Greco}, {Green}, {Groot}, {Grote}, {Grunewald},
  {Guidi}, {Guo}, {Gupta}, {Gupta}, {Gushwa}, {Gustafson}, {Gustafson},
  {Hacker}, {Hall}, {Hall}, {Hammond}, {Haney}, {Hanke}, {Hanks}, {Hanna},
  {Hannam}, {Hanson}, {Hardwick}, {Harms}, {Harry}, {Harry}, {Hart}, {Hartman},
  {Haster}, {Haughian}, {Heidmann}, {Heintze}, {Heitmann}, {Hello}, {Hemming},
  {Hendry}, {Heng}, {Hennig}, {Henry}, {Heptonstall}, {Heurs}, {Hild}, {Hoak},
  {Hofman}, {Holt}, {Holz}, {Hopkins}, {Hough}, {Houston}, {Howell}, {Hu},
  {Huang}, {Huerta}, {Huet}, {Hughey}, {Husa}, {Huttner}, {Huynh-Dinh},
  {Indik}, {Ingram}, {Inta}, {Isa}, {Isac}, {Isi}, {Isogai}, {Iyer}, {Izumi},
  {Jacqmin}, {Jang}, {Jani}, {Jaranowski}, {Jawahar}, {Jian},
  {Jim{\'e}nez-Forteza}, {Johnson}, {Jones}, {Jones}, {Jonker}, {Ju}, {K},
  {Kalaghatgi}, {Kalogera}, {Kandhasamy}, {Kang}, {Kanner}, {Kapadia}, {Karki},
  {Karvinen}, {Kasprzack}, {Katsavounidis}, {Katzman}, {Kaufer}, {Kaur},
  {Kawabe}, {K{\'e}f{\'e}lian}, {Kehl}, {Keitel}, {Kelley}, {Kells}, {Kennedy},
  {Key}, {Khalili}, {Khan}, {Khan}, {Khan}, {Khazanov}, {Kijbunchoo}, {Kim},
  {Kim}, {Kim}, {Kim}, {Kim}, {Kim}, {Kim}, {Kimbrell}, {King}, {King},
  {Kissel}, {Klein}, {Kleybolte}, {Klimenko}, {Koehlenbeck}, {Koley},
  {Kondrashov}, {Kontos}, {Korobko}, {Korth}, {Kowalska}, {Kozak}, {Kringel},
  {Krishnan}, {Kr{\'o}lak}, {Krueger}, {Kuehn}, {Kumar}, {Kumar}, {Kuo},
  {Kutynia}, {Lackey}, {Landry}, {Lange}, {Lantz}, {Lasky}, {Laxen},
  {Lazzarini}, {Lazzaro}, {Leaci}, {Leavey}, {Lebigot}, {Lee}, {Lee}, {Lee},
  {Lee}, {Lenon}, {Leonardi}, {Leong}, {Leroy}, {Letendre}, {Levin}, {Lewis},
  {Li}, {Libson}, {Littenberg}, {Lockerbie}, {Lombardi}, {London}, {Lord},
  {Lorenzini}, {Loriette}, {Lormand}, {Losurdo}, {Lough}, {L{\"u}ck},
  {Lundgren}, {Lynch}, {Ma}, {Machenschalk}, {MacInnis}, {Macleod},
  {Maga{\~n}a-Sandoval}, {Maga{\~n}a Zertuche}, {Magee}, {Majorana},
  {Maksimovic}, {Malvezzi}, {Man}, {Mandic}, {Mangano}, {Mansell}, {Manske},
  {Mantovani}, {Marchesoni}, {Marion}, {M{\'a}rka}, {M{\'a}rka}, {Markosyan},
  {Maros}, {Martelli}, {Martellini}, {Martin}, {Martynov}, {Marx}, {Mason},
  {Masserot}, {Massinger}, {Masso-Reid}, {Mastrogiovanni}, {Matichard},
  {Matone}, {Mavalvala}, {Mazumder}, {McCarthy}, {McClelland}, {McCormick},
  {McGuire}, {McIntyre}, {McIver}, {McManus}, {McRae}, {McWilliams}, {Meacher},
  {Meadors}, {Meidam}, {Melatos}, {Mendell}, {Mercer}, {Merilh}, {Merzougui},
  {Meshkov}, {Messenger}, {Messick}, {Metzdorff}, {Meyers}, {Mezzani}, {Miao},
  {Michel}, {Middleton}, {Mikhailov}, {Milano}, {Miller}, {Miller}, {Miller},
  {Miller}, {Millhouse}, {Minenkov}, {Ming}, {Mirshekari}, {Mishra}, {Mitra},
  {Mitrofanov}, {Mitselmakher}, {Mittleman}, {Moggi}, {Mohan}, {Mohapatra},
  {Montani}, {Moore}, {Moore}, {Moraru}, {Moreno}, {Morriss}, {Mossavi},
  {Mours}, {Mow-Lowry}, {Mueller}, {Muir}, {Mukherjee}, {Mukherjee},
  {Mukherjee}, {Mukund}, {Mullavey}, {Munch}, {Murphy}, {Murray}, {Mytidis},
  {Nardecchia}, {Naticchioni}, {Nayak}, {Nedkova}, {Nelemans}, {Nelson},
  {Neri}, {Neunzert}, {Newton}, {Nguyen}, {Nielsen}, {Nissanke}, {Nitz},
  {Nocera}, {Nolting}, {Normandin}, {Nuttall}, {Oberling}, {Ochsner}, {O'Dell},
  {Oelker}, {Ogin}, {Oh}, {Oh}, {Ohme}, {Oliver}, {Oppermann}, {Oram},
  {O'Reilly}, {O'Shaughnessy}, {Ottaway}, {Overmier}, {Owen}, {Pai}, {Pai},
  {Palamos}, {Palashov}, {Palomba}, {Pal-Singh}, {Pan}, {Pankow}, {Pannarale},
  {Pant}, {Paoletti}, {Paoli}, {Papa}, {Paris}, {Parker}, {Pascucci},
  {Pasqualetti}, {Passaquieti}, {Passuello}, {Patricelli}, {Patrick},
  {Pearlstone}, {Pedraza}, {Pedurand}, {Pekowsky}, {Pele}, {Penn}, {Perreca},
  {Perri}, {Phelps}, {Piccinni}, {Pichot}, {Piergiovanni}, {Pierro}, {Pillant},
  {Pinard}, {Pinto}, {Pitkin}, {Poe}, {Poggiani}, {Popolizio}, {Post},
  {Powell}, {Prasad}, {Predoi}, {Prestegard}, {Price}, {Prijatelj}, {Principe},
  {Privitera}, {Prix}, {Prodi}, {Prokhorov}, {Puncken}, {Punturo}, {Puppo},
  {P{\"u}rrer}, {Qi}, {Qin}, {Qiu}, {Quetschke}, {Quintero}, {Quitzow-James},
  {Raab}, {Rabeling}, {Radkins}, {Raffai}, {Raja}, {Rajan}, {Rakhmanov},
  {Rapagnani}, {Raymond}, {Razzano}, {Re}, {Read}, {Reed}, {Regimbau}, {Rei},
  {Reid}, {Reitze}, {Rew}, {Reyes}, {Ricci}, {Riles}, {Rizzo}, {Robertson},
  {Robie}, {Robinet}, {Rocchi}, {Rolland}, {Rollins}, {Roma}, {Romano},
  {Romanov}, {Romie}, {Rosi{\'n}ska}, {Rowan}, {R{\"u}diger}, {Ruggi}, {Ryan},
  {Sachdev}, {Sadecki}, {Sadeghian}, {Sakellariadou}, {Salconi}, {Saleem},
  {Salemi}, {Samajdar}, {Sammut}, {Sanchez}, {Sandberg}, {Sandeen}, {Sanders},
  {Sassolas}, {Sathyaprakash}, {Saulson}, {Sauter}, {Savage}, {Sawadsky},
  {Schale}, {Schilling}, {Schmidt}, {Schmidt}, {Schnabel}, {Schofield},
  {Sch{\"o}nbeck}, {Schreiber}, {Schuette}, {Schutz}, {Scott}, {Scott},
  {Sellers}, {Sengupta}, {Sentenac}, {Sequino}, {Sergeev}, {Setyawati},
  {Shaddock}, {Shaffer}, {Shahriar}, {Shaltev}, {Shapiro}, {Shawhan},
  {Sheperd}, {Shoemaker}, {Shoemaker}, {Siellez}, {Siemens}, {Sieniawska},
  {Sigg}, {Silva}, {Singer}, {Singer}, {Singh}, {Singh}, {Singhal}, {Sintes},
  {Slagmolen}, {Smith}, {Smith}, {Smith}, {Son}, {Sorazu}, {Sorrentino},
  {Souradeep}, {Srivastava}, {Staley}, {Steinke}, {Steinlechner},
  {Steinlechner}, {Steinmeyer}, {Stephens}, {Stone}, {Strain}, {Straniero},
  {Stratta}, {Strauss}, {Strigin}, {Sturani}, {Stuver}, {Summerscales}, {Sun},
  {Sunil}, {Sutton}, {Swinkels}, {Szczepa{\'n}czyk}, {Tacca}, {Talukder},
  {Tanner}, {T{\'a}pai}, {Tarabrin}, {Taracchini}, {Taylor}, {Theeg},
  {Thirugnanasambandam}, {Thomas}, {Thomas}, {Thomas}, {Thorne}, {Thrane},
  {Tiwari}, {Tiwari}, {Tokmakov}, {Toland}, {Tomlinson}, {Tonelli}, {Tornasi},
  {Torres}, {Torrie}, {T{\"o}yr{\"a}}, {Travasso}, {Traylor}, {Trifir{\`o}},
  {Tringali}, {Trozzo}, {Tse}, {Turconi}, {Tuyenbayev}, {Ugolini},
  {Unnikrishnan}, {Urban}, {Usman}, {Vahlbruch}, {Vajente}, {Valdes}, {van
  Bakel}, {van Beuzekom}, {van den Brand}, {Van Den Broeck}, {Vander-Hyde},
  {van der Schaaf}, {van Heijningen}, {van Veggel}, {Vardaro}, {Vass},
  {Vas{\'u}th}, {Vaulin}, {Vecchio}, {Vedovato}, {Veitch}, {Veitch},
  {Venkateswara}, {Verkindt}, {Vetrano}, {Vicer{\'e}}, {Vinciguerra}, {Vine},
  {Vinet}, {Vitale}, {Vo}, {Vocca}, {Vorvick}, {Voss}, {Vousden}, {Vyatchanin},
  {Wade}, {Wade}, {Wade}, {Walker}, {Wallace}, {Walsh}, {Wang}, {Wang}, {Wang},
  {Wang}, {Wang}, {Ward}, {Warner}, {Was}, {Weaver}, {Wei}, {Weinert},
  {Weinstein}, {Weiss}, {Wen}, {We{\ss}els}, {Westphal}, {Wette}, {Whelan},
  {Whiting}, {Williams}, {Williamson}, {Willis}, {Willke}, {Wimmer}, {Winkler},
  {Wipf}, {Wittel}, {Woan}, {Woehler}, {Worden}, {Wright}, {Wu}, {Wu},
  {Yablon}, {Yam}, {Yamamoto}, {Yancey}, {Yu}, {Yvert}, {Zadro{\.z}ny},
  {Zangrando}, {Zanolin}, {Zendri}, {Zevin}, {Zhang}, {Zhang}, {Zhang}, {Zhao},
  {Zhou}, {Zhou}, {Zhu}, {Zucker}, {Zuraw}, {Zweizig}, {LIGO Scientific
  Collaboration}, \& {Virgo Collaboration}}]{2016ApJ...832L..21A}
{Abbott}, B.~P., {Abbott}, R., {Abbott}, T.~D., {et~al.} 2016, \apjl, 832, L21

\bibitem[{{Abbott} {et~al.}(2019){Abbott}, {Abbott}, {Abbott}, {Abraham},
  {Acernese}, {Ackley}, {Adams}, {Adhikari}, {Adya}, {Affeldt}, {Agathos},
  {Agatsuma}, {Aggarwal}, {Aguiar}, {Aiello}, {Ain}, {Ajith}, {Allen},
  {Allocca}, {Aloy}, {Altin}, {Amato}, {Ananyeva}, {Anderson}, {Anderson},
  {Angelova}, {Antier}, {Appert}, {Arai}, {Araya}, {Areeda}, {Ar{\`e}ne},
  {Arnaud}, {Ascenzi}, {Ashton}, {Aston}, {Astone}, {Aubin}, {Aufmuth},
  {AultONeal}, {Austin}, {Avendano}, {Avila-Alvarez}, {Babak}, {Bacon},
  {Badaracco}, {Bader}, {Bae}, {Baker}, {Baldaccini}, {Ballardin}, {Ballmer},
  {Banagiri}, {Barayoga}, {Barclay}, {Barish}, {Barker}, {Barkett}, {Barnum},
  {Barone}, {Barr}, {Barsotti}, {Barsuglia}, {Barta}, {Bartlett}, {Bartos},
  {Bassiri}, {Basti}, {Bawaj}, {Bayley}, {Bazzan}, {B{\'e}csy}, {Bejger},
  {Belahcene}, {Bell}, {Beniwal}, {Berger}, {Bergmann}, {Bernuzzi}, {Bero},
  {Berry}, {Bersanetti}, {Bertolini}, {Betzwieser}, {Bhandare}, {Bidler},
  {Bilenko}, {Bilgili}, {Billingsley}, {Birch}, {Birney}, {Birnholtz},
  {Biscans}, {Biscoveanu}, {Bisht}, {Bitossi}, {Bizouard}, {Blackburn},
  {Blair}, {Blair}, {Blair}, {Bloemen}, {Bode}, {Boer}, {Boetzel}, {Bogaert},
  {Bondu}, {Bonilla}, {Bonnand}, {Booker}, {Boom}, {Booth}, {Bork}, {Boschi},
  {Bose}, {Bossie}, {Bossilkov}, {Bosveld}, {Bouffanais}, {Bozzi},
  {Bradaschia}, {Brady}, {Bramley}, {Branchesi}, {Brau}, {Briant}, {Briggs},
  {Brighenti}, {Brillet}, {Brinkmann}, {Brisson}, {Brockill}, {Brooks},
  {Brown}, {Brunett}, {Buikema}, {Bulik}, {Bulten}, {Buonanno}, {Buskulic},
  {Buy}, {Byer}, {Cabero}, {Cadonati}, {Cagnoli}, {Cahillane}, {Calder{\'o}n
  Bustillo}, {Callister}, {Calloni}, {Camp}, {Campbell}, {Canepa}, {Cannon},
  {Cao}, {Cao}, {Capocasa}, {Carbognani}, {Caride}, {Carney}, {Carullo},
  {Casanueva Diaz}, {Casentini}, {Caudill}, {Cavagli{\`a}}, {Cavalier},
  {Cavalieri}, {Cella}, {Cerd{\'a}-Dur{\'a}n}, {Cerretani}, {Cesarini},
  {Chaibi}, {Chakravarti}, {Chamberlin}, {Chan}, {Chao}, {Charlton}, {Chase},
  {Chassande-Mottin}, {Chatterjee}, {Chaturvedi}, {Cheeseboro}, {Chen}, {Chen},
  {Chen}, {Cheng}, {Cheong}, {Chia}, {Chincarini}, {Chiummo}, {Cho}, {Cho},
  {Cho}, {Christensen}, {Chu}, {Chua}, {Chung}, {Chung}, {Ciani}, {Ciobanu},
  {Ciolfi}, {Cipriano}, {Cirone}, {Clara}, {Clark}, {Clearwater}, {Cleva},
  {Cocchieri}, {Coccia}, {Cohadon}, {Cohen}, {Colgan}, {Colleoni}, {Collette},
  {Collins}, {Cominsky}, {Constancio}, {Conti}, {Cooper}, {Corban}, {Corbitt},
  {Cordero-Carri{\'o}n}, {Corley}, {Cornish}, {Corsi}, {Cortese}, {Costa},
  {Cotesta}, {Coughlin}, {Coughlin}, {Coulon}, {Countryman}, {Couvares},
  {Covas}, {Cowan}, {Coward}, {Cowart}, {Coyne}, {Coyne}, {Creighton},
  {Creighton}, {Cripe}, {Croquette}, {Crowder}, {Cullen}, {Cumming},
  {Cunningham}, {Cuoco}, {Dal Canton}, {D{\'a}lya}, {Danilishin}, {D'Antonio},
  {Danzmann}, {Dasgupta}, {Da Silva Costa}, {Datrier}, {Dattilo}, {Dave},
  {Davier}, {Davis}, {Daw}, {DeBra}, {Deenadayalan}, {Degallaix}, {De
  Laurentis}, {Del{\'e}glise}, {Del Pozzo}, {DeMarchi}, {Demos}, {Dent}, {De
  Pietri}, {Derby}, {De Rosa}, {De Rossi}, {DeSalvo}, {de Varona},
  {Dhurandhar}, {D{\'\i}az}, {Dietrich}, {Di Fiore}, {Di Giovanni}, {Di
  Girolamo}, {Di Lieto}, {Ding}, {Di Pace}, {Di Palma}, {Di Renzo}, {Dmitriev},
  {Doctor}, {Donovan}, {Dooley}, {Doravari}, {Dorrington}, {Downes}, {Drago},
  {Driggers}, {Du}, {Ducoin}, {Dupej}, {Dwyer}, {Easter}, {Edo}, {Edwards},
  {Effler}, {Ehrens}, {Eichholz}, {Eikenberry}, {Eisenmann}, {Eisenstein},
  {Essick}, {Estelles}, {Estevez}, {Etienne}, {Etzel}, {Evans}, {Evans},
  {Fafone}, {Fair}, {Fairhurst}, {Fan}, {Farinon}, {Farr}, {Farr},
  {Fauchon-Jones}, {Favata}, {Fays}, {Fazio}, {Fee}, {Feicht}, {Fejer}, {Feng},
  {Fernandez-Galiana}, {Ferrante}, {Ferreira}, {Ferreira}, {Ferrini},
  {Fidecaro}, {Fiori}, {Fiorucci}, {Fishbach}, {Fisher}, {Fishner},
  {Fitz-Axen}, {Flaminio}, {Fletcher}, {Flynn}, {Fong}, {Font}, {Forsyth},
  {Fournier}, {Frasca}, {Frasconi}, {Frei}, {Freise}, {Frey}, {Frey},
  {Fritschel}, {Frolov}, {Fulda}, {Fyffe}, {Gabbard}, {Gadre}, {Gaebel},
  {Gair}, {Gammaitoni}, {Ganija}, {Gaonkar}, {Garcia},
  {Garc{\'\i}a-Quir{\'o}s}, {Garufi}, {Gateley}, {Gaudio}, {Gaur}, {Gayathri},
  {Gemme}, {Genin}, {Gennai}, {George}, {George}, {Gergely}, {Germain},
  {Ghonge}, {Ghosh}, {Ghosh}, {Ghosh}, {Giacomazzo}, {Giaime}, {Giardina},
  {Giazotto}, {Gill}, {Giordano}, {Glover}, {Godwin}, {Goetz}, {Goetz},
  {Goncharov}, {Gonz{\'a}lez}, {Gonzalez Castro}, {Gopakumar}, {Gorodetsky},
  {Gossan}, {Gosselin}, {Gouaty}, {Grado}, {Graef}, {Granata}, {Grant}, {Gras},
  {Grassia}, {Gray}, {Gray}, {Greco}, {Green}, {Green}, {Gretarsson}, {Groot},
  {Grote}, {Grunewald}, {Gruning}, {Guidi}, {Gulati}, {Guo}, {Gupta}, {Gupta},
  {Gustafson}, {Gustafson}, {Haegel}, {Halim}, {Hall}, {Hall}, {Hamilton},
  {Hammond}, {Haney}, {Hanke}, {Hanks}, {Hanna}, {Hannam}, {Hannuksela},
  {Hanson}, {Hardwick}, {Haris}, {Harms}, {Harry}, {Harry}, {Haster},
  {Haughian}, {Hayes}, {Healy}, {Heidmann}, {Heintze}, {Heitmann}, {Hello},
  {Hemming}, {Hendry}, {Heng}, {Hennig}, {Heptonstall}, {Hernandez Vivanco},
  {Heurs}, {Hild}, {Hinderer}, {Hoak}, {Hochheim}, {Hofman}, {Holgado},
  {Holland}, {Holt}, {Holz}, {Hopkins}, {Horst}, {Hough}, {Howell}, {Hoy},
  {Hreibi}, {Huerta}, {Huet}, {Hughey}, {Hulko}, {Husa}, {Huttner},
  {Huynh-Dinh}, {Idzkowski}, {Iess}, {Ingram}, {Inta}, {Intini}, {Irwin},
  {Isa}, {Isac}, {Isi}, {Iyer}, {Izumi}, {Jacqmin}, {Jadhav}, {Jani},
  {Janthalur}, {Jaranowski}, {Jenkins}, {Jiang}, {Johnson}, {Jones}, {Jones},
  {Jones}, {Jonker}, {Ju}, {Junker}, {Kalaghatgi}, {Kalogera}, {Kamai},
  {Kandhasamy}, {Kang}, {Kanner}, {Kapadia}, {Karki}, {Karvinen}, {Kashyap},
  {Kasprzack}, {Katsanevas}, {Katsavounidis}, {Katzman}, {Kaufer}, {Kawabe},
  {Keerthana}, {K{\'e}f{\'e}lian}, {Keitel}, {Kennedy}, {Key}, {Khalili},
  {Khan}, {Khan}, {Khan}, {Khan}, {Khazanov}, {Khursheed}, {Kijbunchoo}, {Kim},
  {Kim}, {Kim}, {Kim}, {Kim}, {Kim}, {Kimball}, {King}, {King},
  {Kinley-Hanlon}, {Kirchhoff}, {Kissel}, {Kleybolte}, {Klika}, {Klimenko},
  {Knowles}, {Koch}, {Koehlenbeck}, {Koekoek}, {Koley}, {Kondrashov}, {Kontos},
  {Koper}, {Korobko}, {Korth}, {Kowalska}, {Kozak}, {Kringel}, {Krishnendu},
  {Kr{\'o}lak}, {Kuehn}, {Kumar}, {Kumar}, {Kumar}, {Kumar}, {Kuo}, {Kutynia},
  {Kwang}, {Lackey}, {Lai}, {Lam}, {Landry}, {Lane}, {Lang}, {Lange}, {Lantz},
  {Lanza}, {Lartaux-Vollard}, {Lasky}, {Laxen}, {Lazzarini}, {Lazzaro},
  {Leaci}, {Leavey}, {Lecoeuche}, {Lee}, {Lee}, {Lee}, {Lee}, {Lee}, {Lee},
  {Lehmann}, {Lenon}, {Leroy}, {Letendre}, {Levin}, {Li}, {Li}, {Li}, {Li},
  {Lin}, {Linde}, {Linker}, {Littenberg}, {Liu}, {Liu}, {Lo}, {Lockerbie},
  {London}, {Longo}, {Lorenzini}, {Loriette}, {Lormand}, {Losurdo}, {Lough},
  {Lousto}, {Lovelace}, {Lower}, {L{\"u}ck}, {Lumaca}, {Lundgren}, {Lynch},
  {Ma}, {Macas}, {Macfoy}, {MacInnis}, {Macleod}, {Macquet},
  {Maga{\~n}a-Sandoval}, {Maga{\~n}a Zertuche}, {Magee}, {Majorana},
  {Maksimovic}, {Malik}, {Man}, {Mandic}, {Mangano}, {Mansell}, {Manske},
  {Mantovani}, {Marchesoni}, {Marion}, {M{\'a}rka}, {M{\'a}rka}, {Markakis},
  {Markosyan}, {Markowitz}, {Maros}, {Marquina}, {Marsat}, {Martelli},
  {Martin}, {Martin}, {Martynov}, {Mason}, {Massera}, {Masserot}, {Massinger},
  {Masso-Reid}, {Mastrogiovanni}, {Matas}, {Matichard}, {Matone}, {Mavalvala},
  {Mazumder}, {McCann}, {McCarthy}, {McClelland}, {McCormick}, {McCuller},
  {McGuire}, {McIver}, {McManus}, {McRae}, {McWilliams}, {Meacher}, {Meadors},
  {Mehmet}, {Mehta}, {Meidam}, {Melatos}, {Mendell}, {Mercer}, {Mereni},
  {Merilh}, {Merzougui}, {Meshkov}, {Messenger}, {Messick}, {Metzdorff},
  {Meyers}, {Miao}, {Michel}, {Middleton}, {Mikhailov}, {Milano}, {Miller},
  {Miller}, {Millhouse}, {Mills}, {Milovich-Goff}, {Minazzoli}, {Minenkov},
  {Mishkin}, {Mishra}, {Mistry}, {Mitra}, {Mitrofanov}, {Mitselmakher},
  {Mittleman}, {Mo}, {Moffa}, {Mogushi}, {Mohapatra}, {Montani}, {Moore},
  {Moraru}, {Moreno}, {Morisaki}, {Mours}, {Mow-Lowry}, {Mukherjee},
  {Mukherjee}, {Mukherjee}, {Mukund}, {Mullavey}, {Munch}, {Mu{\~n}iz},
  {Muratore}, {Murray}, {Nagar}, {Nardecchia}, {Naticchioni}, {Nayak},
  {Neilson}, {Nelemans}, {Nelson}, {Nery}, {Neunzert}, {Ng}, {Ng}, {Nguyen},
  {Nichols}, {Nissanke}, {Nocera}, {North}, {Nuttall}, {Obergaulinger},
  {Oberling}, {O'Brien}, {O'Dea}, {Ogin}, {Oh}, {Oh}, {Ohme}, {Ohta}, {Okada},
  {Oliver}, {Oppermann}, {Oram}, {O'Reilly}, {Ormiston}, {Ortega},
  {O'Shaughnessy}, {Ossokine}, {Ottaway}, {Overmier}, {Owen}, {Pace}, {Pagano},
  {Page}, {Pai}, {Pai}, {Palamos}, {Palashov}, {Palomba}, {Pal-Singh}, {Pan},
  {Pang}, {Pang}, {Pankow}, {Pannarale}, {Pant}, {Paoletti}, {Paoli}, {Parida},
  {Parker}, {Pascucci}, {Pasqualetti}, {Passaquieti}, {Passuello}, {Patil},
  {Patricelli}, {Pearlstone}, {Pedersen}, {Pedraza}, {Pedurand}, {Pele},
  {Penn}, {Perez}, {Perreca}, {Pfeiffer}, {Phelps}, {Phukon}, {Piccinni},
  {Pichot}, {Piergiovanni}, {Pillant}, {Pinard}, {Pirello}, {Pitkin},
  {Poggiani}, {Pong}, {Ponrathnam}, {Popolizio}, {Porter}, {Powell},
  {Prajapati}, {Prasad}, {Prasai}, {Prasanna}, {Pratten}, {Prestegard},
  {Privitera}, {Prodi}, {Prokhorov}, {Puncken}, {Punturo}, {Puppo},
  {P{\"u}rrer}, {Qi}, {Quetschke}, {Quinonez}, {Quintero}, {Quitzow-James},
  {Raab}, {Radkins}, {Radulescu}, {Raffai}, {Raja}, {Rajan}, {Rajbhandari},
  {Rakhmanov}, {Ramirez}, {Ramos-Buades}, {Rana}, {Rao}, {Rapagnani},
  {Raymond}, {Razzano}, {Read}, {Regimbau}, {Rei}, {Reid}, {Reitze}, {Ren},
  {Ricci}, {Richardson}, {Richardson}, {Ricker}, {Riles}, {Rizzo}, {Robertson},
  {Robie}, {Robinet}, {Rocchi}, {Rolland}, {Rollins}, {Roma}, {Romanelli},
  {Romano}, {Romel}, {Romie}, {Rose}, {Rosi{\'n}ska}, {Rosofsky}, {Ross},
  {Rowan}, {R{\"u}diger}, {Ruggi}, {Rutins}, {Ryan}, {Sachdev}, {Sadecki},
  {Sakellariadou}, {Salconi}, {Saleem}, {Samajdar}, {Sammut}, {Sanchez},
  {Sanchez}, {Sanchis-Gual}, {Sandberg}, {Sanders}, {Santiago}, {Sarin},
  {Sassolas}, {Saulson}, {Sauter}, {Savage}, {Schale}, {Scheel}, {Scheuer},
  {Schmidt}, {Schnabel}, {Schofield}, {Sch{\"o}nbeck}, {Schreiber}, {Schulte},
  {Schutz}, {Schwalbe}, {Scott}, {Scott}, {Seidel}, {Sellers}, {Sengupta},
  {Sennett}, {Sentenac}, {Sequino}, {Sergeev}, {Setyawati}, {Shaddock},
  {Shaffer}, {Shahriar}, {Shaner}, {Shao}, {Sharma}, {Shawhan}, {Shen},
  {Shink}, {Shoemaker}, {Shoemaker}, {ShyamSundar}, {Siellez}, {Sieniawska},
  {Sigg}, {Silva}, {Singer}, {Singh}, {Singhal}, {Sintes}, {Sitmukhambetov},
  {Skliris}, {Slagmolen}, {Slaven-Blair}, {Smith}, {Smith}, {Somala}, {Son},
  {Sorazu}, {Sorrentino}, {Souradeep}, {Sowell}, {Spencer}, {Srivastava},
  {Srivastava}, {Staats}, {Stachie}, {Standke}, {Steer}, {Steinke},
  {Steinlechner}, {Steinlechner}, {Steinmeyer}, {Stevenson}, {Stocks}, {Stone},
  {Stops}, {Strain}, {Stratta}, {Strigin}, {Strunk}, {Sturani}, {Stuver},
  {Sudhir}, {Summerscales}, {Sun}, {Sunil}, {Suresh}, {Sutton}, {Swinkels},
  {Szczepa{\'n}czyk}, {Tacca}, {Tait}, {Talbot}, {Talukder}, {Tanner},
  {T{\'a}pai}, {Taracchini}, {Tasson}, {Taylor}, {Thies}, {Thomas}, {Thomas},
  {Thondapu}, {Thorne}, {Thrane}, {Tiwari}, {Tiwari}, {Tiwari}, {Toland},
  {Tonelli}, {Tornasi}, {Torres-Forn{\'e}}, {Torrie}, {T{\"o}yr{\"a}},
  {Travasso}, {Traylor}, {Tringali}, {Trovato}, {Trozzo}, {Trudeau}, {Tsang},
  {Tse}, {Tso}, {Tsukada}, {Tsuna}, {Tuyenbayev}, {Ueno}, {Ugolini},
  {Unnikrishnan}, {Urban}, {Usman}, {Vahlbruch}, {Vajente}, {Valdes}, {van
  Bakel}, {van Beuzekom}, {van den Brand}, {Van Den Broeck}, {Vander-Hyde},
  {van Heijningen}, {van der Schaaf}, {van Veggel}, {Vardaro}, {Varma}, {Vass},
  {Vas{\'u}th}, {Vecchio}, {Vedovato}, {Veitch}, {Veitch}, {Venkateswara},
  {Venugopalan}, {Verkindt}, {Vetrano}, {Vicer{\'e}}, {Viets}, {Vine}, {Vinet},
  {Vitale}, {Vo}, {Vocca}, {Vorvick}, {Vyatchanin}, {Wade}, {Wade}, {Wade},
  {Walet}, {Walker}, {Wallace}, {Walsh}, {Wang}, {Wang}, {Wang}, {Wang},
  {Wang}, {Ward}, {Warden}, {Warner}, {Was}, {Watchi}, {Weaver}, {Wei},
  {Weinert}, {Weinstein}, {Weiss}, {Wellmann}, {Wen}, {Wessel}, {We{\ss}els},
  {Westhouse}, {Wette}, {Whelan}, {Whiting}, {Whittle}, {Wilken}, {Williams},
  {Williamson}, {Willis}, {Willke}, {Wimmer}, {Winkler}, {Wipf}, {Wittel},
  {Woan}, {Woehler}, {Wofford}, {Worden}, {Wright}, {Wu}, {Wysocki}, {Xiao},
  {Yamamoto}, {Yancey}, {Yang}, {Yap}, {Yazback}, {Yeeles}, {Yu}, {Yu}, {Yuen},
  {Yvert}, {Zadro{\.Z}ny}, {Zanolin}, {Zelenova}, {Zendri}, {Zevin}, {Zhang},
  {Zhang}, {Zhang}, {Zhao}, {Zhou}, {Zhou}, {Zhu}, {Zucker}, {Zweizig}, {Dunn},
  {Suvorova}, {Evans}, {Moran}, {LIGO Scientific Collaboration}, \& {Virgo
  Collaboration}}]{2019PhRvD.100l2002A}
{Abbott}, B.~P., {Abbott}, R., {Abbott}, T.~D., {et~al.} 2019, \prd, 100,
  122002

\bibitem[{{Abdul-Masih}(2023)}]{Abdul-Masih2023}
{Abdul-Masih}, M. 2023, \aap, 669, L11

\bibitem[{{Asplund} {et~al.}(2005){Asplund}, {Grevesse}, \& {Sauval}}]{asp1:05}
{Asplund}, M., {Grevesse}, N., \& {Sauval}, A.~J. 2005, in Astronomical Society
  of the Pacific Conference Series, Vol. 336, Cosmic Abundances as Records of
  Stellar Evolution and Nucleosynthesis, ed. {T.~G.~Barnes III \& F.~N.~Bash}
  (San Francisco: Astronomical Society of the Pacific), 25

\bibitem[{{Batten}(1989)}]{1989SSRv...50....1B}
{Batten}, A.~H. 1989, \ssr, 50, 1

\bibitem[{{Baym} {et~al.}(2018){Baym}, {Hatsuda}, {Kojo}, {Powell}, {Song}, \&
  {Takatsuka}}]{2018RPPh...81e6902B}
{Baym}, G., {Hatsuda}, T., {Kojo}, T., {et~al.} 2018, Reports on Progress in
  Physics, 81, 056902

\bibitem[{{Bj{\"o}rklund} {et~al.}(2022){Bj{\"o}rklund}, {Sundqvist}, {Singh},
  {Puls}, \& {Najarro}}]{2022arXiv220308218B}
{Bj{\"o}rklund}, R., {Sundqvist}, J.~O., {Singh}, S.~M., {Puls}, J., \&
  {Najarro}, F. 2022, arXiv e-prints, arXiv:2203.08218

\bibitem[{{B{\"o}hm-Vitense}(1958)}]{boe1:58}
{B{\"o}hm-Vitense}, E. 1958, \zap, 46, 108

\bibitem[{{Bonanos} {et~al.}(2010){Bonanos}, {Lennon}, {K{\"o}hlinger}, {van
  Loon}, {Massa}, {Sewilo}, {Evans}, {Panagia}, {Babler}, {Block}, {Bracker},
  {Engelbracht}, {Gordon}, {Hora}, {Indebetouw}, {Meade}, {Meixner}, {Misselt},
  {Robitaille}, {Shiao}, \& {Whitney}}]{2010AJ....140..416B}
{Bonanos}, A.~Z., {Lennon}, D.~J., {K{\"o}hlinger}, F., {et~al.} 2010, \aj,
  140, 416

\bibitem[{{Brott} {et~al.}(2011){Brott}, {de Mink}, {Cantiello}, {Langer}, {de
  Koter}, {Evans}, {Hunter}, {Trundle}, \& {Vink}}]{bro1:11}
{Brott}, I., {de Mink}, S.~E., {Cantiello}, M., {et~al.} 2011, \aap, 530, A115

\bibitem[{{Burbidge} {et~al.}(1957){Burbidge}, {Burbidge}, {Fowler}, \&
  {Hoyle}}]{1957RvMP...29..547B}
{Burbidge}, E.~M., {Burbidge}, G.~R., {Fowler}, W.~A., \& {Hoyle}, F. 1957,
  Reviews of Modern Physics, 29, 547

\bibitem[{{Cassinelli} \& {Olson}(1979)}]{cas1:79}
{Cassinelli}, J.~P. \& {Olson}, G.~L. 1979, \apj, 229, 304

\bibitem[{{Conroy} {et~al.}(2020){Conroy}, {Kochoska}, {Hey}, {Pablo},
  {Hambleton}, {Jones}, {Giammarco}, {Abdul-Masih}, \& {Pr{\v{s}}a}}]{con1:20}
{Conroy}, K.~E., {Kochoska}, A., {Hey}, D., {et~al.} 2020, \apjs, 250, 34

\bibitem[{{Conti} \& {Alschuler}(1971)}]{1971ApJ...170..325C}
{Conti}, P.~S. \& {Alschuler}, W.~R. 1971, \apj, 170, 325

\bibitem[{{Crowther} {et~al.}(2002){Crowther}, {Hillier}, {Evans}, {Fullerton},
  {De Marco}, \& {Willis}}]{2002ApJ...579..774C}
{Crowther}, P.~A., {Hillier}, D.~J., {Evans}, C.~J., {et~al.} 2002, \apj, 579,
  774

\bibitem[{{Danforth} {et~al.}(2003){Danforth}, {Sankrit}, {Blair}, {Howk}, \&
  {Chu}}]{2003ApJ...586.1179D}
{Danforth}, C.~W., {Sankrit}, R., {Blair}, W.~P., {Howk}, J.~C., \& {Chu},
  Y.-H. 2003, \apj, 586, 1179

\bibitem[{{de Mink} {et~al.}(2007){de Mink}, {Pols}, \&
  {Hilditch}}]{2007A&A...467.1181D}
{de Mink}, S.~E., {Pols}, O.~R., \& {Hilditch}, R.~W. 2007, \aap, 467, 1181

\bibitem[{{Dufton} {et~al.}(2019){Dufton}, {Evans}, {Hunter}, {Lennon}, \&
  {Schneider}}]{2019AA...626A..50D}
{Dufton}, P.~L., {Evans}, C.~J., {Hunter}, I., {Lennon}, D.~J., \& {Schneider},
  F.~R.~N. 2019, \aap, 626, A50

\bibitem[{{Foreman-Mackey} {et~al.}(2013){Foreman-Mackey}, {Hogg}, {Lang}, \&
  {Goodman}}]{for1:13}
{Foreman-Mackey}, D., {Hogg}, D.~W., {Lang}, D., \& {Goodman}, J. 2013, \pasp,
  125, 306

\bibitem[{{Gaia Collaboration} {et~al.}(2022){Gaia Collaboration}, {Vallenari},
  {Brown}, {Prusti}, {de Bruijne}, {Arenou}, {Babusiaux}, {Biermann},
  {Creevey}, {Ducourant}, {Evans}, {Eyer}, {Guerra}, {Hutton}, {Jordi},
  {Klioner}, {Lammers}, {Lindegren}, {Luri}, {Mignard}, {Panem}, {Pourbaix},
  {Randich}, {Sartoretti}, {Soubiran}, {Tanga}, {Walton}, {Bailer-Jones},
  {Bastian}, {Drimmel}, {Jansen}, {Katz}, {Lattanzi}, {van Leeuwen}, {Bakker},
  {Cacciari}, {Casta{\~n}eda}, {De Angeli}, {Fabricius}, {Fouesneau},
  {Fr{\'e}mat}, {Galluccio}, {Guerrier}, {Heiter}, {Masana}, {Messineo},
  {Mowlavi}, {Nicolas}, {Nienartowicz}, {Pailler}, {Panuzzo}, {Riclet}, {Roux},
  {Seabroke}, {Sordo{\o}rcit}, {Th{\'e}venin}, {Gracia-Abril}, {Portell},
  {Teyssier}, {Altmann}, {Andrae}, {Audard}, {Bellas-Velidis}, {Benson},
  {Berthier}, {Blomme}, {Burgess}, {Busonero}, {Busso}, {C{\'a}novas}, {Carry},
  {Cellino}, {Cheek}, {Clementini}, {Damerdji}, {Davidson}, {de Teodoro},
  {Nu{\~n}ez Campos}, {Delchambre}, {Dell'Oro}, {Esquej},
  {Fern{\'a}ndez-Hern{\'a}ndez}, {Fraile}, {Garabato}, {Garc{\'\i}a-Lario},
  {Gosset}, {Haigron}, {Halbwachs}, {Hambly}, {Harrison}, {Hern{\'a}ndez},
  {Hestroffer}, {Hodgkin}, {Holl}, {Jan{\ss}en}, {Jevardat de Fombelle},
  {Jordan}, {Krone-Martins}, {Lanzafame}, {L{\"o}ffler}, {Marchal}, {Marrese},
  {Moitinho}, {Muinonen}, {Osborne}, {Pancino}, {Pauwels}, {Recio-Blanco},
  {Reyl{\'e}}, {Riello}, {Rimoldini}, {Roegiers}, {Rybizki}, {Sarro}, {Siopis},
  {Smith}, {Sozzetti}, {Utrilla}, {van Leeuwen}, {Abbas}, {{\'A}brah{\'a}m},
  {Abreu Aramburu}, {Aerts}, {Aguado}, {Ajaj}, {Aldea-Montero}, {Altavilla},
  {{\'A}lvarez}, {Alves}, {Anders}, {Anderson}, {Anglada Varela}, {Antoja},
  {Baines}, {Baker}, {Balaguer-N{\'u}{\~n}ez}, {Balbinot}, {Balog}, {Barache},
  {Barbato}, {Barros}, {Barstow}, {Bartolom{\'e}}, {Bassilana}, {Bauchet},
  {Becciani}, {Bellazzini}, {Berihuete}, {Bernet}, {Bertone}, {Bianchi},
  {Binnenfeld}, {Blanco-Cuaresma}, {Blazere}, {Boch}, {Bombrun}, {Bossini},
  {Bouquillon}, {Bragaglia}, {Bramante}, {Breedt}, {Bressan}, {Brouillet},
  {Brugaletta}, {Bucciarelli}, {Burlacu}, {Butkevich}, {Buzzi}, {Caffau},
  {Cancelliere}, {Cantat-Gaudin}, {Carballo}, {Carlucci}, {Carnerero},
  {Carrasco}, {Casamiquela}, {Castellani}, {Castro-Ginard}, {Chaoul},
  {Charlot}, {Chemin}, {Chiaramida}, {Chiavassa}, {Chornay}, {Comoretto},
  {Contursi}, {Cooper}, {Cornez}, {Cowell}, {Crifo}, {Cropper}, {Crosta},
  {Crowley}, {Dafonte}, {Dapergolas}, {David}, {David}, {de Laverny}, {De
  Luise}, {De March}, {De Ridder}, {de Souza}, {de Torres}, {del Peloso}, {del
  Pozo}, {Delbo}, {Delgado}, {Delisle}, {Demouchy}, {Dharmawardena}, {Di
  Matteo}, {Diakite}, {Diener}, {Distefano}, {Dolding}, {Edvardsson}, {Enke},
  {Fabre}, {Fabrizio}, {Faigler}, {Fedorets}, {Fernique}, {Fienga}, {Figueras},
  {Fournier}, {Fouron}, {Fragkoudi}, {Gai}, {Garcia-Gutierrez},
  {Garcia-Reinaldos}, {Garc{\'\i}a-Torres}, {Garofalo}, {Gavel}, {Gavras},
  {Gerlach}, {Geyer}, {Giacobbe}, {Gilmore}, {Girona}, {Giuffrida}, {Gomel},
  {Gomez}, {Gonz{\'a}lez-N{\'u}{\~n}ez}, {Gonz{\'a}lez-Santamar{\'\i}a},
  {Gonz{\'a}lez-Vidal}, {Granvik}, {Guillout}, {Guiraud},
  {Guti{\'e}rrez-S{\'a}nchez}, {Guy}, {Hatzidimitriou}, {Hauser}, {Haywood},
  {Helmer}, {Helmi}, {Sarmiento}, {Hidalgo}, {Hilger}, {H{\l}adczuk}, {Hobbs},
  {Holland}, {Huckle}, {Jardine}, {Jasniewicz}, {Jean-Antoine Piccolo},
  {Jim{\'e}nez-Arranz}, {Jorissen}, {Juaristi Campillo}, {Julbe}, {Karbevska},
  {Kervella}, {Khanna}, {Kontizas}, {Kordopatis}, {Korn}, {K{\'o}sp{\'a}l},
  {Kostrzewa-Rutkowska}, {Kruszy{\'n}ska}, {Kun}, {Laizeau}, {Lambert},
  {Lanza}, {Lasne}, {Le Campion}, {Lebreton}, {Lebzelter}, {Leccia}, {Leclerc},
  {Lecoeur-Taibi}, {Liao}, {Licata}, {Lindstr{\o}m}, {Lister}, {Livanou},
  {Lobel}, {Lorca}, {Loup}, {Madrero Pardo}, {Magdaleno Romeo}, {Managau},
  {Mann}, {Manteiga}, {Marchant}, {Marconi}, {Marcos}, {Marcos Santos},
  {Mar{\'\i}n Pina}, {Marinoni}, {Marocco}, {Marshall}, {Polo},
  {Mart{\'\i}n-Fleitas}, {Marton}, {Mary}, {Masip}, {Massari},
  {Mastrobuono-Battisti}, {Mazeh}, {McMillan}, {Messina}, {Michalik}, {Millar},
  {Mints}, {Molina}, {Molinaro}, {Moln{\'a}r}, {Monari}, {Mongui{\'o}},
  {Montegriffo}, {Montero}, {Mor}, {Mora}, {Morbidelli}, {Morel}, {Morris},
  {Muraveva}, {Murphy}, {Musella}, {Nagy}, {Noval}, {Oca{\~n}a}, {Ogden},
  {Ordenovic}, {Osinde}, {Pagani}, {Pagano}, {Palaversa}, {Palicio},
  {Pallas-Quintela}, {Panahi}, {Payne-Wardenaar}, {Pe{\~n}alosa Esteller},
  {Penttil{\"a}}, {Pichon}, {Piersimoni}, {Pineau}, {Plachy}, {Plum}, {Poggio},
  {Pr{\v{s}}a}, {Pulone}, {Racero}, {Ragaini}, {Rainer}, {Raiteri}, {Rambaux},
  {Ramos}, {Ramos-Lerate}, {Re Fiorentin}, {Regibo}, {Richards}, {Rios Diaz},
  {Ripepi}, {Riva}, {Rix}, {Rixon}, {Robichon}, {Robin}, {Robin}, {Roelens},
  {Rogues}, {Rohrbasser}, {Romero-G{\'o}mez}, {Rowell}, {Royer}, {Ruz Mieres},
  {Rybicki}, {Sadowski}, {S{\'a}ez N{\'u}{\~n}ez}, {Sagrist{\`a} Sell{\'e}s},
  {Sahlmann}, {Salguero}, {Samaras}, {Sanchez Gimenez}, {Sanna},
  {Santove{\~n}a}, {Sarasso}, {Schultheis}, {Sciacca}, {Segol}, {Segovia},
  {S{\'e}gransan}, {Semeux}, {Shahaf}, {Siddiqui}, {Siebert}, {Siltala},
  {Silvelo}, {Slezak}, {Slezak}, {Smart}, {Snaith}, {Solano}, {Solitro},
  {Souami}, {Souchay}, {Spagna}, {Spina}, {Spoto}, {Steele},
  {Steidelm{\"u}ller}, {Stephenson}, {S{\"u}veges}, {Surdej}, {Szabados},
  {Szegedi-Elek}, {Taris}, {Taylo}, {Teixeira}, {Tolomei}, {Tonello}, {Torra},
  {Torra}, {Torralba Elipe}, {Trabucchi}, {Tsounis}, {Turon}, {Ulla}, {Unger},
  {Vaillant}, {van Dillen}, {van Reeven}, {Vanel}, {Vecchiato}, {Viala},
  {Vicente}, {Voutsinas}, {Weiler}, {Wevers}, {Wyrzykowski}, {Yoldas}, {Yvard},
  {Zhao}, {Zorec}, {Zucker}, \& {Zwitter}}]{2022arXiv220800211G}
{Gaia Collaboration}, {Vallenari}, A., {Brown}, A.~G.~A., {et~al.} 2022, arXiv
  e-prints, arXiv:2208.00211

\bibitem[{{Georgy} {et~al.}(2013){Georgy}, {Ekstr{\"o}m}, {Eggenberger},
  {Meynet}, {Haemmerl{\'e}}, {Maeder}, {Granada}, {Groh}, {Hirschi}, {Mowlavi},
  {Yusof}, {Charbonnel}, {Decressin}, \& {Barblan}}]{2013A&A...558A.103G}
{Georgy}, C., {Ekstr{\"o}m}, S., {Eggenberger}, P., {et~al.} 2013, \aap, 558,
  A103

\bibitem[{{Gr{\"a}fener} {et~al.}(2002){Gr{\"a}fener}, {Koesterke}, \&
  {Hamann}}]{2002A&A...387..244G}
{Gr{\"a}fener}, G., {Koesterke}, L., \& {Hamann}, W.~R. 2002, \aap, 387, 244

\bibitem[{{Hainich} {et~al.}(2015){Hainich}, {Pasemann}, {Todt}, {Shenar},
  {Sand er}, \& {Hamann}}]{2015A&A...581A..21H}
{Hainich}, R., {Pasemann}, D., {Todt}, H., {et~al.} 2015, \aap, 581, A21

\bibitem[{{Hainich} {et~al.}(2014){Hainich}, {R{\"u}hling}, {Todt}, {Oskinova},
  {Liermann}, {Gr{\"a}fener}, {Foellmi}, {Schnurr}, \&
  {Hamann}}]{2014A&A...565A..27H}
{Hainich}, R., {R{\"u}hling}, U., {Todt}, H., {et~al.} 2014, \aap, 565, A27

\bibitem[{{Hamann} \& {Gr{\"a}fener}(2004)}]{2004A&A...427..697H}
{Hamann}, W.~R. \& {Gr{\"a}fener}, G. 2004, \aap, 427, 697

\bibitem[{{Heger} {et~al.}(2003){Heger}, {Fryer}, {Woosley}, {Langer}, \&
  {Hartmann}}]{2003ApJ...591..288H}
{Heger}, A., {Fryer}, C.~L., {Woosley}, S.~E., {Langer}, N., \& {Hartmann},
  D.~H. 2003, \apj, 591, 288

\bibitem[{{Heger} {et~al.}(2000){Heger}, {Langer}, \& {Woosley}}]{Heg1:00}
{Heger}, A., {Langer}, N., \& {Woosley}, S.~E. 2000, \apj, 528, 368

\bibitem[{{Hilditch} {et~al.}(2005){Hilditch}, {Howarth}, \&
  {Harries}}]{2005MNRAS.357..304H}
{Hilditch}, R.~W., {Howarth}, I.~D., \& {Harries}, T.~J. 2005, \mnras, 357, 304

\bibitem[{{Hillier} \& {Miller}(1999)}]{hil1:99}
{Hillier}, D.~J. \& {Miller}, D.~L. 1999, \apj, 519, 354

\bibitem[{{Hollenbach} \& {Tielens}(1999)}]{1999RvMP...71..173H}
{Hollenbach}, D.~J. \& {Tielens}, A.~G.~G.~M. 1999, Reviews of Modern Physics,
  71, 173

\bibitem[{{Horvat} {et~al.}(2018){Horvat}, {Conroy}, {Pablo}, {Hambleton},
  {Kochoska}, {Giammarco}, \& {Pr{\v{s}}a}}]{hor1:18}
{Horvat}, M., {Conroy}, K.~E., {Pablo}, H., {et~al.} 2018, \apjs, 237, 26

\bibitem[{{Hunter} {et~al.}(2007){Hunter}, {Dufton}, {Smartt}, {Ryans},
  {Evans}, {Lennon}, {Trundle}, {Hubeny}, \& {Lanz}}]{hun1:07}
{Hunter}, I., {Dufton}, P.~L., {Smartt}, S.~J., {et~al.} 2007, \aap, 466, 277

\bibitem[{{Jones} {et~al.}(2020){Jones}, {Conroy}, {Horvat}, {Giammarco},
  {Kochoska}, {Pablo}, {Brown}, {Sowicka}, \& {Pr{\v{s}}a}}]{jon1:20}
{Jones}, D., {Conroy}, K.~E., {Horvat}, M., {et~al.} 2020, \apjs, 247, 63

\bibitem[{{Kajino} {et~al.}(2019){Kajino}, {Aoki}, {Balantekin}, {Diehl},
  {Famiano}, \& {Mathews}}]{2019PrPNP.107..109K}
{Kajino}, T., {Aoki}, W., {Balantekin}, A.~B., {et~al.} 2019, Progress in
  Particle and Nuclear Physics, 107, 109

\bibitem[{{Kasen} {et~al.}(2017){Kasen}, {Metzger}, {Barnes}, {Quataert}, \&
  {Ramirez-Ruiz}}]{2017Natur.551...80K}
{Kasen}, D., {Metzger}, B., {Barnes}, J., {Quataert}, E., \& {Ramirez-Ruiz}, E.
  2017, \nat, 551, 80

\bibitem[{{Kippenhahn} {et~al.}(1980){Kippenhahn}, {Ruschenplatt}, \&
  {Thomas}}]{kip1:80}
{Kippenhahn}, R., {Ruschenplatt}, G., \& {Thomas}, H.~C. 1980, \aap, 91, 175

\bibitem[{Kipping(2013)}]{10.1093/mnrasl/slt075}
Kipping, D.~M. 2013, Monthly Notices of the Royal Astronomical Society:
  Letters, 434, L51

\bibitem[{{Kurt} \& {Dufour}(1998)}]{1998RMxAC...7..202K}
{Kurt}, C.~M. \& {Dufour}, R.~J. 1998, in Revista Mexicana de Astronomia y
  Astrofisica Conference Series, Vol.~7, Revista Mexicana de Astronomia y
  Astrofisica Conference Series, ed. R.~J. {Dufour} \& S.~{Torres-Peimbert},
  202

\bibitem[{{Langer} {et~al.}(1983){Langer}, {Fricke}, \& {Sugimoto}}]{lan1:83}
{Langer}, N., {Fricke}, K.~J., \& {Sugimoto}, D. 1983, \aap, 126, 207

\bibitem[{{Li} {et~al.}(2022){Li}, {Qian}, \& {Liao}}]{2022AJ....163..203L}
{Li}, F.~X., {Qian}, S.~B., \& {Liao}, W.~P. 2022, \aj, 163, 203

\bibitem[{{Mac Low} {et~al.}(2005){Mac Low}, {Balsara}, {Kim}, \& {de
  Avillez}}]{2005ApJ...626..864M}
{Mac Low}, M.-M., {Balsara}, D.~S., {Kim}, J., \& {de Avillez}, M.~A. 2005,
  \apj, 626, 864

\bibitem[{{Marchant} {et~al.}(2016){Marchant}, {Langer}, {Podsiadlowski},
  {Tauris}, \& {Moriya}}]{mar1:16}
{Marchant}, P., {Langer}, N., {Podsiadlowski}, P., {Tauris}, T.~M., \&
  {Moriya}, T.~J. 2016, \aap, 588, A50

\bibitem[{{Martins}(2018)}]{2018A&A...616A.135M}
{Martins}, F. 2018, \aap, 616, A135

\bibitem[{{Massey} {et~al.}(2021){Massey}, {Neugent}, {Dorn-Wallenstein},
  {Eldridge}, {Stanway}, \& {Levesque}}]{2021ApJ...922..177M}
{Massey}, P., {Neugent}, K.~F., {Dorn-Wallenstein}, T.~Z., {et~al.} 2021, \apj,
  922, 177

\bibitem[{{Massey} {et~al.}(1989){Massey}, {Parker}, \&
  {Garmany}}]{1989AJ.....98.1305M}
{Massey}, P., {Parker}, J.~W., \& {Garmany}, C.~D. 1989, \aj, 98, 1305

\bibitem[{{Matzner}(2002)}]{2002ApJ...566..302M}
{Matzner}, C.~D. 2002, \apj, 566, 302

\bibitem[{{Miller-Jones} {et~al.}(2021){Miller-Jones}, {Bahramian}, {Orosz},
  {Mandel}, {Gou}, {Maccarone}, {Neijssel}, {Zhao}, {Zi{\'o}{\l}kowski},
  {Reid}, {Uttley}, {Zheng}, {Byun}, {Dodson}, {Grinberg}, {Jung}, {Kim},
  {Marcote}, {Markoff}, {Rioja}, {Rushton}, {Russell}, {Sivakoff}, {Tetarenko},
  {Tudose}, \& {Wilms}}]{2021Sci...371.1046M}
{Miller-Jones}, J. C.~A., {Bahramian}, A., {Orosz}, J.~A., {et~al.} 2021,
  Science, 371, 1046

\bibitem[{{Moe} \& {Di Stefano}(2017)}]{2017ApJS..230...15M}
{Moe}, M. \& {Di Stefano}, R. 2017, \apjs, 230, 15

\bibitem[{{Mokiem} {et~al.}(2007){Mokiem}, {de Koter}, {Evans}, {Puls},
  {Smartt}, {Crowther}, {Herrero}, {Langer}, {Lennon}, {Najarro}, {Villamariz},
  \& {Vink}}]{2007A&A...465.1003M}
{Mokiem}, M.~R., {de Koter}, A., {Evans}, C.~J., {et~al.} 2007, \aap, 465, 1003

\bibitem[{{Nascimbeni} {et~al.}(2016){Nascimbeni}, {Piotto}, {Ortolani},
  {Giuffrida}, {Marrese}, {Magrin}, {Ragazzoni}, {Pagano}, {Rauer}, {Cabrera},
  {Pollacco}, {Heras}, {Deleuil}, {Gizon}, \& {Granata}}]{2016MNRAS.463.4210N}
{Nascimbeni}, V., {Piotto}, G., {Ortolani}, S., {et~al.} 2016, \mnras, 463,
  4210

\bibitem[{{Naz{\'e}} {et~al.}(2002){Naz{\'e}}, {Hartwell}, {Stevens},
  {Corcoran}, {Chu}, {Koenigsberger}, {Moffat}, \&
  {Niemela}}]{2002ApJ...580..225N}
{Naz{\'e}}, Y., {Hartwell}, J.~M., {Stevens}, I.~R., {et~al.} 2002, \apj, 580,
  225

\bibitem[{{Nebot G{\'o}mez-Mor{\'a}n} \&
  {Oskinova}(2018)}]{2018A&A...620A..89N}
{Nebot G{\'o}mez-Mor{\'a}n}, A. \& {Oskinova}, L.~M. 2018, \aap, 620, A89

\bibitem[{{Nelson} \& {Eggleton}(2001)}]{2001ApJ...552..664N}
{Nelson}, C.~A. \& {Eggleton}, P.~P. 2001, \apj, 552, 664

\bibitem[{{Niemela} \& {Gamen}(2004)}]{2004NewAR..48..727N}
{Niemela}, V. \& {Gamen}, R. 2004, \nar, 48, 727

\bibitem[{{Niemela} {et~al.}(1986){Niemela}, {Marraco}, \&
  {Cabanne}}]{1986PASP...98.1133N}
{Niemela}, V.~S., {Marraco}, H.~G., \& {Cabanne}, M.~L. 1986, \pasp, 98, 1133

\bibitem[{{Nieuwenhuijzen} \& {de Jager}(1990)}]{nie1:90}
{Nieuwenhuijzen}, H. \& {de Jager}, C. 1990, \aap, 231, 134

\bibitem[{{Orosz} {et~al.}(2011){Orosz}, {McClintock}, {Aufdenberg},
  {Remillard}, {Reid}, {Narayan}, \& {Gou}}]{2011ApJ...742...84O}
{Orosz}, J.~A., {McClintock}, J.~E., {Aufdenberg}, J.~P., {et~al.} 2011, \apj,
  742, 84

\bibitem[{{Oskinova} {et~al.}(2011){Oskinova}, {Todt}, {Ignace}, {Brown},
  {Cassinelli}, \& {Hamann}}]{2011MNRAS.416.1456O}
{Oskinova}, L.~M., {Todt}, H., {Ignace}, R., {et~al.} 2011, \mnras, 416, 1456

\bibitem[{{Paczy{\'n}ski}(1967)}]{1967AcA....17..355P}
{Paczy{\'n}ski}, B. 1967, \actaa, 17, 355

\bibitem[{{Paczy{\'n}ski}(1971)}]{1971ARA&A...9..183P}
{Paczy{\'n}ski}, B. 1971, \araa, 9, 183

\bibitem[{{Pauldrach} {et~al.}(1986){Pauldrach}, {Puls}, \&
  {Kudritzki}}]{pau1:86}
{Pauldrach}, A., {Puls}, J., \& {Kudritzki}, R.~P. 1986, \aap, 164, 86

\bibitem[{{Pauli} {et~al.}(2022{\natexlab{a}}){Pauli}, {Langer},
  {Aguilera-Dena}, {Wang}, \& {Marchant}}]{2022A&A...667A..58P}
{Pauli}, D., {Langer}, N., {Aguilera-Dena}, D.~R., {Wang}, C., \& {Marchant},
  P. 2022{\natexlab{a}}, \aap, 667, A58

\bibitem[{{Pauli} {et~al.}(2023){Pauli}, {Oskinova}, {Hamann}, {Bowman},
  {Todt}, {Shenar}, {Sander}, {Erba}, {G{\'o}mez-Gonz{\'a}lez}, {Kehrig},
  {Klencki}, {Kuiper}, {Mehner}, {de Mink}, {Oey}, {Ramachandran},
  {Schootemeijer}, {Reyero Serantes}, \& {Wofford}}]{pauli2023}
{Pauli}, D., {Oskinova}, L.~M., {Hamann}, W.~R., {et~al.} 2023, arXiv e-prints,
  arXiv:2303.03989

\bibitem[{{Pauli} {et~al.}(2022{\natexlab{b}}){Pauli}, {Oskinova}, {Hamann},
  {Ramachandran}, {Todt}, {Sander}, {Shenar}, {Rickard}, {Ma{\'\i}z
  Apell{\'a}niz}, \& {Prinja}}]{2022A&A...659A...9P}
{Pauli}, D., {Oskinova}, L.~M., {Hamann}, W.~R., {et~al.} 2022{\natexlab{b}},
  \aap, 659, A9

\bibitem[{{Paxton} {et~al.}(2011){Paxton}, {Bildsten}, {Dotter}, {Herwig},
  {Lesaffre}, \& {Timmes}}]{Paxton2011}
{Paxton}, B., {Bildsten}, L., {Dotter}, A., {et~al.} 2011, \apjs, 192, 3

\bibitem[{{Paxton} {et~al.}(2013){Paxton}, {Cantiello}, {Arras}, {Bildsten},
  {Brown}, {Dotter}, {Mankovich}, {Montgomery}, {Stello}, {Timmes}, \&
  {Townsend}}]{Paxton2013}
{Paxton}, B., {Cantiello}, M., {Arras}, P., {et~al.} 2013, \apjs, 208, 4

\bibitem[{{Paxton} {et~al.}(2015){Paxton}, {Marchant}, {Schwab}, {Bauer},
  {Bildsten}, {Cantiello}, {Dessart}, {Farmer}, {Hu}, {Langer}, {Townsend},
  {Townsley}, \& {Timmes}}]{Paxton2015}
{Paxton}, B., {Marchant}, P., {Schwab}, J., {et~al.} 2015, \apjs, 220, 15

\bibitem[{{Paxton} {et~al.}(2018){Paxton}, {Schwab}, {Bauer}, {Bildsten},
  {Blinnikov}, {Duffell}, {Farmer}, {Goldberg}, {Marchant}, {Sorokina},
  {Thoul}, {Townsend}, \& {Timmes}}]{Paxton2018}
{Paxton}, B., {Schwab}, J., {Bauer}, E.~B., {et~al.} 2018, \apjs, 234, 34

\bibitem[{{Paxton} {et~al.}(2019){Paxton}, {Smolec}, {Schwab}, {Gautschy},
  {Bildsten}, {Cantiello}, {Dotter}, {Farmer}, {Goldberg}, {Jermyn}, {Kanbur},
  {Marchant}, {Thoul}, {Townsend}, {Wolf}, {Zhang}, \& {Timmes}}]{Paxton2019}
{Paxton}, B., {Smolec}, R., {Schwab}, J., {et~al.} 2019, \apjs, 243, 10

\bibitem[{{Peters}(1964)}]{peters1964}
{Peters}, P.~C. 1964, Physical Review, 136, 1224

\bibitem[{{Pignatari} {et~al.}(2010){Pignatari}, {Gallino}, {Heil}, {Wiescher},
  {K{\"a}ppeler}, {Herwig}, \& {Bisterzo}}]{2010ApJ...710.1557P}
{Pignatari}, M., {Gallino}, R., {Heil}, M., {et~al.} 2010, \apj, 710, 1557

\bibitem[{{Pols}(1994)}]{1994A&A...290..119P}
{Pols}, O.~R. 1994, \aap, 290, 119

\bibitem[{{Price-Whelan} {et~al.}(2017){Price-Whelan}, {Hogg},
  {Foreman-Mackey}, \& {Rix}}]{2017ApJ...837...20P}
{Price-Whelan}, A.~M., {Hogg}, D.~W., {Foreman-Mackey}, D., \& {Rix}, H.-W.
  2017, \apj, 837, 20

\bibitem[{{Pr{\v{s}}a} {et~al.}(2016){Pr{\v{s}}a}, {Conroy}, {Horvat}, {Pablo},
  {Kochoska}, {Bloemen}, {Giammarco}, {Hambleton}, \& {Degroote}}]{prs1:16}
{Pr{\v{s}}a}, A., {Conroy}, K.~E., {Horvat}, M., {et~al.} 2016, \apjs, 227, 29

\bibitem[{{Pr{\v{s}}a} \& {Zwitter}(2005)}]{prs1:05}
{Pr{\v{s}}a}, A. \& {Zwitter}, T. 2005, \apj, 628, 426

\bibitem[{{Ramachandran} {et~al.}(2018){Ramachandran}, {Hamann}, {Hainich},
  {Oskinova}, {Shenar}, {Sander}, {Todt}, \& {Gallagher}}]{2018A&A...615A..40R}
{Ramachandran}, V., {Hamann}, W.~R., {Hainich}, R., {et~al.} 2018, \aap, 615,
  A40

\bibitem[{{Ramachandran} {et~al.}(2019){Ramachandran}, {Hamann}, {Oskinova},
  {Gallagher}, {Hainich}, {Shenar}, {Sand er}, {Todt}, \&
  {Fulmer}}]{2019A&A...625A.104R}
{Ramachandran}, V., {Hamann}, W.~R., {Oskinova}, L.~M., {et~al.} 2019, \aap,
  625, A104

\bibitem[{{Rauw} \& {Naz{\'e}}(2016)}]{rau1:16}
{Rauw}, G. \& {Naz{\'e}}, Y. 2016, Advances in Space Research, 58, 761

\bibitem[{{Rickard} {et~al.}(2022){Rickard}, {Hainich}, {Hamann}, {Oskinova},
  {Prinja}, {Ramachandran}, {Pauli}, {Todt}, {Sander}, {Shenar}, {Chu}, \&
  {Gallagher}}]{2022RICKARDa}
{Rickard}, M.~J., {Hainich}, R., {Hamann}, W.~R., {et~al.} 2022, \aap, 666,
  A189

\bibitem[{{Rogers} \& {Pittard}(2013)}]{2013MNRAS.431.1337R}
{Rogers}, H. \& {Pittard}, J.~M. 2013, \mnras, 431, 1337

\bibitem[{{Sabbi} {et~al.}(2007){Sabbi}, {Sirianni}, {Nota}, {Tosi},
  {Gallagher}, {Meixner}, {Oey}, {Walterbos}, {Pasquali}, {Smith}, \&
  {Angeretti}}]{2007AJ....133...44S}
{Sabbi}, E., {Sirianni}, M., {Nota}, A., {et~al.} 2007, \aj, 133, 44

\bibitem[{{Sana} {et~al.}(2013){Sana}, {de Koter}, {de Mink}, {Dunstall},
  {Evans}, {H{\'e}nault-Brunet}, {Ma{\'\i}z Apell{\'a}niz},
  {Ram{\'\i}rez-Agudelo}, {Taylor}, {Walborn}, {Clark}, {Crowther}, {Herrero},
  {Gieles}, {Langer}, {Lennon}, \& {Vink}}]{2013A&A...550A.107S}
{Sana}, H., {de Koter}, A., {de Mink}, S.~E., {et~al.} 2013, \aap, 550, A107

\bibitem[{{Sana} {et~al.}(2012){Sana}, {de Mink}, {de Koter}, {Langer},
  {Evans}, {Gieles}, {Gosset}, {Izzard}, {Le Bouquin}, \&
  {Schneider}}]{2012Sci...337..444S}
{Sana}, H., {de Mink}, S.~E., {de Koter}, A., {et~al.} 2012, Science, 337, 444

\bibitem[{{Sana} {et~al.}(2014){Sana}, {Le Bouquin}, {Lacour}, {Berger},
  {Duvert}, {Gauchet}, {Norris}, {Olofsson}, {Pickel}, {Zins}, {Absil}, {de
  Koter}, {Kratter}, {Schnurr}, \& {Zinnecker}}]{2014ApJS..215...15S}
{Sana}, H., {Le Bouquin}, J.~B., {Lacour}, S., {et~al.} 2014, \apjs, 215, 15

\bibitem[{{Sander} {et~al.}(2015){Sander}, {Shenar}, {Hainich},
  {G{\'\i}menez-Garc{\'\i}a}, {Todt}, \& {Hamann}}]{2015A&A...577A..13S}
{Sander}, A., {Shenar}, T., {Hainich}, R., {et~al.} 2015, \aap, 577, A13

\bibitem[{{Schootemeijer} \& {Langer}(2018)}]{2018A&A...611A..75S}
{Schootemeijer}, A. \& {Langer}, N. 2018, \aap, 611, A75

\bibitem[{{Schootemeijer} {et~al.}(2019){Schootemeijer}, {Langer}, {Grin}, \&
  {Wang}}]{sch1:19}
{Schootemeijer}, A., {Langer}, N., {Grin}, N.~J., \& {Wang}, C. 2019, \aap,
  625, A132

\bibitem[{{Scott} {et~al.}(2015){Scott}, {Grevesse}, {Asplund}, {Sauval},
  {Lind}, {Takeda}, {Collet}, {Trampedach}, \& {Hayek}}]{sco1:15}
{Scott}, P., {Grevesse}, N., {Asplund}, M., {et~al.} 2015, \aap, 573, A25

\bibitem[{{Shenar} {et~al.}(2020{\natexlab{a}}){Shenar}, {Gilkis}, {Vink},
  {Sana}, \& {Sander}}]{2020A&A...634A..79S}
{Shenar}, T., {Gilkis}, A., {Vink}, J.~S., {Sana}, H., \& {Sander}, A.~A.~C.
  2020{\natexlab{a}}, \aap, 634, A79

\bibitem[{{Shenar} {et~al.}(2014){Shenar}, {Hamann}, \&
  {Todt}}]{2014A&A...562A.118S}
{Shenar}, T., {Hamann}, W.~R., \& {Todt}, H. 2014, \aap, 562, A118

\bibitem[{{Shenar} {et~al.}(2015){Shenar}, {Oskinova}, {Hamann}, {Corcoran},
  {Moffat}, {Pablo}, {Richardson}, {Waldron}, {Huenemoerder}, {Ma{\'\i}z
  Apell{\'a}niz}, {Nichols}, {Todt}, {Naz{\'e}}, {Hoffman}, {Pollock}, \&
  {Negueruela}}]{2015ApJ...809..135S}
{Shenar}, T., {Oskinova}, L., {Hamann}, W.~R., {et~al.} 2015, \apj, 809, 135

\bibitem[{{Shenar} {et~al.}(2019){Shenar}, {Sablowski}, {Hainich}, {Todt},
  {Moffat}, {Oskinova}, {Ramachandran}, {Sana}, {Sander}, {Schnurr},
  {St-Louis}, {Vanbeveren}, {G{\"o}tberg}, \& {Hamann}}]{she1:19}
{Shenar}, T., {Sablowski}, D.~P., {Hainich}, R., {et~al.} 2019, \aap, 627, A151

\bibitem[{{Shenar} {et~al.}(2020{\natexlab{b}}){Shenar}, {Sablowski},
  {Hainich}, {Todt}, {Moffat}, {Oskinova}, {Ramachandran}, {Sana}, {Sander},
  {Schnurr}, {St-Louis}, {Vanbeveren}, {G{\"o}tberg}, \& {Hamann}}]{she2:19}
{Shenar}, T., {Sablowski}, D.~P., {Hainich}, R., {et~al.} 2020{\natexlab{b}},
  \aap, 641, C2

\bibitem[{{Sim{\'o}n-D{\'\i}az} \& {Herrero}(2014)}]{2014A&A...562A.135S}
{Sim{\'o}n-D{\'\i}az}, S. \& {Herrero}, A. 2014, \aap, 562, A135

\bibitem[{{Smith}(2014)}]{2014ARA&A..52..487S}
{Smith}, N. 2014, \araa, 52, 487

\bibitem[{{Sota} {et~al.}(2014){Sota}, {Ma{\'\i}z Apell{\'a}niz}, {Morrell},
  {Barb{\'a}}, {Walborn}, {Gamen}, {Arias}, \& {Alfaro}}]{2014ApJS..211...10S}
{Sota}, A., {Ma{\'\i}z Apell{\'a}niz}, J., {Morrell}, N.~I., {et~al.} 2014,
  \apjs, 211, 10

\bibitem[{{Sota} {et~al.}(2011){Sota}, {Ma{\'\i}z Apell{\'a}niz}, {Walborn},
  {Alfaro}, {Barb{\'a}}, {Morrell}, {Gamen}, \& {Arias}}]{2011ApJS..193...24S}
{Sota}, A., {Ma{\'\i}z Apell{\'a}niz}, J., {Walborn}, N.~R., {et~al.} 2011,
  \apjs, 193, 24

\bibitem[{{Spruit}(2002)}]{spr1:02}
{Spruit}, H.~C. 2002, \aap, 381, 923

\bibitem[{{Thielemann} {et~al.}(2011){Thielemann}, {Arcones}, {K{\"a}ppeli},
  {Liebend{\"o}rfer}, {Rauscher}, {Winteler}, {Fr{\"o}hlich}, {Dillmann},
  {Fischer}, {Martinez-Pinedo}, {Langanke}, {Farouqi}, {Kratz}, {Panov}, \&
  {Korneev}}]{2011PrPNP..66..346T}
{Thielemann}, F.~K., {Arcones}, A., {K{\"a}ppeli}, R., {et~al.} 2011, Progress
  in Particle and Nuclear Physics, 66, 346

\bibitem[{{Trundle} {et~al.}(2007){Trundle}, {Dufton}, {Hunter}, {Evans},
  {Lennon}, {Smartt}, \& {Ryans}}]{tru1:07}
{Trundle}, C., {Dufton}, P.~L., {Hunter}, I., {et~al.} 2007, \aap, 471, 625

\bibitem[{{Udalski} {et~al.}(2015){Udalski}, {Szyma{\'n}ski}, \&
  {Szyma{\'n}ski}}]{2015AcA....65....1U}
{Udalski}, A., {Szyma{\'n}ski}, M.~K., \& {Szyma{\'n}ski}, G. 2015, \actaa, 65,
  1

\bibitem[{{Vanbeveren} \& {Conti}(1980)}]{1980A&A....88..230V}
{Vanbeveren}, D. \& {Conti}, P.~S. 1980, \aap, 88, 230

\bibitem[{{Vanbeveren} {et~al.}(1998){Vanbeveren}, {De Loore}, \& {Van
  Rensbergen}}]{1998A&ARv...9...63V}
{Vanbeveren}, D., {De Loore}, C., \& {Van Rensbergen}, W. 1998, \aapr, 9, 63

\bibitem[{{Venn}(1999)}]{1999ApJ...518..405V}
{Venn}, K.~A. 1999, \apj, 518, 405

\bibitem[{{Vida{\~n}a}(2018)}]{2018EPJP..133..445V}
{Vida{\~n}a}, I. 2018, European Physical Journal Plus, 133, 445

\bibitem[{{Vink} {et~al.}(2001){Vink}, {de Koter}, \&
  {Lamers}}]{2001A&A...369..574V}
{Vink}, J.~S., {de Koter}, A., \& {Lamers}, H.~J.~G.~L.~M. 2001, \aap, 369, 574

\bibitem[{{Walborn}(1978)}]{1978ApJ...224L.133W}
{Walborn}, N.~R. 1978, \apjl, 224, L133

\bibitem[{{Walborn} {et~al.}(2002){Walborn}, {Howarth}, {Lennon}, {Massey},
  {Oey}, {Moffat}, {Skalkowski}, {Morrell}, {Drissen}, \&
  {Parker}}]{2002AJ....123.2754W}
{Walborn}, N.~R., {Howarth}, I.~D., {Lennon}, D.~J., {et~al.} 2002, \aj, 123,
  2754

\bibitem[{{Walborn} {et~al.}(2000){Walborn}, {Lennon}, {Heap}, {Lindler},
  {Smith}, {Evans}, \& {Parker}}]{2000PASP..112.1243W}
{Walborn}, N.~R., {Lennon}, D.~J., {Heap}, S.~R., {et~al.} 2000, \pasp, 112,
  1243

\bibitem[{{Weisskopf} {et~al.}(2000){Weisskopf}, {Tananbaum}, {Van Speybroeck},
  \& {O'Dell}}]{cxo2000}
{Weisskopf}, M.~C., {Tananbaum}, H.~D., {Van Speybroeck}, L.~P., \& {O'Dell},
  S.~L. 2000, in Society of Photo-Optical Instrumentation Engineers (SPIE)
  Conference Series, Vol. 4012, X-Ray Optics, Instruments, and Missions III,
  ed. J.~E. {Truemper} \& B.~{Aschenbach}, 2--16

\bibitem[{{Wellstein} \& {Langer}(1999)}]{1999A&A...350..148W}
{Wellstein}, S. \& {Langer}, N. 1999, \aap, 350, 148

\bibitem[{{Wilson}(1941)}]{1941ApJ....93...29W}
{Wilson}, O.~C. 1941, \apj, 93, 29

\bibitem[{{Wilson}(1990)}]{wil1:90}
{Wilson}, R.~E. 1990, \apj, 356, 613

\bibitem[{{Zahn} {et~al.}(2010){Zahn}, {Ranc}, \& {Morel}}]{zahn2010}
{Zahn}, J.~P., {Ranc}, C., \& {Morel}, P. 2010, \aap, 517, A7

\end{thebibliography}

%-join the .bib files when you upload your source files
%-------------------------------------------------------------------

\begin{appendix}
\onecolumn
\section{Additional tables}

    \begin{table}[h]
        \footnotesize
        \center
        \caption{Radial velocities of different lines associated with the primary.}
        \begin{tabular}{ c | c c c c c c | c } 
            \hline
            \hline
            \rule{0cm}{2.8ex}
            \centering
                    MJD & {\NIVd{3478{\,--\,}3485}} & {\NIV{4058}}             & {\NIV{6381}}            & {\NIVd{7103{\,--\,}7129}} & {\NVd{4604{,\,}4620}}    & {\HeII{6683}$^{(a)}$} & {mean RV$^{(b)}$}\\
                        &{$[\si{km\,s^{-1}}]$}    &{$[\si{km\,s^{-1}}]$} &{$[\si{km\,s^{-1}}]$} &{$[\si{km\,s^{-1}}]$} &{$[\si{km\,s^{-1}}]$} &{$[\si{km\,s^{-1}}]$} &{$[\si{km\,s^{-1}}]$}\\
                    \hline \rule{0cm}{2.8ex}%
                    \rule{0cm}{2.4ex} 52249.01198 & $\,\,\,\,\,\,7 \pm \,\,\,6$ & $\,\,\,18 \pm 7$   & $\,\,\,\,\,\,9 \pm 15$           &     \,\,\,---                      &     \,\,\,---                      &     \,\,\,---                      &  $\,\,\,11 \pm 10$       \\
                    \rule{0cm}{2.4ex} 53277.25739 &     \,\,\,---               & $\,\,\,56 \pm 7$   &     \,\,\,---                    &     \,\,\,---                      &     \,\,\,---                      &     \,\,\,---                      &  $\,\,\,56 \pm \,\,\,7$  \\
                    \rule{0cm}{2.4ex} 53277.25939 &     \,\,\,---               & $\,\,\,51 \pm 7$   &     \,\,\,---                    &     \,\,\,---                      &     \,\,\,---                      &     \,\,\,---                      &  $\,\,\,51 \pm \,\,\,7$  \\
                    \rule{0cm}{2.4ex} 53277.26022 &     \,\,\,---               & $\,\,\,42 \pm 6$   &     \,\,\,---                    &     \,\,\,---                      &     \,\,\,---                      &     \,\,\,---                      &  $\,\,\,42 \pm \,\,\,6$  \\
                    \rule{0cm}{2.4ex} 53277.28112 &     \,\,\,---              & $\,\,\,48 \pm 6$   &     \,\,\,---                    &     \,\,\,---                      &     \,\,\,---                      &     \,\,\,---                      &  $\,\,\,48 \pm \,\,\,6$  \\
                    \rule{0cm}{2.4ex} 53280.19821 &     \,\,\,---              &     \,\,\,---      &     \,\,\,---                    &     \,\,\,---                      & $\,\,\,18 \pm 15$                &     \,\,\,---                      &  $\,\,\,18 \pm 15$       \\
                    \rule{0cm}{2.4ex} 53292.06195 &     \,\,\,---              &     \,\,\,---      &     \,\,\,---                    &     \,\,\,---                      & $-57 \pm 15$                     &     \,\,\,---                      & $-57 \pm 15$                 \\
                    \rule{0cm}{2.4ex} 57611.32930 &     \,\,\,---              &     \,\,\,---      & $\,\,\,97 \pm 15$                & $100 \pm 15$                       &     \,\,\,---                      & $\,\,\,96 \pm 13$                  &  $\,\,\,98 \pm 14$       \\
                    \rule{0cm}{2.4ex} 57613.24630 &     \,\,\,---              &     \,\,\,---      & $-74 \pm 15$                     & $-75 \pm 15$                       &     \,\,\,---                      & $-86 \pm 14$                       & $-78 \pm 15$                 \\
                    \rule{0cm}{2.4ex} 57613.28829 &     \,\,\,---              &     \,\,\,---      & $-82 \pm 15$                     & $-75 \pm 15$                       &     \,\,\,---                      & $-73 \pm 13$                       & $-77 \pm 14$                 \\
                    \rule{0cm}{2.4ex} 57617.23463 &     \,\,\,---              &     \,\,\,---      & $\,\,\,54 \pm 15$                & $\,\,\,61 \pm 14$                  &     \,\,\,---                      & $\,\,\,51 \pm \,\,\,8$             &  $\,\,\,55 \pm 13$       \\
                    \rule{0cm}{2.4ex} 57618.10327 &     \,\,\,---              &     \,\,\,---      & $\,\,\,84 \pm 15$                & $\,\,\,61 \pm 15$                  &     \,\,\,---                      & $\,\,\,70 \pm \,\,\,8$             &  $\,\,\,72 \pm 13$       \\
                    \rule{0cm}{2.4ex} 57620.14865 &     \,\,\,---              &     \,\,\,---      & $\,\,\,39 \pm 15$                & $\,\,\,38 \pm 14$                  &     \,\,\,---                      & $\,\,\,31 \pm \,\,\,8$             &  $\,\,\,36 \pm 13$       \\
                    \rule{0cm}{2.4ex} 57620.19042 &     \,\,\,---              &     \,\,\,---      & $\,\,\,40 \pm 15$                & $\,\,\,45 \pm 14$                  &     \,\,\,---                      & $\,\,\,32 \pm \,\,\,8$             &  $\,\,\,39 \pm 13$       \\
                    \rule{0cm}{2.4ex} 57620.23262 &     \,\,\,---              &     \,\,\,---      & $\,\,\,45 \pm 15$                & $\,\,\,46 \pm \,\,\,8$             &     \,\,\,---                      & $\,\,\,41 \pm \,\,\,8$             &  $\,\,\,43 \pm 11$       \\
                    \rule{0cm}{2.4ex} 57621.20640 &     \,\,\,---              &     \,\,\,---      & $\,\,\,62 \pm 15$                & $\,\,\,68 \pm 15$                  &     \,\,\,---                      & $\,\,\,62 \pm \,\,\,8$             &  $\,\,\,64 \pm 13$       \\
                    \rule{0cm}{2.4ex} 57621.24834 &     \,\,\,---             &     \,\,\,---      & $\,\,\,55 \pm 15$                & $\,\,\,45 \pm 10$                  &     \,\,\,---                       & $\,\,\,48 \pm \,\,\,8$             &  $\,\,\,49 \pm 11$       \\
                    \rule{0cm}{2.4ex} 57622.14718 &     \,\,\,---             &     \,\,\,---      & $-81 \pm 15$                     & $-80 \pm 15$                       &     \,\,\,---                       & $-94 \pm \,\,\,9$                  & $-85 \pm 13$             \\
                    \rule{0cm}{2.4ex} 59216.11133 & $\,\,\,98 \pm \,\,\,6$    & $\,\,\,97 \pm 6$   &  $\,\,\,82 \pm 22$               & $\,\,\,93 \pm 22$                  & $\,\,\,92 \pm 16$                & $\,\,\,95 \pm \,\,\,8$             &  $\,\,\,92 \pm 15$       \\
                    \hline
            \end{tabular}
                \rule{0cm}{2.8ex}%
                \tablefoot{
                        \ignorespaces
                        $^{(a)}$ The line shows a clear but distinguishable contribution of the secondary. It can be used as a cross-check for the quality and self-consistency of the secondary's RV fit.}
        \label{table:RVs_prim}
    \end{table}

    \begin{table}
        \footnotesize
        \center
        \caption{Radial velocities of different lines associated with the secondary.}
        \begin{tabular}{ c | c c c c | c } 
            \hline
            \hline
            \rule{0cm}{2.8ex}
            \centering
                    MJD & {\HeI{4471}} & {\HeI{4713}}             & {\HeII{6683}$^{(a)}$}            & {\CIVd{5801{,\,}5812}$^{(a)}$} & {mean RV$^{(b)}$}\\
                        &{$[\si{km\,s^{-1}}]$}    &{$[\si{km\,s^{-1}}]$} &{$[\si{km\,s^{-1}}]$} &{$[\si{km\,s^{-1}}]$} &{$[\si{km\,s^{-1}}]$}\\
                    \hline \rule{0cm}{2.8ex}%
                    \rule{0cm}{2.4ex} 52249.01198 & $\,\,\,18 \pm \,\,\,9$ & $\,\,\,\,\,\,6 \pm 31$ & \,\,\,---              & $\,\,\,\,\,\,7 \pm 16$ &  $10 \pm 21$      \\
                    \rule{0cm}{2.4ex} 53277.25739 & $-34 \pm \,\,\,7$      & \,\,\,---              & \,\,\,---              & \,\,\,---              & $-34 \pm \,\,\,7$ \\
                    \rule{0cm}{2.4ex} 53277.25939 & $-32 \pm \,\,\,9$      & \,\,\,---              & \,\,\,---              & \,\,\,---              & $-32 \pm \,\,\,9$ \\
                    \rule{0cm}{2.4ex} 53277.26022 & $-39 \pm \,\,\,7$      & \,\,\,---              & \,\,\,---              & \,\,\,---              & $-39 \pm \,\,\,7$ \\
                    \rule{0cm}{2.4ex} 53277.28112 & $-39 \pm \,\,\,7$      & \,\,\,---              & \,\,\,---              & \,\,\,---              & $-39 \pm \,\,\,7$ \\
                    \rule{0cm}{2.4ex} 53280.19821 & \,\,\,---              & $-12 \pm 29$           & \,\,\,---              & \,\,\,---              & $-12 \pm 29$      \\
                    \rule{0cm}{2.4ex} 53292.06195 & \,\,\,---              & $\,\,\,41 \pm 27$      & \,\,\,---              & \,\,\,---              & $41 \pm 27$       \\
                    \rule{0cm}{2.4ex} 57611.32930 & \,\,\,---              & $-48 \pm 22$           & $-62 \pm 20$           & $-60 \pm 13$           & $-57 \pm 19$ \\
                    \rule{0cm}{2.4ex} 57613.24630 & \,\,\,---              & $\,\,\,57 \pm 28$      & $\,\,\,69 \pm 25$      & $\,\,\,77 \pm 14$      &  $67 \pm 23$ \\
                    \rule{0cm}{2.4ex} 57613.28829 & \,\,\,---              & $\,\,\,71 \pm 21$      & $\,\,\,71 \pm 19$      & $\,\,\,77 \pm 23$      &  $73 \pm 21$ \\
                    \rule{0cm}{2.4ex} 57617.23463 & \,\,\,---              & $-19 \pm 20$           & $-24 \pm 23$           & $-22 \pm 14$           & $-22 \pm 19$ \\
                    \rule{0cm}{2.4ex} 57618.10327 & \,\,\,---              & $-30 \pm 25$           & $-35 \pm 23$           & $-38 \pm 17$           & $-34 \pm 22$ \\
                    \rule{0cm}{2.4ex} 57620.14865 & \,\,\,---              & $-34 \pm 25$           & $-32 \pm 23$           & $-35 \pm 16$           & $-34 \pm 22$ \\
                    \rule{0cm}{2.4ex} 57620.19042 & \,\,\,---              & $-32 \pm 23$           & $-40 \pm 23$           & $-29 \pm 14$           & $-34 \pm 22$ \\
                    \rule{0cm}{2.4ex} 57620.23262 & \,\,\,---              & $-32 \pm 18$           & $-31 \pm 31$           & $-34 \pm 14$           & $-32 \pm 22$ \\
                    \rule{0cm}{2.4ex} 57621.20640 & \,\,\,---              & $-21 \pm 26$           & $-29 \pm 22$           & $-22 \pm 16$           & $-24 \pm 22$ \\
                    \rule{0cm}{2.4ex} 57621.24834 & \,\,\,---              & $-28 \pm 25$           & $-30 \pm 20$           & $-31 \pm 16$           & $-30 \pm 20$ \\
                    \rule{0cm}{2.4ex} 57622.14718 & \,\,\,---              & $\,\,\,71 \pm 22$      & $\,\,\,70 \pm 25$      & $\,\,\,75 \pm 16$      &  $72 \pm 21$ \\
                    \rule{0cm}{2.4ex} 59216.11133 & $-55 \pm \,\,\,8$      & $-57 \pm 25$           & $-55 \pm 25$           & $-52 \pm 11$           & $-55 \pm 19$ \\
                    \hline
            \end{tabular}
                \rule{0cm}{2.8ex}%
                \tablefoot{
                        \ignorespaces
                        $^{(a)}$ The line shows a clear but distinguishable contribution of the primary.}
            \label{table:RVs_sec}
    \end{table}
    
    \begin{table}
        \small
        \center
        \caption{List of used lines. Includes which component the lines are attributed to based on relative position to other lines and shifts at different epochs.}
        \begin{tabular}{ c c c } 
            \hline\hline\rule{0cm}{2.8ex}
            \rule{0.3cm}{0cm}Line\rule{0.3cm}{0cm}& Wavelength ($\AA$) &Components source and description\\
            \hline\rule{0cm}{2.8ex}
            \rule{0cm}{2.4ex}O\,\textsc{vi} & 1031.9, 1037.6 & Absorption, predominately from primary\\
            \rule{0cm}{2.4ex}P\,\textsc{v} & 1118.0 & Both components show absorption \\
            \rule{0cm}{2.4ex}P\,\textsc{v} & 1128.0 & Both components show absorption \\
            \rule{0cm}{2.4ex}C\,\textsc{iv} & 1169.0 & Both components show absorption \\
            \rule{0cm}{2.4ex}C\,\textsc{iii}& 1175.3 & Secondary absorption\\
            \rule{0cm}{2.4ex}N\,\textsc{v} & 1238.8, 1242.8 doublet & Resonance wind lines with absorption and emission for both\\
            \rule{0cm}{2.4ex}O\,\textsc{iv} & 1338.6, 1343.0, 1343.5 triplet &Both components show absorption \\
            \rule{0cm}{2.4ex}O\,\textsc{v} & 1371.0 & Both components show absorption \\
            \rule{0cm}{2.4ex}O\,\textsc{iii} & 1408.6, 1409.8, 1410.8, 1410.9, 1412.1 & Both components show absorption \\
            \rule{0cm}{2.4ex}Fe\,\textsc{v-vi} & Range $\sim 1412 - 1475$ & Iron forest lines, both components show absorption\\
            \rule{0cm}{2.4ex}N\,\textsc{iv} & 1483.3, 1486.5 & Both components show absorption\\
            \rule{0cm}{2.4ex}Si\,\textsc{v} & 1502.0 & Both components show absorption \\
            \rule{0cm}{2.4ex}C\,\textsc{iv} & 1548.2, 1550.8 doublet & Resonant wind line showing contribution of both components \\
            \rule{0cm}{2.4ex}He\,\textsc{ii} & 1640.5 & Primary P-Cygni with emission, secondary absorption \\
            \ion{N}{iv} & 3478.7, 3483.0, 3484.9 triplet & Both components show absorption \\
            \rule{0cm}{2.4ex}He\,\textsc{i} & 3964.7 & Primary absorption\\
            \rule{0cm}{2.4ex}He\,\textsc{ii} & 3970.1 & Both components show absorption \\
            
            \rule{0cm}{2.4ex}He\,\textsc{ii} & 4025.6& Overlapping with \HeI{4026.2} for both primary and secondary\\
            \rule{0cm}{2.4ex}He\,\textsc{i} & 4026.2 & Overlapping with \HeII{4025.6} for both primary and secondary\\
            \rule{0cm}{2.4ex}N\,\textsc{iv} & 4057.8 & Primary emission \\

            \rule{0cm}{2.4ex}He\,\textsc{ii} &4100.7 & Both components show absorption \\
            \rule{0cm}{2.4ex}He\,\textsc{ii} &4199.9 & Both components show absorption\\
            \rule{0cm}{2.4ex}He\,\textsc{ii} &4338.7 & Both components show absorption\\
            \rule{0cm}{2.4ex}He\,\textsc{i} & 4471.5 & Secondary absorption \\
            
            \rule{0cm}{2.4ex}He\,\textsc{ii} &4541.6 & Both components show absorption\\
            \rule{0cm}{2.4ex}N\,\textsc{v} & 4603.8, 4619.9 doublet & Absorption, predominately from primary\\
            \rule{0cm}{2.4ex}N\,\textsc{iii} &4634.1, 4640.6 doublet & Both components show emission \\
            \rule{0cm}{2.4ex}O\,\textsc{ii} & 4650.0 & Primary emission, overlapped by primary He\,\textsc{ii} $\,\lambda\,4686\AA$ emission \\
            \rule{0cm}{2.4ex}He\,\textsc{ii} &4685.8 & Primary emission, secondary absorption\\
            \rule{0cm}{2.4ex}He\,\textsc{i} &4713.2 & Secondary absorption\\
            \rule{0cm}{2.4ex}He\,\textsc{ii} & 4859.3 & Both components show absorption \\
            \rule{0cm}{2.4ex}He\,\textsc{i} & 4921.9 & Secondary absorption\\
            \rule{0cm}{2.4ex}He\,\textsc{i} &5015.7 & Absorption, predominately secondary \\
            \rule{0cm}{2.4ex}He\,\textsc{ii} &5411.5 & Both components show absorption\\
            \rule{0cm}{2.4ex}C\,\textsc{iv} &5801.3, 5812.0 doublet & Primary emission, secondary absorption \\
            \rule{0cm}{2.4ex}He\,\textsc{i} &5875.4, 5875.5, 5875.8 triplet & Both components show absorption\\
            \rule{0cm}{2.4ex}He\,\textsc{i} &6170.7 & Both components show absorption\\
            \rule{0cm}{2.4ex}N\,\textsc{iv} & 6212.4, 6215.5, 6219.9 triplet & No observations. Model of primary shows emission\\
            \rule{0cm}{2.4ex}He\,\textsc{i} &6233.8 & Both components show absorption\\
            \rule{0cm}{2.4ex}N\,\textsc{iv} & 6380.8 & Primary absorption\\
            \rule{0cm}{2.4ex}He\,\textsc{ii} &6527.1 & Both components show absorption\\
            \rule{0cm}{2.4ex}He\,\textsc{ii} &6560.1 & Primary emission, secondary absorption \\
            \rule{0cm}{2.4ex}He\,\textsc{ii} &6678.1, 6679.6 doublet &Secondary absorption\\
            \rule{0cm}{2.4ex}He\,\textsc{ii} &6683.2 & Both components show absorption\\
            \rule{0cm}{2.4ex}He\,\textsc{ii} &7065.3 & Absorption, predominately Secondary \\
            \rule{0cm}{2.4ex}N\,\textsc{iv} & ${7102{-}7129}$ multiplet & Emission, predominately from primary\vspace{0.15cm}\\
            \hline
        \end{tabular}
        \label{table:lines_considered}
    \end{table}

\clearpage

\twocolumn

\section{Light curve modelling with PHOEBE}
\label{app:light_curve}

    Although it is  in the survey region of the Optical Gravitational Lensing Experiment \citep[OGLE, ][]{2015AcA....65....1U} catalogue, to date no light curve of this object has been reported.    
    In addition, the resolution of the Transiting Exoplanet Survey Satellite (TESS) is $\SI{21}{\arcsecond\,pixel^{-1}}$, not fine enough to disentangle the light curve of our target from the other O stars in NGC\,346. To demonstrate if detectable photometric variations based on our results are expected at all, we employed the PHOEBE code (used in Sect.~\ref{sec:PHOEBE}) to create a synthetic light curve.

\begin{figure}[t]
        \centering
        \includegraphics[width=\hsize]{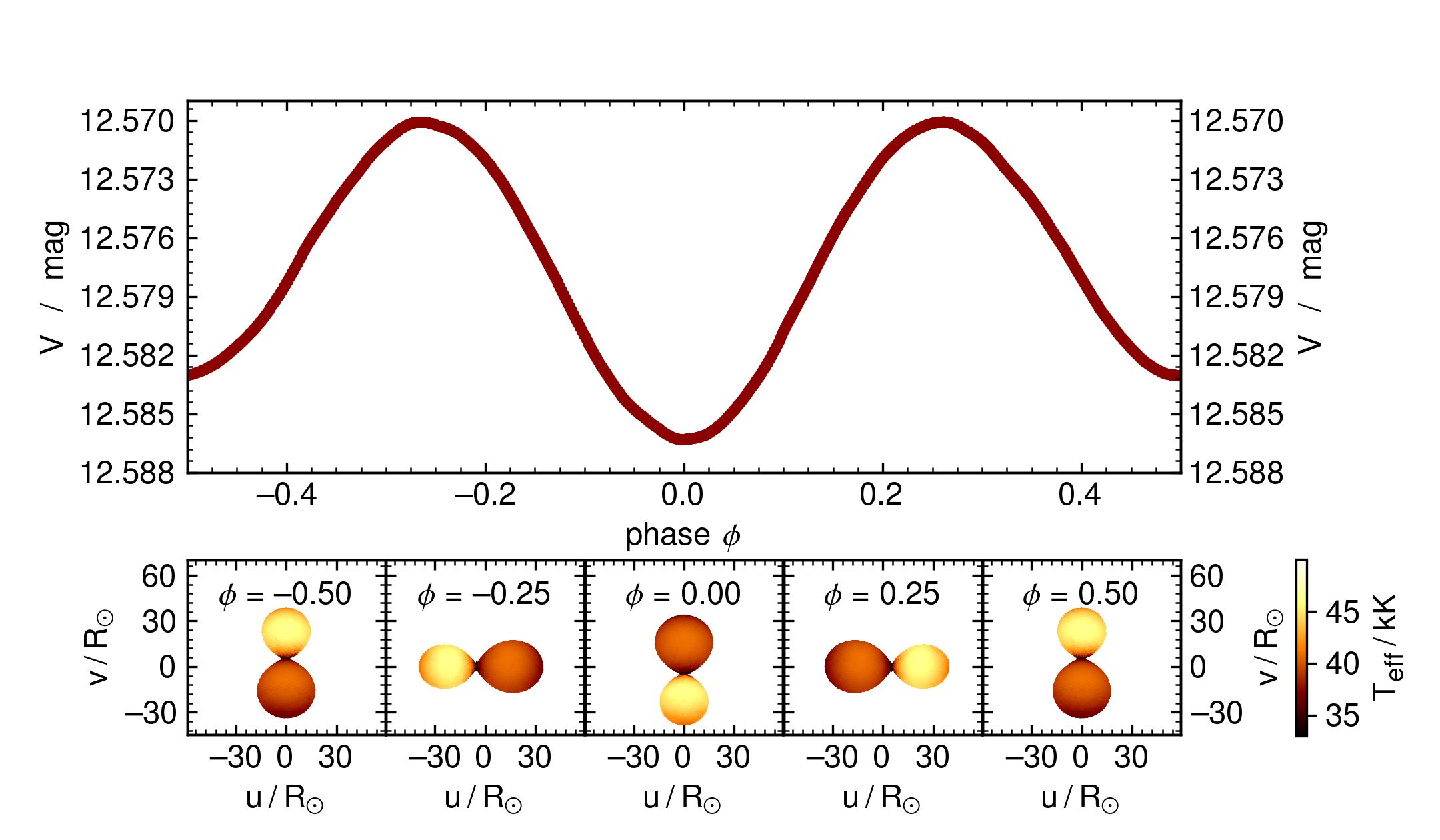}
        \caption{Synthetic predicted light curves of SSN\,7 for an inclination angle of $i=\SI{16}{^\circ}$ and the orbital configuration of the binary components in the plane of sky at different phases. \textit{Upper panel:} Phased synthetic light curve in the Johnson V band. \textit{Lower panel:} Orbital configuration of the binary components at phases $\phi=-0.5$, $-0.25$, $0.0$, $0.25$, and $0.5$. The models are colour-coded by   effective temperature, taking into account the physical effect mentioned in the text.}
        \label{fig:pred_lc}
    \end{figure}

    Typically light curves are observed in the optical. Hence, we decided to create the synthetic light curve (shown in Figure ~\ref{fig:pred_lc}) in the Johnson V band. In the PHOEBE model we used the stellar parameters of the binary components obtained from our spectroscopic analysis (see Sect.~\ref{sec:analysis_PoWR}) as input parameters. Furthermore, we assume that gravitational darkening can be approximated by a power law with a coefficient of $\beta_\mathrm{grav} = 1$, as typically used for hot stars with radiative envelopes. The atmospheres of both components are approximated by a black body, when compared to the SED of our atmospheric model it is a valid approximation for the V-band flux. Limb darkening is modelled by using a quadratic approximation and using coefficients ${a_1=0.2694}$ and ${b_1=-0.003}$ for the primary and ${a_2=0.1751}$ and ${b_2=0.0871}$ for the secondary, obtained from the emergent flux distribution of our stellar atmosphere models. The effect of reflections and resulting heating in the binary is modelled using the build in the Horvat scheme that includes Lambert scattering as well as the reflection scheme of \citet{wil1:90}. It is assumed that $90\%$ of the incident flux will be reflected.
    
    For a better understanding on how the different photometric variations are linked to the orbital configuration of the binary, we depict the orbital configurations of the stars in the lower part of Fig.~\ref{fig:pred_lc}. It is evident that the expected maximum photometric variation of the binary components is $\SI{18}{mmag}$. The shape of the light curve is only caused by ellipsoidal variability and not by eclipses.

    Given the expected low variability, we conclude that only the upcoming powerful telescopes, such as the ELT, will be able to detect this kind of variability. Hence, we can say that our derived low inclination angle is consistent with the lack of published light curve data.

    \begin{minipage}[h][5cm]{\textwidth}
    \end{minipage}
    
\end{appendix}

\end{document}